\documentclass[prx,twocolumn,amsmath,amssymb,superscriptaddress,floatfix]{revtex4-2}
\pdfoutput=1
\usepackage{bm,psfrag,graphicx,setspace}
\usepackage[T1]{fontenc}
\usepackage[left]{lineno}
\usepackage{xcolor}
\usepackage{braket}
\usepackage[draft]{changes}
\usepackage{bbold}
\usepackage{quantikz}
\usepackage{hyperref}
\usepackage{pdfpages} 
\usepackage{pgffor} 

\makeatletter
\AtBeginDocument{\let\LS@rot\@undefined}
\makeatother

\def\supplementfilename{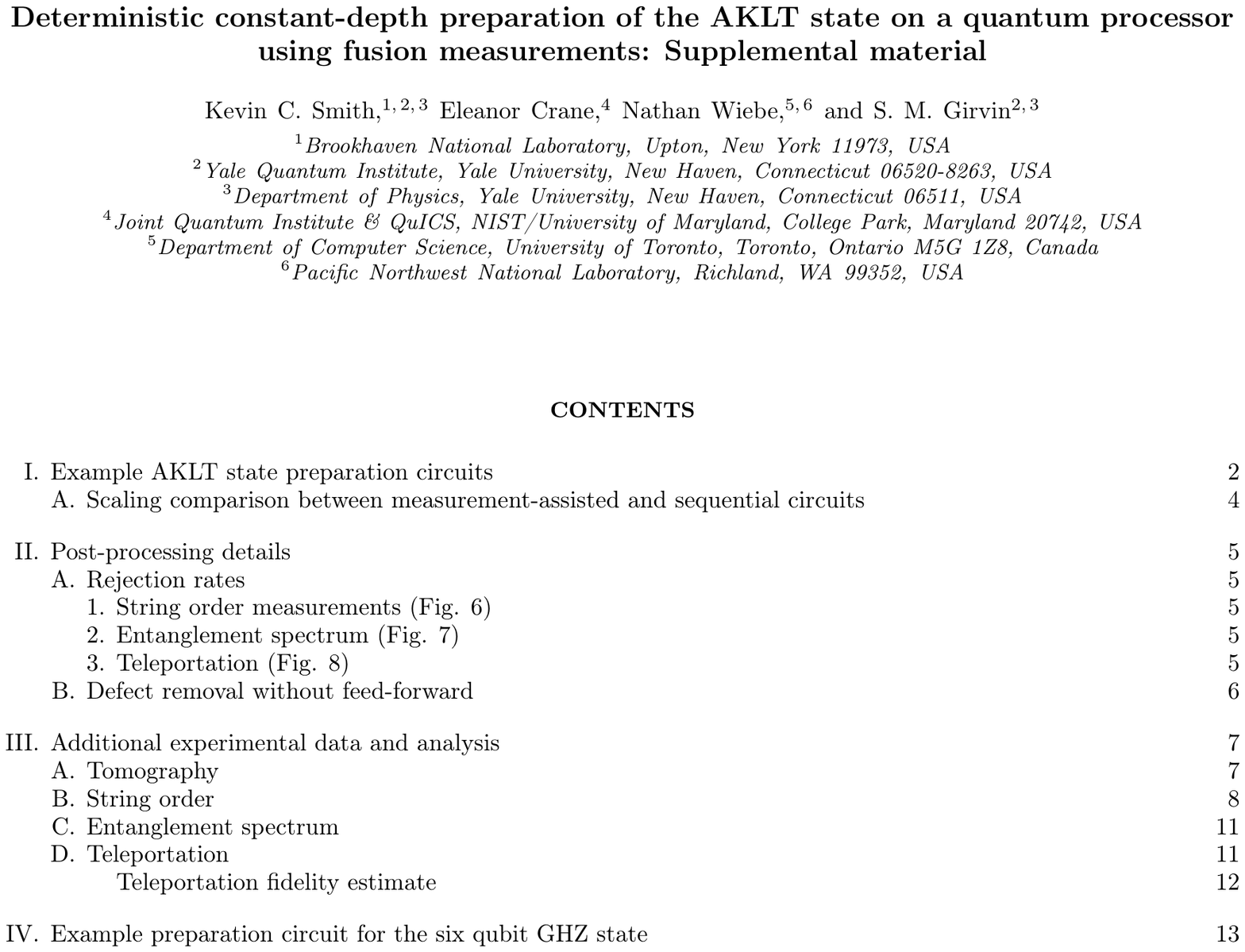}

\pdfximage{\supplementfilename}
\def\numbersupplementpages{\the\pdflastximagepages}

\newif\ifarXiv
\arXivtrue 

\definecolor{tealgreen}{HTML}{00856b}
\hypersetup{colorlinks=true, allcolors=tealgreen}
\selectfont

\begin{document}
\pagenumbering{arabic}
\title{Deterministic constant-depth preparation of the AKLT state on a quantum processor using fusion measurements}
\author{Kevin C. Smith}
\affiliation{Brookhaven National Laboratory, Upton, New York 11973, USA}
\affiliation{Yale Quantum Institute, Yale University, New Haven, Connecticut 06520-8263, USA}
\affiliation{Department of Physics, Yale University, New Haven, Connecticut 06511, USA}
\email{kevin.smith@yale.edu}
\author{Eleanor Crane}
\affiliation{Joint Quantum Institute \& QuICS, NIST/University of Maryland, College Park, Maryland 20742, USA}
\author{Nathan Wiebe}
\affiliation{Department of Computer Science, University of Toronto, Toronto, Ontario M5G 1Z8, Canada}
\affiliation{Pacific Northwest National Laboratory, Richland, WA 99352, USA}
\author{S. M. Girvin}
\affiliation{Yale Quantum Institute, Yale University, New Haven, Connecticut 06520-8263, USA}
\affiliation{Department of Physics, Yale University, New Haven, Connecticut 06511, USA}

\begin{abstract}
The ground state of the spin-1 Affleck, Kennedy, Lieb and Tasaki (AKLT) model is a paradigmatic example of both a matrix product state and a symmetry-protected topological phase, and additionally holds promise as a resource state for measurement-based quantum computation. Having a nonzero correlation length, the AKLT state cannot be exactly prepared by a constant-depth unitary circuit composed of local gates. In this work, we demonstrate that this no-go limit can be evaded by augmenting a constant-depth circuit with fusion measurements, such that the total preparation time is independent of system size and entirely deterministic. We elucidate our preparation scheme using the language of tensor networks, and furthermore show that the $\mathbb{Z}_2\times\mathbb{Z}_2$ symmetry of the AKLT state directly affords this speed-up over previously known preparation methods. To demonstrate the practical advantage of measurement-assisted preparation on noisy intermediate-scale quantum (NISQ) devices, we carry out our protocol on an IBM Quantum processor. We measure both the string order and entanglement spectrum of prepared AKLT chains and, employing these as metrics, find improved results over the known (purely unitary) sequential preparation approach. We conclude with a demonstration of quantum teleportation using the AKLT state prepared by our measurement-assisted scheme. This work thus serves to provide an efficient strategy to prepare a specific resource in the form of the AKLT state and, more broadly, experimentally demonstrates the possibility for realizable improvement in state preparation afforded by measurement-based circuit depth reduction strategies on NISQ-era devices.
\end{abstract}

\maketitle
\section{Introduction}
The efficient preparation of physically interesting and useful quantum states is a vital ingredient across quantum information science, from the fundamental study of quantum phases of matter \cite{Altman_2021, Zeng2019} to varied applications such as measurement-based quantum computation (MBQC) \cite{Raussendorf2001,Briegel_2009}, quantum error correction \cite{Terhal_2015}, and quantum sensing \cite{Degen_2017}. Consequently, the development of resource-efficient state preparation routines is a task of great importance, particularly for implementation on noisy intermediate-scale quantum (NISQ) devices. Especially advantageous are protocols capable of deterministically preparing nontrivial entangled resource states in constant-time, i.e, with a total runtime independent of the size of the state.

However, many states of interest -- particularly those with nontrivial topological properties -- cannot be prepared from a product state by a constant-depth circuit composed of unitary local gates, instead requiring a circuit whose depth scales with system size \cite{Bravyi2006,Chen_2010}. This presents a major roadblock for the preparation, study, and application utility of classically intractable many-body states on quantum computers, particularly in the NISQ era where quantum processors are limited to relatively shallow circuits due to decoherence.

A notable instance of a state that cannot be prepared exactly by a constant-depth local unitary circuit is the ground state of the spin-1 Affleck-Kennedy-Lieb-Tasaki (AKLT) model. Here referred to simply as the AKLT state, it is a paradigmatic example of both a matrix product state (MPS) and a symmetry-protected topological (SPT) phase \cite{Pollmann2009}, and additionally holds promise as a resource for MBQC \cite{Verstraete2004,Gross2007a,Brennen2008,Kaltenbaek2010,Wei2011}. While its SPT order prohibits its constant-time preparation from a product state with local unitary gates preserving symmetry, because the AKLT state has a nonzero correlation length, a stronger statement can be made: it cannot be prepared by \emph{any} constant-depth (local) unitary circuit. This is due to the fact that, under evolution of a constant-depth circuit, the correlator of any two local operators vanishes outside of a causal light cone \cite{Bravyi2006}. Known methods to prepare the AKLT state include sequential unitary, \cite{Schoen2005,Huang2015}, dissipative \cite{Zhou2021}, adiabatic \cite{Wei2022c}, and non-deterministic measurement-based \cite{Kaltenbaek2010, Murta2022, Chen2022} schemes, all of which require preparation times that scale with system size. In this work, we overcome this limitation.

Though purely unitary circuits provide a natural framework for the preparation (and classification) of quantum phases of matter from a quantum many-body perspective \cite{Chen_2010}, general quantum computational tasks leverage a broader toolbox consisting of local operations and classical communication (LOCC), and efforts to classify states under this paradigm are under way \cite{Piroli_2021,Tantivasadakarn2022a}. To that end, several recent theoretical proposals have shown that measurement, in addition to unitary evolution, can aid in the preparation of certain long-range entangled states including the Greenberger-Horne-Zeilinger (GHZ) state, toric code, and states with non-Abelian topological orders \cite{Tantivasadakarn2021,Verresen2021,Tantivasadakarn2022,Lu2022}.

Here, we blend together LOCC-assisted circuits and tensor network representations to demonstrate that the spin-1 AKLT state can be prepared using a constant-depth circuit augmented by measurements, such that the total preparation time is independent of system size. The core idea behind our scheme is to use a variant of the sequential MPS preparation protocol outlined in Ref.\ \cite{Schoen2005} to prepare small AKLT chains, and subsequently ``fuse'' them together using Bell measurements, similar in spirit to recent proposals for fusion-based quantum computation \cite{Bartolucci2021}. While the individual measurement outcomes are inherently probabilistic, we show that, remarkably, the protocol can be made entirely deterministic by leveraging the on-site $\mathbb{Z}_2\times\mathbb{Z}_2$ symmetry of the AKLT state, thus standing in contrast to known, probabilistic measurement-based schemes to prepare the AKLT state \cite{Kaltenbaek2010, Murta2022, Chen2022}.

We demonstrate the practical advantage of our measurement-based approach to state preparation by carrying out experiments on IBM Quantum processors. We prepare AKLT states of up to $N=6$ spin-1 sites using our measurement-assisted preparation scheme (requiring 18 total qubits) and, for comparison, carry out companion experiments using a purely unitary, sequential approach. We thoroughly validate prepared states by measuring both their string order and entanglement spectra. We find that our measurement-assisted scheme outperforms the purely unitary preparation, even for the small system sizes studied here. To further validate the prepared state and demonstrate that our measurement-assisted scheme leaves ample circuit depth for post-preparation utility, we carry out a quantum teleportation protocol using the prepared AKLT state. For our longest prepared chains, we demonstrate teleportation fidelities above 76\% for all target states, surpassing the classical limit of 2/3 \cite{Massar_1995}. Finally, we show that teleportation fidelities can be further enhanced to exceed 99\% when combined with purification techniques in post-processing.

The aim of this paper is twofold. First, we propose and experimentally demonstrate the measurement-assisted, constant-time preparation of a specific, topologically nontrivial resource state, useful for applications ranging from quantum teleportation and MBQC \cite{Verstraete2004,Gross2007a, Brennen2008, Kaltenbaek2010} to blind quantum computation \cite{morimae2015ground} and remote state preparation \cite{liu2014controlled}. Second, and more broadly, this work serves as an experimental demonstration of the practical utility and readily available advantage that measurement-assisted preparation schemes -- a topic of rapidly advancing theoretical interest \cite{Tantivasadakarn2021,Verresen2021,Tantivasadakarn2022,Lu2022} --  can provide over their purely unitary counterparts, even for relatively small system sizes.

The remainder of this work is organized as follows: In Section \ref{sec:reviewAKLT}, we briefly review the AKLT model and discuss the properties of its ground state. In Section \ref{sec:prepare}, we present our measurement-assisted scheme to prepare the AKLT state. We begin with a detailed discussion of the sequential preparation of a generic, translationally invariant MPS (Section \ref{ssec:generalsequential}) and follow this with a sequential preparation scheme specific to the AKLT state (Section \ref{ssec:sequentialprep}). We build upon these results to demonstrate how individually prepared matrix product states can be fused together using Bell measurements with probabilistic outcomes. We follow with a recipe that leverages $\mathbb{Z}_2\times\mathbb{Z}_2$ symmetry to convert this probabilistic approach into one that is entirely deterministic for the AKLT state, thereby enabling its preparation with a constant-depth circuit augmented by measurements (Section \ref{ssec:measurementprep}). 

In Section \ref{sec:experiments}, we present our experiments on IBM devices, beginning with the preparation of our ``building-blocks'' -- AKLT chains of two sites (Section \ref{ssec:tomography_small}). We follow this with a presentation of experiments preparing longer chains up to $N=6$ sites. We use string order and entanglement spectrum measurements to validate the preparation, and furthermore compare results from the measurement-assisted and sequential approaches (Section \ref{ssec:longchain}). Finally, we carry out quantum teleportation experiments on an IBM device using the AKLT state prepared by our measurement-assisted approach as a resource (Section \ref{ssec:teleportation}). We conclude in Section \ref{sec:conclusion} with a discussion of our findings and some brief comments on future directions, including possible extension to higher dimensions.

\section{Review of the AKLT model}
\label{sec:reviewAKLT}
The AKLT model describes a one-dimensional spin-1 chain imbued with both bilinear and biquadratic nearest-neighbor interactions,
\begin{equation}
    H = \sum_i \vec{S}_i\cdot \vec{S}_{i+1} + \frac{1}{3}(\vec{S}_i\cdot\vec{S}_{i+1})^2.
    \label{eq:Hamiltonian}
\end{equation}
Initially proposed to gain insight into Haldane's conjecture that the Heisenberg model has a finite energy gap and exponentially decaying correlations for integer spin, the AKLT model has remained of considerable interest since its introduction more than three decades ago.

The AKLT model provides a paradigmatic example of SPT order \cite{Pollmann2009, Cirac_2021}, i.e., its ground state (the AKLT state) cannot be prepared by a constant-depth local unitary circuit that preserves its protecting symmetry. In the case of the Haldane phase (which encompasses the AKLT model), it has been shown that any one of time-reversal symmetry, spatial reflection symmetry, or dihedral $\mathbb{Z}_2\times\mathbb{Z}_2$ symmetry (corresponding to local $\pi-$rotations about two orthogonal axes at every site) is sufficient for protection \cite{Pollmann2009, Pollmann2010}. Consequently, the AKLT state displays features emblematic of a SPT phase, such as fractionalized excitations at the edges and a hidden string order. 

Not unrelated to its SPT order \cite{Else2012, Stephen2017}, the AKLT state has been shown to be a useful resource state for quantum information science. Most notably, it was one of the first discovered entangled resources for MBQC beyond the cluster state model \cite{Verstraete2004, Gross2007a}. As the AKLT state is the exact gapped ground state of a Hamiltonian with two-body interactions (in contrast to the parent Hamiltonian of the cluster state which has higher-body interactions \cite{nielsen2006cluster}), it carries significant potential as a practically realizable resource with protection against local, symmetry-preserving Hamiltonian noise \cite{Brennen2008,Kaltenbaek2010}. Though a single 1D AKLT state encodes just a single qubit and is therefore not sufficient for universal MBQC, Ref.\ \cite{Brennen2008} proposed a hybrid scheme that combines multiple 1D AKLT states with  circuit-model logical two-qubit gates to achieve universal computation. Furthermore, the spin-3/2 AKLT state on a honeycomb lattice has been shown to be universal \cite{Wei2011,Wei2012a}.

\begin{figure}
\centering
\includegraphics[width=0.95\linewidth]{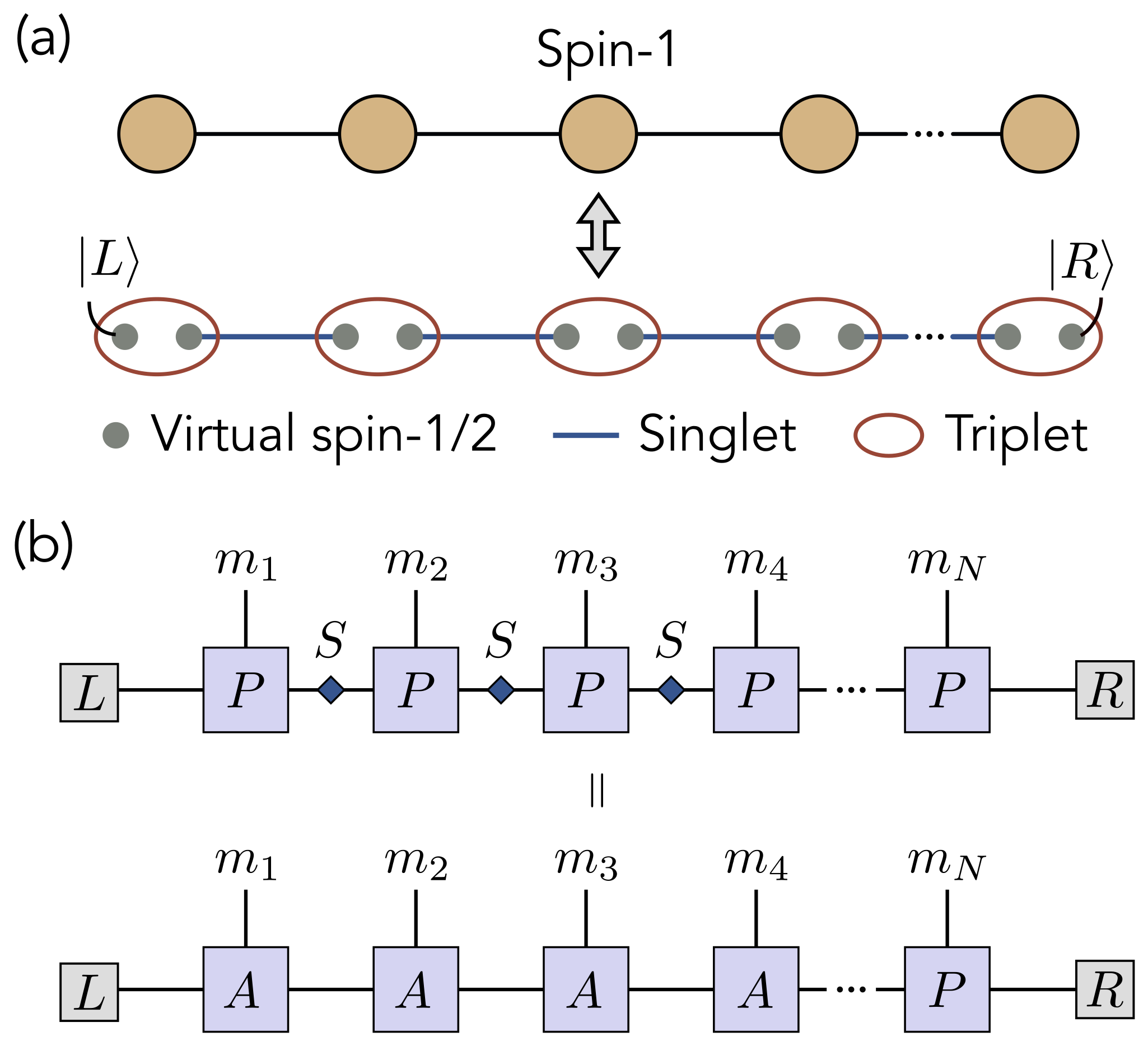}
\caption{(a) Illustration of the spin-1 AKLT state as a valence bond solid. The ground state of the AKLT state is most intuitively constructed by considering each spin-1 site as being composed of virtual spin-1/2 pairs (gray circles), with inter-site pairs in the singlet state and intra-site pairs projected onto the triplet subspace. (b) Diagram of the matrix product state representation of the AKLT state, with left- and right-boundary conditions $\ket{L}$ and $\ket{R}$ corresponding to the state of the free edge qubits. For each tensor, vertical legs represent physical indices, and horizontal legs represent virtual indices. Connected horizontal lines between adjacent tensors indicates contraction.}
\label{fig:f1}
\end{figure}

As shown in Fig.\ \ref{fig:f1}, the spin-1 AKLT state can be understood as a chain of virtual spin-1/2 pairs (i.e., subsystems with dimension $D$=2) alternating between a singlet and triplet configuration, with the latter corresponding to individual spin-1 sites. Using this intuition, the AKLT state can be written exactly as a matrix product state (MPS), here with open boundary conditions:
\begin{equation}
    \begin{split}
        |\Psi\rangle &= \sum_{\vec{m}} \bra{L}P^{m_1}SP^{m_2}S\ldots P^{m_N}\ket{R} |\vec{m}\rangle, \\
        &= \sum_{\vec{m}} \bra{L}A^{m_1}A^{m_2}\ldots A^{m_{N-1}}P^{m_N}\ket{R} |\vec{m}\rangle.
    \end{split}
    \label{eq:AKLTopen}
\end{equation}

In the above, $S$ and $P^{m_i}$ are $2\times2$ matrices encoding coefficients of singlet and triplet states (e.g., such that $\ket{s} = \sum_{ij}S_{ij}\ket{ij}$ is the singlet state for the qubits labeled $i$ and $j$), and $\vec{m}$ is shorthand for the ensemble of physical indices $m_i\in \{+,0,-\}$ corresponding to the three distinct eigenstates of $S^z$ with eigenvalue $+1$, $0$ and $-1$, respectively. In the second line, we have defined $A^{m_i}=P^{m_i} S$, where $A^-=-\sqrt{2/3}\,\sigma^-$, $A^0=-\sqrt{1/3}\,\sigma^z$ and $A^+=\sqrt{2/3}\,\sigma^+$. The AKLT state $\ket{\Psi}$ is thus an MPS of bond dimension $D=2$ and physical dimension $d=3$, the former referring to the dimensions of the contracted indices and the latter to the physical spin-1 degrees of freedom.

For the case of open boundary conditions, the ground state of the AKLT model is fourfold degenerate, here understood through the invariance of $\braket{\Psi|H|\Psi}$ upon particular choice of boundary conditions $\bra{L}$ and $\ket{R}$. This freedom is a manifestation of  fractionalized spin-1/2 degrees of freedom at the edges, a phenomenon symptomatic of the underlying SPT order \cite{Chen2013, Zeng2019}. Remarkably, these degrees of freedom are physical and observable \cite{Hagiwara_1990, Mishra_2021}, despite the uncoupled system consisting of only spin-1 particles. In the case of periodic boundary conditions, the ground state is unique and corresponds to the edge spins being in the singlet state,
\begin{equation}
    |\Psi\rangle = \sum_{\vec{m}} \textrm{Tr}\left(A^{m_1}A^{m_2}\ldots A^{m_{N-1}}A^{m_N}\right) |\vec{m}\rangle.
    \label{eq:AKLTperiodic}
\end{equation}

Though the exponential decay of spin-spin correlations $\braket{S_i^z S_{i+\ell}^z}\sim e^{-\ell/\xi}$ suggests an apparent featurelessness of the AKLT chain, further inspection reveals a hidden antiferromagnetic ordering of the spin-1 sites, explained as follows: using $+$, $-$, and $0$ to denote the spin-1 eigenstate of $S^z$ with eigenvalue $+1$, $-1$ and $0$, the AKLT state is composed only of configurations where $+$ and $-$ alternate, with any number of intermediate sites in the state $0$. For example, the configuration $|+0-0\,0+-\rangle$ is allowed, whereas $|++-0\,0+-\rangle$ is disallowed. This hidden order is characterized by the string order parameter \cite{Nijs1989},
\begin{equation}
    \braket{O_{\textrm{str}}^z}_{i,i+\ell} = \braket{S^z_i\prod_{j=i+1}^{i+\ell-1} e^{i\pi S^z_{j}}S^z_{i+\ell}}
    \label{eq:strorder}
\end{equation} 
which takes on a finite value throughout the Haldane phase, approaching $-4/9$ for $\ell\to\infty$ at the AKLT point \footnote{Because the bulk operators merely alter the sign of $\langle O_{\textrm{str}}^z\rangle$ in the $S^z$ basis, the particular value of $-4/9$ can be understood as arising from the roughly equal probability for each edge spin-1 site to be in the $+1$, $0$, or $-1$ eigenstate of $S^z$ when $N\gg 1$: out of the $9$ possible combinations, the four with $\pm 1$ at both edges contribute $-1$ to $\langle O_{\textrm{str}}^z\rangle$. All others have a vanishing eigenvalue.}.

\section{Preparing the AKLT state}\label{sec:prepare}

In this section, we introduce our measurement-assisted procedure. For completeness, we note that several approaches to prepare the AKLT state are known, including sequential unitary, \cite{Schoen2005,Huang2015}, dissipative \cite{Zhou2021}, adiabatic \cite{Wei2022c}, and non-deterministic measurement-based \cite{Kaltenbaek2010, Murta2022, Chen2022} strategies, all of which require preparation times that scale with system size. 

A particularly straightforward (non-deterministic) approach involves preparation of inter-site qubit pairs in the singlet state, and subsequent probabilistic application of the (non-unitary) triplet subspace projector to intra-site pairs, in analogy to the virtual spin-1/2 construction of the AKLT state in Fig. \ref{fig:f1}. This strategy was originally employed in Ref. \cite{Kaltenbaek2010} in a photonic platform, and more recently implemented on an IBM Quantum device in Ref. \cite{Chen2022} by invoking ancilla qubits to explicitly block-encode the triplet projectors. However, the success probability of this approach exponentially decays as $\sim(3/4)^N$, introducing a significant repetition overhead for all but the smallest chains. This approach therefore stands in stark contrast to the method introduced and experimentally demonstrated here, which is entirely deterministic with a preparation time independent of $N$.

 We begin by reviewing the sequential preparation of a generic, translationally invariant MPS \cite{Schoen2005} and, following this, we illustrate its application to the spin-1 AKLT state. Building upon this framework, we show how Bell measurements, together with the $\mathbb{Z}_2\times\mathbb{Z}_2$ symmetry of the AKLT state, can be leveraged to deterministically fuse individual AKLT chains and ultimately convert the sequential (linear-depth) preparation algorithm into one that is constant-depth and deterministic.

\subsection{Sequential preparation of a generic MPS}
\label{ssec:generalsequential}
 For both completeness and clarity, it is simplest to first consider the sequential generation of an arbitrary, translationally invariant MPS with bond dimension $D$ of the form
\begin{equation}
    \ket{\Psi} = \sum_{\vec{m}}\bra{L}A^{m_1}A^{m_2}A^{m_3}\ldots A^{m_N}\ket{R}\ket{\vec{m}},
    \label{eq:genericMPS}
\end{equation}
 describing a 1D chain of spin-$s$ particles. While $\ket{R}$ and $\bra{L}$ correspond to right and left boundary conditions of the spin chain, it is helpful to reinterpret them as the initial and final state of some subsystem in a fictitious Hilbert space of dimension $D$, often termed the bond, virtual, or correlation space. Through this lens, the matrices $A^m$ appear as (possibly non-unitary) operators acting on a virtual subsystem.

This analogy is made stronger by tracing out the spin-$s$ degrees of freedom, revealing the reduced density matrix of the virtual subsystem to evolve according to the map $\Gamma(\rho) = \sum_m A^m \rho A^{m\dagger}$. We note that any MPS can be cast into canonical form such that $\sum_m A^{m\dagger} A^m = \mathbb{1}$ (which we assume to be the case without loss of generality in the remainder of this work), in which case $\Gamma$ becomes a completely-positive trace-preserving (CPTP) map, or quantum channel, and $A^m$ the corresponding Kraus operators.

Because its evolution is described by a CPTP map, it is always possible to embed the bond space subsystem in a larger Hilbert space that evolves unitarily (via the Stinespring dilation theorem). We take as our larger Hilbert space that of the bond space subsystem and the spin-$s$ chain, and assume all spin-$s$ sites to be initialized in the state $\ket{\psi_0}$. We note that the particular choice of $\ket{\psi_0}$ is arbitrary, and therefore may be chosen for convenience according to the experimental platform. 

We next define a unitary $U$ such that it transforms the input spin-$s$ state $\ket{\psi_0}$ (and arbitrary bond-space state) according to $U\ket{\psi_0}\bra{\psi_0} = \sum_{m} A^{m} \otimes \ket{m}\bra{\psi_0}$. The action of $U$ on other spin-$s$ states can be chosen freely without loss of generality. Sequential application of $U$ to the initial state $\ket{\Psi_0} = \ket{R}\otimes \ket{\psi_0}^{\otimes N}$ yields

\begin{equation}
    \begin{split}
        U_1U_2\ldots U_N \ket{\Psi_0} =  \sum_{\vec{m}}A^{m_1}A^{m_2}\ldots A^{m_N}\ket{R}\ket{\vec{m}},
    \end{split}
    \label{eq:sequnitaries}
\end{equation}
 where $U_i$ denotes application to the bond space ancilla, initially in the state $\ket{R}$, and the $i$th ``physical'' spin-$s$ site. In this scheme, the ancilla serves as a memory that propagates correlations between the physical sites, allowing us to prepare a generic MPS without joint operations between the physical spin sites. Crucially, the required size of the memory is given by the bond dimension, highlighting an intuitive correspondence between the complexity of the memory and the maximum entanglement entropy contributed per cut bond, $\log(D)$ \cite{Schuch2007,Orus2013}.

In general, the bond space ancilla will be left entangled with the physical degrees of freedom upon application of the sequence of unitaries in Eq.~{\ref{eq:sequnitaries}}. Notably, the bipartite entanglement content between the physical and bond subsystems is physically equivalent to that of a semi-infinite chain with an entanglement cut adjacent to the last prepared physical site \cite{Gopalakrishnan2019,Foss_Feig_2022}. This feature will be exploited later in this work to experimentally characterize and validate preparation of the AKLT state via measurements of its entanglement spectrum. 

For the present goal of preparing a finite chain, the physical and bond subsystems may be disentangled by measuring the memory ancilla (equivalent to a left-projection of the form $\ket{L}\bra{L}$), collapsing the physical degrees of freedom onto the state 
Eq.~\ref{eq:genericMPS} for some left boundary condition $\ket{L}$. While the determination of $\ket{L}$ is probabilistic for a generic MPS, a slight variation of this approach can achieve specific boundary conditions -- see Ref.\ \cite{Schoen2005} for details.

\subsection{Sequential preparation of the AKLT state}\label{ssec:sequentialprep}

\begin{figure*}
\centering
\includegraphics[width=1\linewidth]{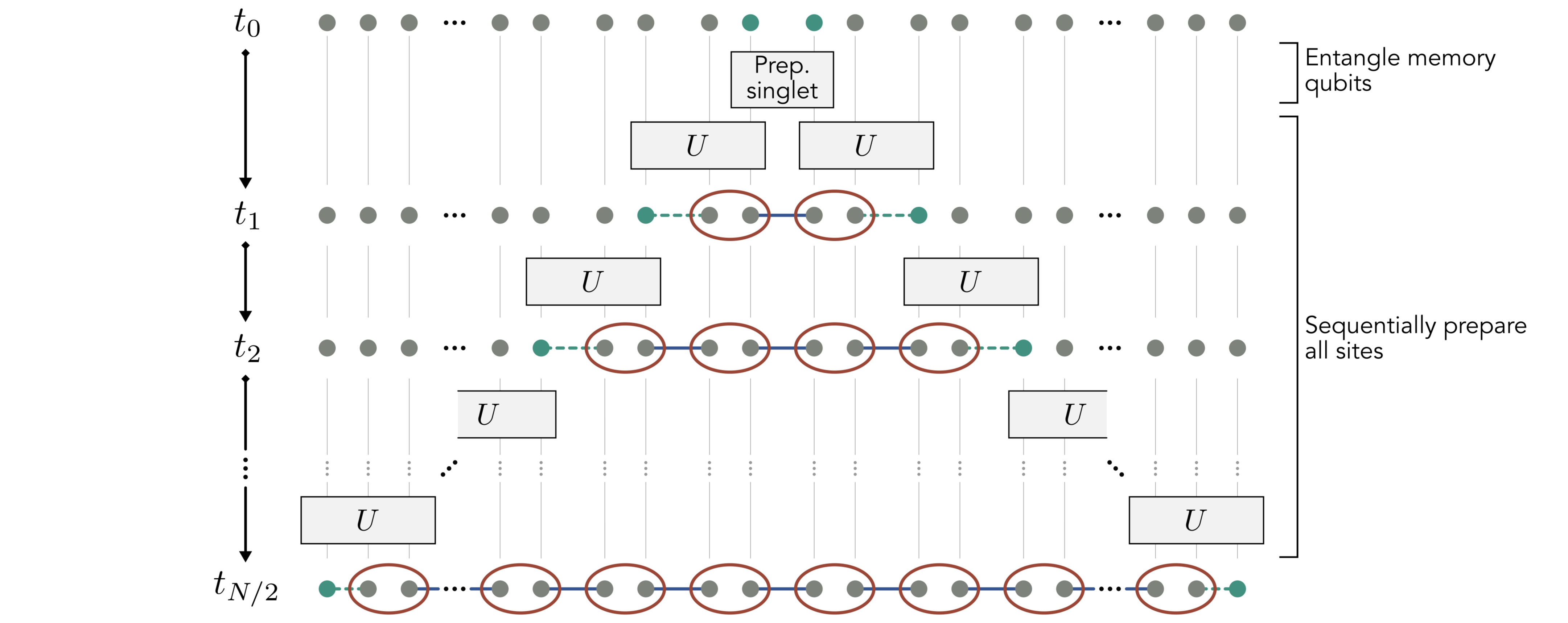}
\caption{Sequential preparation of the AKLT state with a linear depth circuit. Each time step consists of applying the site preparation unitary $U$ to two qubits composing a spin-1 site (gray circles) and a memory qubit (green circles). The latter is responsible for propagating correlations between spin-1 sites. To reduce the overall circuit depth, our sequential procedure takes advantage of spatial inversion symmetry to grow the AKLT state simultaneously at both ends, using the two memory qubits in parallel.}
\label{fig:f2}
\end{figure*}

We now turn to the particular details of the sequential preparation of the AKLT state. As the AKLT state consists of physical spin-1 sites, we take inspiration from its virtual spin-1/2 representation (see Fig.\ \ref{fig:f1}a), and use pairs of qubits to compose each spin-1 site. For the remainder of this work, we will adopt the notation $\ket{+}$, $\ket{\bar{0}}$, and $\ket{-}$ for the spin-1 eigenstates of $S^z$ with corresponding eigenvalues $+1$, $0$, and $-1$, respectively, and $\ket{s}$ for the singlet state. Here, the bar in the spin-1 state $\ket{\bar{0}}$ is used to differentiate from the qubit state $\ket{0}$ wherever clarification is needed.

Recalling that the AKLT state is an MPS of bond dimension $D=2$, a single ``memory qubit'' (terminology which we will adopt for the bond space subsystem throughout the remainder of this manuscript) is sufficient to prepare the AKLT state with a definite right-boundary condition $\ket{R}$ and indefinite left-boundary condition $\bra{L}$; choice of the former is equivalent to choosing a particular initial state for the memory, while the latter is determined upon measurement of the memory qubit after all spin-1 sites have been prepared.

Here, we alter this strategy in order to reduce circuit depth. Taking advantage of the spatial inversion symmetry of the AKLT state (see Appendix\ \ref{sec:app:parallelmemory}), we utilize two initially entangled memory qubits -- one which ultimately encodes the left-boundary condition $\bra{L}$ and the other the right-boundary condition $\ket{R}$. We emphasize that these boundary conditions reflect more than just mathematical choice: they encode the physical state of the left and right edge modes of the AKLT chain. As will become clear, the action of enforcing a particular boundary condition via measurement, albeit probabilistically, is a crucial ingredient of our measurement-aided constant-time preparation scheme.

Fig.\ \ref{fig:f2} summarizes our scheme to sequentially prepare an arbitrary length AKLT chain with a linear-depth circuit. The memory qubits are first prepared in the singlet state $\ket{s}=(\ket{01} - \ket{10})/\sqrt{2}$, and we choose as an initial starting state for each spin-1 site $\ket{\psi_0}=\ket{\bar{0}}$. Though not yet explicitly enforced, we will later choosing an encoding such that $\ket{\bar{0}}$ corresponds to both qubits being in the ground state. As in the generic MPS case, the primary ingredient of the circuit is the unitary $U$ that prepares each site and imprints this history on the memory. Recalling the composing matrices of the MPS representation of the AKLT state $A^+=\sqrt{2/3}\,\sigma^+$, $A^0=-\sqrt{1/3}\,\sigma^z$, and $A^-=-\sqrt{2/3}\,\sigma^-$, this unitary takes the form
\begin{equation}
    U = \sqrt{\frac{1}{3}}(\sigma^+ S^+ - \sigma^- S^- - \sigma^z I) + C,
\end{equation}
where $S^+$, $S^-$ and $S^z$ are raising, lowering and $z$-component spin-1 operators, and $C$ is a unitary completion operator which obeys $C\ket{\bar{0}} = 0$ and enforces unitarity of $U$, but is otherwise arbitrary. 

As explained above, this particular form of $U$ derives from the MPS representation of the AKLT state, though it may also be intuitively appreciated from the perspective of the well-known hidden antiferromagnetic ordering of the AKLT state (see the discussion in Section \ref{sec:reviewAKLT}). To see this, note that the first (second) term will prepare a site in the $\ket{+}$ ($\ket{-}$) state, and paired with this is the raising (lowering) of the memory qubit via $\sigma^+$ ($\sigma^-$). Because $\sigma^+$ and $\sigma^-$ each return 0 when acted twice, preparing a site in the state $\ket{+}$ must necessarily be followed by either preparation of $\ket{-}$ or $\ket{\bar{0}}$. Furthermore, preparation of a second, subsequent site in the state $\ket{+}$ cannot occur until a site has been prepared in the state $\ket{-}$, lowering the state of the memory. Repeated application of $U$ to a chain of sites therefore results in exactly the hidden antiferromagnetic ordering of the AKLT state, where spins of alternating polarization are diluted with sites in the state $\ket{\bar{0}}$.

After preparation of all $N$ sites, the state of the composite system consisting of both the spin chain and memory qubits is
\begin{equation}
    \ket{\Psi_{\textrm{seq}}} = \sum_{ij}\sum_{\vec{m}} \braket{i|A^{m_1}A^{m_2}\ldots A^{m_{N-1}}P^{m_N}|j}\ket{ij}\ket{\vec{m}},
    \label{eq:AKLTseq}
\end{equation}
the AKLT state with boundary conditions entangled with the two memory qubits, labeled by indices $i$ and $j$. Similar to the case of a single memory qubit, particular boundary conditions are enforced upon measurement of memory qubits. Here, however, both left \emph{and} right boundary conditions are enforced upon final measurement of memory qubits, as opposed to the latter corresponding to the initial state of a single memory qubit. As will be shown, this feature plays an important role for our measurement-assisted preparation scheme.

We additionally note the possibility to measure the two memory qubits in a basis such that they become entangled, consequently entangling the AKLT edge spin-1/2s via entanglement swapping. For example, the SWAP test can be used to measure the exchange symmetry of the two memory qubits. In the case where the asymmetric singlet state is measured, a singlet bond is teleported to the pair of edge spin-1/2s, and the spin chain collapses onto the AKLT state with periodic boundary conditions, Eq.~\ref{eq:AKLTperiodic}. Interestingly, failure to measure the singlet does not prohibit successful preparation of the AKLT state with periodic boundary conditions, as one can recover upon projection onto the symmetric subspace -- such a possibility is due to the SPT order of the AKLT state, and in particular its $\mathbb{Z}_2\times\mathbb{Z}_2$ symmetry (see Appendix \ref{sec:app:periodicboundaries} for details). In the following section, we will take advantage of this same symmetry to show that AKLT chains can be \emph{deterministically} fused together using Bell measurements, thereby enabling the constant-time preparation of arbitrarily large chains.

\subsection{Measurement-assisted preparation of the AKLT state in constant-time}\label{ssec:measurementprep}
We now demonstrate that by augmenting the above procedure with measurements, it is possible to deterministically prepare the AKLT state using a constant-depth circuit. The core idea is to prepare multiple small AKLT states and, using the fact that the memory qubits encode boundary conditions, subsequently stitch them together with fusion measurements. To illustrate our procedure most clearly, we first consider the parallel preparation of two independent AKLT chains each of length $n\equiv  N/2$. We then generalize to the most efficient scenario of fusing $N/2$ independent two-site chains. While here our focus is the AKLT state, we note that the general procedure follows in an equivalent fashion for other matrix product states with appropriate symmetries. For more examples, see Appendix \ref{sec:app:GHZcluster}.

\subsubsection{Probabilistic fusion of two AKLT chains}
Following the procedure outlined in the previous section, two AKLT chains (with boundary conditions still entangled with memory qubits) can be written as

\begin{widetext}
\begin{equation}
    \ket{\Psi} = \sum_{ijk\ell}\sum_{\vec{m}}\braket{i|A^{m_1}A^{m_2}\ldots A^{m_{n-1}}P^{m_n}|j}\braket{k|A^{m_{n+1}}A^{m_{n+2}}\ldots A^{m_{N-1}}P^{m_{N}}|\ell}\ket{ijk\ell}\ket{\vec{m}},
\end{equation}
where indices $i$, $j$, $k$, and $\ell$ denote the state of the four memory qubits, and $\vec{m}=\{m_1, m_2,\ldots m_N\}$ is a composite index for the two chains. We note that while we choose to group together all $N$ physical indices of the MPS, the two independent chains of length $n$ are, at this stage, unentangled.

By projecting the memory qubits labeled $j$ and $k$ onto a Bell pair, a maximally entangled bond is teleported to the edge qubits of the two AKLT chains. To see this, it is helpful to imagine measuring the middle pair of memory qubits in the state $\ket{\phi} =\sum_{ij}\phi_{ij}\ket{ij}$. This collapses the above state onto
\begin{equation}
    \ket{\Psi'} = \sum_{i\ell}\sum_{\vec{m}}\braket{i|A^{m_1}A^{m_2}\ldots A^{m_{n-1}}P^{m_n}MA^{m_{n+1}}A^{m_{n+2}}\ldots A^{m_{N-1}}P^{m_{N}}|\ell}\ket{i\ell}\ket{\vec{m}}\otimes\ket{\phi},
\label{eq:fused_w_M}
\end{equation}
\end{widetext}
where $M \propto \sum_{jk} \phi_{jk}\ket{j}\bra{k}$ and
the ordering of tensor products has been rearranged for clarity. In other words, the state $\ket{\Psi}$ collapses onto an $N$-site MPS with a probabilistically determined matrix $M$ at its center. Recalling the relation $A^m = P^m S$, it is not hard to see that measurement of the antisymmetric (singlet) Bell state $\ket{\Psi^-}~=~(\ket{01}~-~\ket{10})/\sqrt{2}$) yields the desired outcome: $M=S$, and $\ket{\Psi'}$ becomes exactly the $N$-site AKLT state with boundaries still entangled with edge memory qubits (and in a product state with the measured memory qubits). 

\begin{figure}
\centering
\includegraphics[width=1\linewidth]{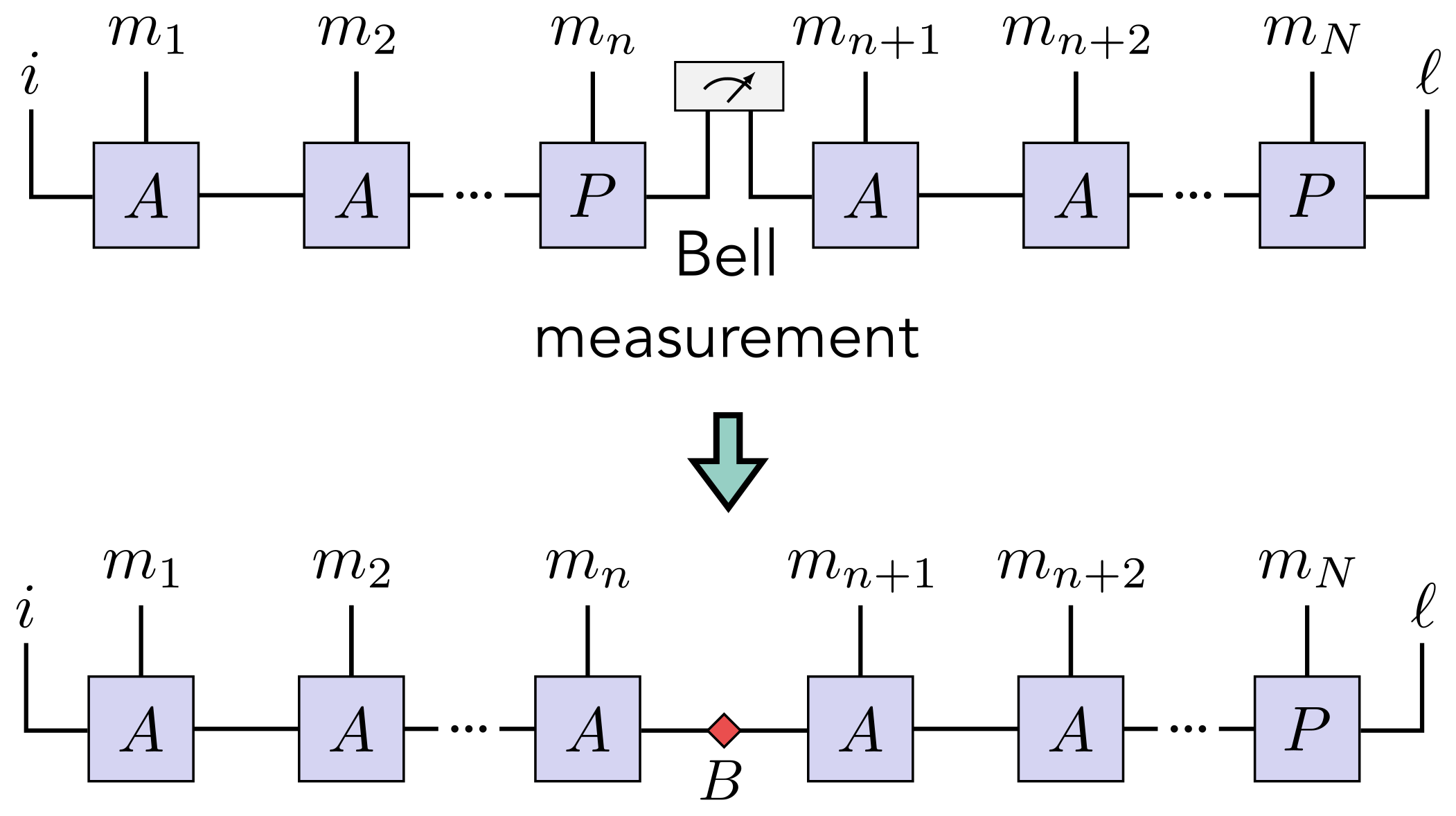}
\caption{Fusion of two AKLT states via Bell measurement. After preparation, the edge spin-1/2s of each MPS are each entangled with a memory qubit. Through entanglement swapping, Bell measurement of memory qubits belonging to different chains ``fuses'' the two independent MPS, teleporting a particular tensor $M = \{I,X,Y,Z\}$ to the intermediate bond. Using the fact that $A^m = P^m S$ and $S\propto Y$, this amounts to a longer AKLT state with the local Pauli defect $B \propto YM$.}
\label{fig:f3}
\end{figure}

However it is not guaranteed that the two memory qubits are measured in a singlet -- each Bell state is equally probable \footnote{This is only true for for Bell measurement of memory qubits belonging to independent chains. If the edge memory qubits of a single chain are measured in the Bell basis, i.e. to enforce boundary conditions, the exact probability of each outcome is dependent on the length of the chain, though tends toward 1/4 for large $N$}, and it is therefore imperative to consider the resulting MPS for each case. Ignoring global phases and normalization coefficients, measurement of each Bell state results in $M$ becoming one of the four Pauli matrices. Noting that $S\propto Y$ and that the Pauli matrices form a closed group under multiplication, this ensures that the state $\ket{\Psi'}$, regardless of measurement result, can always be written in the form
\begin{equation}
    \ket{\Psi'} = \sum_{i\ell \vec{m}}\braket{i|A^{m_1}\ldots A^{m_n}BA^{m_{n+1}}\ldots P^{m_{N}}|\ell}\ket{i\ell}\ket{\vec{m}},
    \label{eq:AKLTdefect}
\end{equation}
where we have discarded the state of the memory qubits $\ket{\phi}$ as they play no further role in the preparation. $\ket{\Psi'}$ is thus exactly the AKLT state, up to the defect matrix $B$. All possible outcomes for the Bell measurement and corresponding matrices $M$ and $B$ are summarized in Table \ref{tab:bell}.

\begin{table}
\centering
\def\arraystretch{2}
{\setlength{\tabcolsep}{4pt}
\begin{tabular}{|| c | c | c||} 
 \hline
 Measurement result $\ket{\phi}$ & $M\propto\sum\limits_{ij}\phi_{ij}\ket{i}\bra{j}$ & Defect $B$ \\ [5pt] 
 \hline\hline
 $\ket{\Phi^+} = \frac{1}{\sqrt{2}}(\ket{00}+\ket{11})$ & $I$ & $Y$\\[4pt]
 \hline
 $\ket{\Phi^-} = \frac{1}{\sqrt{2}}(\ket{00}-\ket{11})$ & $Z$ & $X$\\[4pt]
 \hline
 $\ket{\Psi^+} = \frac{1}{\sqrt{2}}(\ket{01}+\ket{10})$ & $X$ & $Z$  \\[4pt]
 \hline
 $\ket{\Psi^-} = \frac{1}{\sqrt{2}}(\ket{01}-\ket{10})$ & $Y$ & $I$  \\ [4pt]
 \hline
\end{tabular}}
\caption{Possible measurement results and the corresponding matrix $M$ inserted into the MPS. In each case, $M\propto SB$, where $S$ is the desired singlet matrix and $B$ a residual Pauli defect. Global phases have been ignored in defining $M$ and $B$.}
\label{tab:bell}
\end{table}

Naively, the probabilistic outcome of the measurement indicates that this strategy cannot be used to determinsitically prepare the AKLT state. We will now show, however, that one can always recover from a ``failed'' measurement and remove the defect $B$ by leveraging the on-site $\mathbb{Z}_2\times\mathbb{Z}_2$ symmetry of the AKLT state. 

\subsubsection{Leveraging the \texorpdfstring{$\mathbb{Z}_2 \times \mathbb{Z}_2$}{Z2Z2} symmetry of the AKLT state to remove defects}
To explain, let us first discuss the physical intuition of this symmetry and, in particular, how it relates to the edge states of the AKLT chain. We first note that the Hamiltonian in Eq.~\ref{eq:Hamiltonian} is invariant upon rotation of each spin-1 site by an angle $\pi$. To see this, one can conjugate the Hamiltonian by $\prod_{j=1}^N \textrm{exp}(i\pi S_j^x)$ which, in effect, transforms each spin-1 operator as $(S_j^x,S_j^y,S_j^z)\to (S_j^x,-S_j^y,-S_j^z)$, leaving the Hamiltonian unchanged. To establish the corresponding symmetry group, it is sufficient to consider rotation about two orthogonal axes (e.g., $\textrm{exp}(i\pi S_j^x)$ and $\textrm{exp}(i\pi S_j^y)$, as rotation about the third, $\textrm{exp}(i\pi S_j^z)$, can always be generated through their combination). As a result, the Hamiltonian has a global on-site $\mathbb{Z}_2\times\mathbb{Z}_2$ symmetry, also referred to as a dihedral symmetry $D_2$ \cite{Pollmann2009}.

The symmetry of the AKLT ground state itself, however, is more subtle. Here, we wish to merely elucidate a connection between the hidden $\mathbb{Z}_2\times\mathbb{Z}_2$ symmetry-breaking \cite{Kennedy_1992} in the Haldane phase and the presence of edge states, the latter of which play an important role in our preparation scheme. For an accessible in-depth discussion, we refer to Ref.\ \cite{Pollmann2009}. 

Each on-site rotation can be simply written in terms of its action on $S^z$ eigenstates:
\begin{equation}
    \begin{alignedat}{4}
        &U_X &&= e^{i\pi S^x}&& = -&&\ket{+}\bra{-} - \ket{\bar{0}}\bra{\bar{0}} - \ket{-}\bra{+} \\
        &U_Y &&= e^{i\pi S^y}&& = &&\ket{+}\bra{-} - \ket{\bar{0}}\bra{\bar{0}} + \ket{-}\bra{+} \\
        &U_Z &&= e^{i\pi S^z}&& = -&&\ket{+}\bra{+} + \ket{\bar{0}}\bra{\bar{0}} - \ket{-}\bra{-} \\
        &U_I &&= I&& = &&\ket{+}\bra{+} + \ket{\bar{0}}\bra{\bar{0}} + \ket{-}\bra{-}, \\
    \label{eq:U_B}
    \end{alignedat}
\end{equation}
where we have also defined the spin-1 identity operator as $U_I$ for reasons that will soon become clear. Taking $U_X = e^{i\pi S^x}$ as an example, application of this unitary operator to each site of the AKLT state with \emph{periodic} boundary conditions will leave the state invariant up to a global phase -- each configuration of the many-body state has a partner with the same amplitude but with all spins flipped. 

In the case of \emph{open} boundary conditions, however, this same procedure results in a non-identity operation at the edges. To see this, consider application of $U_X$ to the AKLT ground state with edge spin-1/2s together in the $m=+1$ triplet state $\ket{00}$ (i.e., Eq.~\ref{eq:AKLTopen} with $L=R=0$). Because all bulk inter-site pairs are in the singlet state, the entire chain is initially in the $+1$ eigenstate of the total spin operator $S_{\textrm{tot}}^z = \sum_i^N S_i^z$. Application of $U_X$ to each site flips all spin-1s, and the entire chain is consequently transformed to a $-1$ eigenstate of $S_{\textrm{tot}}^z$. Noting that the energy cannot change due to invariance of the Hamiltonian, this final state is necessarily an AKLT ground state with edge spins in the $m=-1$ triplet state $\ket{11}$ (Eq.~\ref{eq:AKLTopen} with $L=R=1$). In other words, application of $U_X$ to every site leaves the bulk invariant, but ultimately corresponds to a nontrivial action on the edge spin-1/2s. 

The core principle behind our preparation algorithm is the ability to correct this nontrivial action on the edges by applying an additional unitary operation to the memory qubits, which effectively serve as a control for the edge states. We will now make this more concrete by directly appealing to the MPS representation of the AKLT state and, in particular, how its $\mathbb{Z}_2\times\mathbb{Z}_2$ symmetry is manifest in the structure of the local tensor $A$. Noting that the Pauli matrices form a projective representation of $\mathbb{Z}_2\times\mathbb{Z}_2$, the tensor $A$ necessarily obeys the relation \cite{Schuch2007, Pollmann2010, Cirac_2021}
\begin{equation}
    \sum_m (U_B)_{mm'} A^{m'} = e^{i \theta_B}B^\dagger A^m B
    \label{eq:SPTfundamental}
\end{equation}
for Pauli operators $B=\{I,X,Y,Z\}$ and corresponding spin-1 operators $U_B$ defined in Eq.~\ref{eq:U_B}. This equality is shown pictorially in Fig.\ \ref{fig:f4}a. We note that the global phase $e^{i\theta_B}$ is unimportant for our discussion, but nonetheless here takes on the value $\pm 1$ and is physically relevant to the degeneracy of the entanglement spectrum \cite{Pollmann2010}.

This property -- that the global on-site $\mathbb{Z}_2\times\mathbb{Z}_2$ symmetry of the AKLT state is reflected in the local tensors $A$ as the ability to ``push'' the operator $U_B$ onto the virtual level -- is a direct consequence of the SPT order of the Haldane phase, and many related properties may be understood through this lens (e.g., string order, hidden symmetry breaking, edge excitations, entanglement spectrum degeneracy) \cite{Schuch2007, Pollmann2009, Pollmann2010, Cirac_2021}. Clarity is particularly added to the prior discussion of symmetry-breaking at the edges, as application of $U_X$ to every site clearly leaves the state invariant up to the modified boundary conditions $\ket{L}\to X\ket{L}$ and $\ket{R}\to X \ket{R}$. We note that similar relations exist for the time-reversal and spatial-inversion symmetries of the Haldane phase \cite{Pollmann2010}. While we do not explicitly rely on the former, the latter enables our use of two memory qubits in parallel, as explained in Appendix \ref{sec:app:parallelmemory}.

Finally, we now describe the procedure for removing a defect $B$ from the fused AKLT chain pair in Eq.~\ref{eq:AKLTdefect}. Recalling Table \ref{tab:bell}, each Bell measurement result yields the AKLT state up to a defect operator $B$ (with boundary conditions entangled with any residual memory qubits). Importantly, this defect $B$ is one of the four Pauli operators. As shown in Fig.\ \ref{fig:f4}c, we exploit the $\mathbb{Z}_2\times\mathbb{Z}_2$ symmetry -- in particular, its manifestation at the level of the tensor $A$ -- to push the operator $B$ from the intermediary bond teleported into the chain through fusion measurement to one of the edge states. Taking the state in 
Eq.~\ref{eq:AKLTdefect} as a starting point, application of the unitary $U_B$ on all sites to the left of the defect translates it from the measured link to the far virtual leg of the left edge qubit, and applying $B^\dagger$ \footnote{While $B^\dagger = B$ in this particular case, we leave the Hermitian adjoint intact for generality, as this strategy is potentially applicable to non-Pauli defects in preparation of other SPTs.} to the left memory qubit removes it entirely:
\begin{equation}
    \begin{split}
        \ket{\Psi''} &= B^\dagger\otimes I \otimes U_B^{\otimes n} \otimes I^{\otimes {N-n}}\ket{\Psi'} \\
        &=\sum_{i\ell \vec{m}}\braket{i|BA^{m_1}\ldots A^{m_{N-1}} P^{m_{N}}|\ell}(B^\dagger\otimes I)\ket{i\ell}\ket{\vec{m}} \\
        &=\sum_{i\ell\vec{m}}\braket{i|A^{m_1}\ldots A^{m_{N-1}} P^{m_{N}}|\ell}\ket{i\ell}\ket{\vec{m}}.
    \end{split}
    \label{eq:AKLTpar}
\end{equation}

In the top line, the first two operators are applied to the memory qubits, and all others to the spin-1 sites. The choice of permuting the defect to the left is arbitrary, and applying $U_B$ to all sites to the right of the defect is equally valid. 

To see that the final equality in Eq.~\ref{eq:AKLTpar} is true, it is helpful to rearrange the second line and note that $\sum_i B^\dagger \ket{i}\bra{i} B = \sum_i \ket{i}\bra{i}$ for any unitary $B$. Alternatively, the final line may be viewed as a change of summation basis with respect to the second line. Nonetheless, the final state $\ket{\Psi''}$ is identical to the sequentially prepared state $\ket{\Psi_{\textrm{seq}}}$ in Eq.~\ref{eq:AKLTseq}, thus demonstrating that it is possible to deterministically fuse two AKLT chains and remove any measurement-induced defects. As described in detail in Section \ref{ssec:sequentialprep}, measurement of the remaining memory qubits enforces boundary conditions for the AKLT chain, with all outcomes yielding a particular state in the fourfold degenerate ground state space of the AKLT Hamiltonian with open boundary conditions.

\subsubsection{Summary of the preparation algorithm}
\begin{figure*}
\centering
\includegraphics[width=1\linewidth]{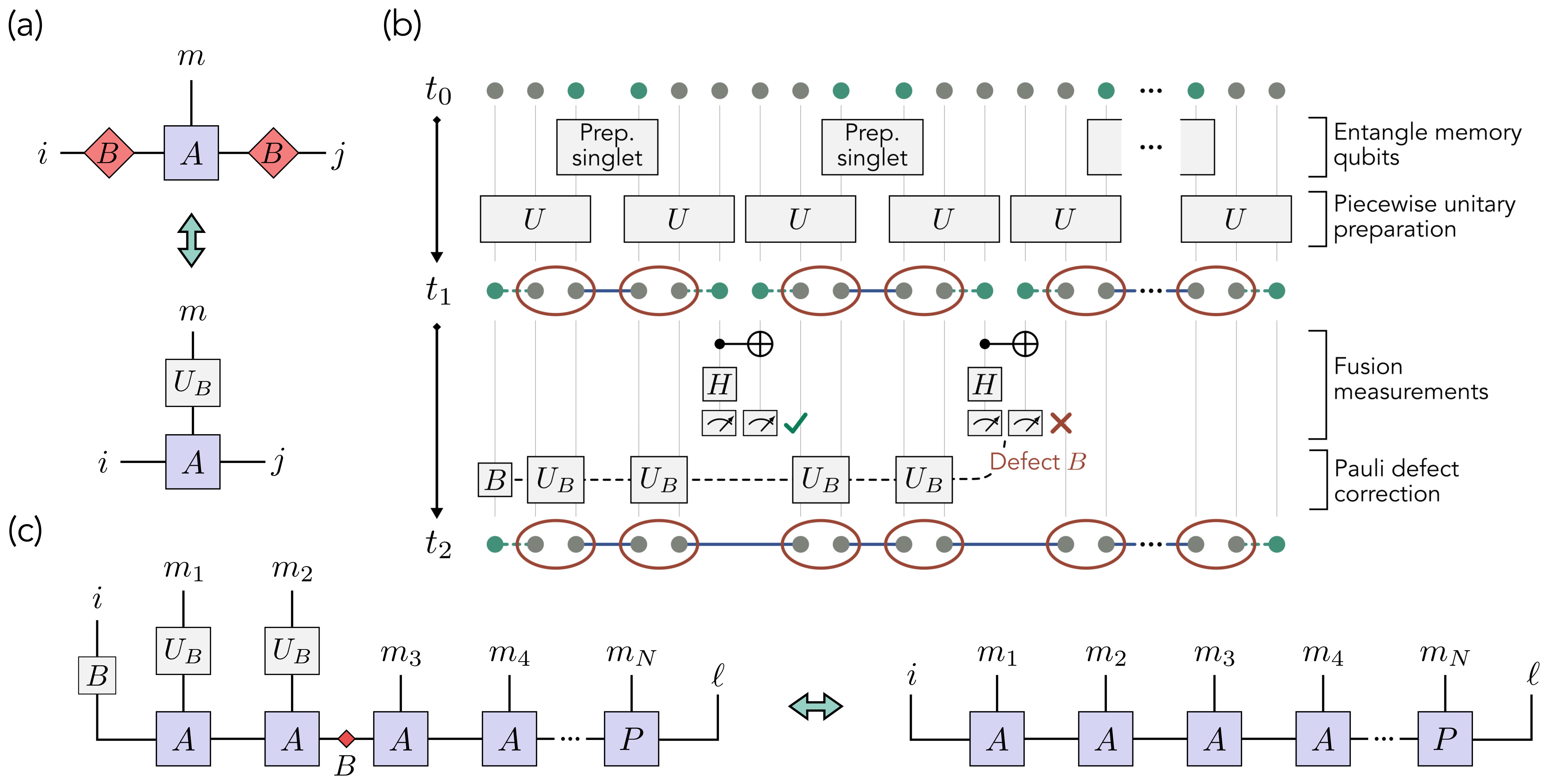}
\caption{Measurement-assisted preparation of the AKLT state. (a) The manifestation of $\mathbb{Z}_2\times \mathbb{Z}_2$ symmetry at the level of the local tensor $A$. Conjugation by $B$ applied to the virtual legs is equivalent to application of $U_B$ on the physical leg. (b) Circuit diagram of our measurement-assisted preparation. We begin by applying a single time step of the sequential preparation, forming many two-site AKLT states in parallel. We then fuse together independent chains through Bell measurement of their memory qubits, introducing the Pauli defect $B$. This defect is then removed by applying the symmetry action of panel (a). Notably, this final step does not have to be carried out on quantum hardware as the unitary $U_B$ merely permutes the encoding of the spin-1 basis states. (c) Tensor network illustration of the removal of the defect $B$ using the $\mathbb{Z}_2\times\mathbb{Z}_2$ symmetry of the AKLT state.} \label{fig:f4}
\end{figure*}
We now combine these ingredients to demonstrate one of the main results of this work: an algorithm to deterministically prepare an $N$-site AKLT chain using a constant-depth circuit augmented by fusion measurements.

Due to the spatial inversion symmetry of the AKLT state, a single chain can be prepared using both memory qubits in parallel (see Appendix \ref{sec:app:parallelmemory}). As a result, the most (depth) efficient preparation of large chains consists of fusing many two-site chains, each requiring a single application of $U$ per memory qubit. In total, we require $3N$ qubits -- two to encode each spin-1 site, plus one additional (memory) qubit per site.

A circuit diagram of the full algorithm is shown in Fig.\ \ref{fig:f4}b. The individual steps are as follows:
\begin{enumerate}
    \item {\bf Initialization.} Initialize all $N$ spin-1 sites in the $\ket{\bar{0}}$ state. Prepare all $N/2$ memory qubit pairs in the singlet state.
    \item {\bf Preparation of two-site chains.} Apply $U$ to each spin-1 site and memory qubit, resulting in $N/2$ two-site chains, each with edge modes (or equivalently, boundary conditions) entangled with unmeasured edge memory qubits.
    \item {\bf Fusion measurement.} Perform a Bell measurement on each pair of inter-site memory qubits, fusing each MPS with a probabilistically determined Pauli bond tensor $M_i$, each reducible to a Pauli defect $B_i$ (see Table \ref{tab:bell}).
    \item {\bf Correct defects (post-processing)}. Remove each Pauli defect $B$ by commuting it to one of the remaining memory qubits via application of $U_B$ to each intermediate spin-1 site (see Fig.\ \ref{fig:f4}c). We emphasize that this step need not be carried out on quantum hardware, and may alternatively be handled in post-processing without the need for classical feed-forward of measurement results (see below).
\end{enumerate}

While the final step seems to require classical feed-forward, the simplicity of the operators $U_B$ (see Eq.~\ref{eq:U_B}) allows for removal of defects through permutation of the encoded basis states. For the encoding we use on IBM-Q processors ($\ket{+} = \ket{10}$, $\ket{-} = \ket{01}$, $\ket{\bar{0}} = \ket{00}$), application of $U_B$ amounts to a Pauli frame change for each individual qubit. Details regarding the post-processing are included in the Supplemental Material \cite{SM}.

Crucially, the circuit depth of this algorithm is independent of $N$, reflecting a significant advantage over the known sequential preparation of matrix product states \cite{Schuch2007}, as well as other recently proposed dissipative \cite{Zhou2021}, adiabatic \cite{Wei2022c}, and non-deterministic measurement-based algorithms \cite{Murta2022, Chen2022} to prepare the AKLT state, each requiring total preparation times that scale with $N$. 

It is important to note that this speed-up does not come without trade-offs: due to the additional memory qubits, $3N + (N \textrm{ mod } 2)$ qubits are required in the measurement-assisted approach, compared to $2N+2$ for its sequential counterpart. Furthermore, because each sub-chain is prepared using the sequential approach, there are no savings in terms of entangling gates; in fact, the measurement-assisted approach requires more entangling operations due to the $\lfloor (N-1)/2\rfloor$ fusion measurements (see Supplemental Material \cite{SM} for optimized CNOT gate counts). Still, these are modest trade-offs for a constant-depth algorithm, particularly in the NISQ-era when processors are limited to fairly shallow circuits.

\section{Preparation of the AKLT state on IBM Quantum processors}\label{sec:experiments}

We now illustrate that the speed-up afforded by our algorithm amounts to a realizable improvement in state preparation on NISQ hardware. To show this, we carry out several experiments on IBM Quantum processors. In all cases, we encode spin-1 $S^z$ eigenstates in the computational basis of a pair of qubits: $\ket{+}=\ket{10}$, $\ket{\bar{0}}=\ket{00}$, and $\ket{-}=\ket{01}$. Such a choice allows for straightforward measurement in the $S^z$ basis, while also lending to a particularly efficient implementation of the site preparation gate $U$ (see Fig.\ \ref{fig:f6}b) and minimizing relaxation errors, as the disallowed ``singlet'' state corresponds to $\ket{s}=\ket{11}$. Importantly, this encoding choice bears no consequence on the symmetry-protected topological order of the state (or its usefulness as a resource for, e.g., MBQC), as a local single-site unitary exists which transforms from the chosen encoding to the more conventional triplet basis $\ket{+}=\ket{00}$, $\ket{-}=\ket{11}$, $\ket{\bar{0}}=(\ket{10}+\ket{01})/\sqrt{2}$

We first demonstrate the high-fidelity preparation of the ``building-blocks'' of the measurement-based protocol. In particular, we carry out quantum state tomography experiments for small chains composed of two and three sites. Following this, we prepare longer chains of up to $N=6$ sites (consisting of 12 qubits) using our measurement-assisted protocol on a 27 qubit device. To benchmark the state preparation, we measure the string order parameter $\langle O_{\textrm{str}}^z\rangle$ (see Eq.~\ref{eq:strorder}) and entanglement spectrum of the prepared chains, demonstrating good agreement with theoretical expectation. For comparison, we carry out companion experiments using the sequential preparation protocol outlined in Fig.\ \ref{fig:f2}, and find that the measurement-assisted scheme outperforms the sequential approach on an IBM Quantum processor, even for the relatively small system sizes studied here. Finally, we demonstrate post-preparation utility of the AKLT state by carrying out quantum teleportation experiments following our measurement-assisted scheme.

We note that the results presented in this section correspond to our best runs for both measurement-assisted and sequential approaches across repeated experiments over the course of several months. Though we find that the quality of experimental results varies between device calibrations, we observe that the measurement-assisted approach reliably outperforms its sequential counterpart. For further examples of representative experimental data across repeated experiments, see the Supplemental Material \cite{SM}.

\begin{figure*}
\centering
\includegraphics[width=1\linewidth]{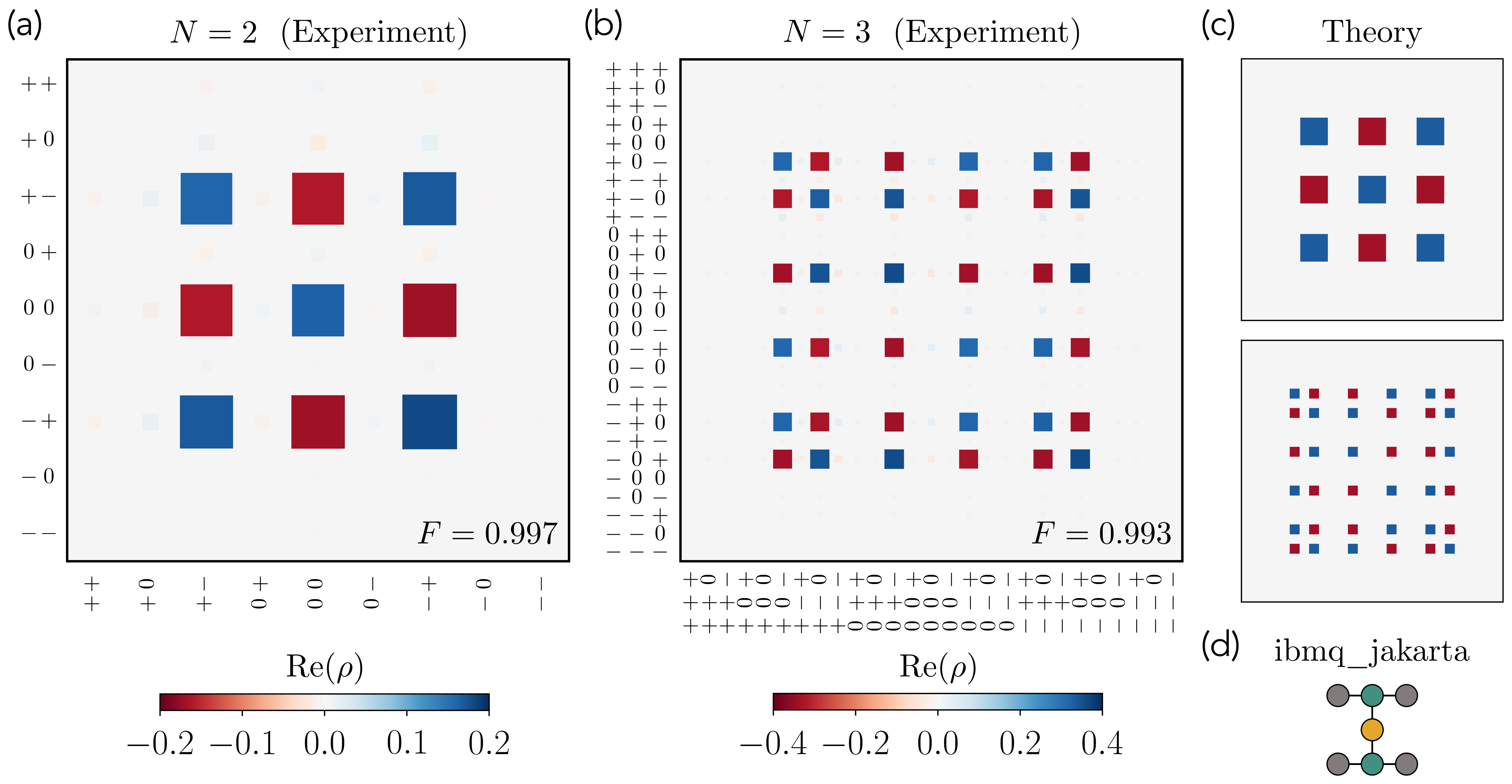}
\caption{Reconstructed density matrix of AKLT chains with periodic boundary conditions prepared on an IBM Quantum processor. For both $N=2$ and $N=3$ sites, we perform a full set of Pauli measurements, each consisting of $10^5$ shots. We post-select on measurement of valid spin-1 states, and use McWeeny purification to obtain a maximum-likelihood pure state $\rho$. The computed likelihood of $\rho$ is 88\% and 65\% for $N=2$ and $N=3$, respectively. In both cases, $\textrm{Im}(\rho)$ is expected to vanish. (a) Hinton diagram of $\textrm{Re}(\rho)$ for $N=2$ sites, with $\textrm{max}(\textrm{Im}(\rho)) = 0.019$ and fidelity $F(\rho,\rho_{\textrm{AKLT}})=0.997$ between the maximum-likelihood pure state $\rho$ and the exact AKLT state $\rho_{\textrm{AKLT}}$. (b) Hinton diagram of $\textrm{Re}(\rho)$ for $N=3$ sites, with $\textrm{max}(\textrm{Im}(\rho)) = 0.034$ and fidelity $F(\rho,\rho_{\textrm{AKLT}})=0.993$. (c) Exact theoretical density matrices for the $N=2$ (top) and $N=3$ (bottom) site AKLT ground state with periodic boundary conditions. (d) Qubit connectivity graph of the 7 qubit device {ibmq\_jakarta}. Preparation circuits are transpiled such that physical qubits in green encode the memory qubits for all but the initial entangling step, while pairs of gray qubits encode a spin-1 site. For $N=3$, we additionally post-select on measurement of the yellow ancillary qubit to ensure a successful swap after entangling the memory qubits, with a rejection rate $\sim2\%$.}

\label{fig:f5}
\end{figure*}

\subsection{Tomography experiments on small chains}\label{ssec:tomography_small}
We first benchmark the preparation of small AKLT chains of two and three sites with periodic boundary conditions on the 7 qubit processor {ibmq\_jakarta}. To reduce the total number of qubits required, we employ a measurement scheme that ultimately incorporates the memory qubits into the chain as an additional spin-1 site. 

In brief, this approach takes advantage of the previously discussed effect that measurement of the two edge memory qubits enforces particular boundary conditions for the chain with which they are entangled. Consequently, one may use a SWAP test to project the memory qubits onto either the singlet (antisymmetric) or triplet (symmetric) subspace. The former leaves the chain in the AKLT ground state with periodic boundary conditions. In contrast, the latter outcome enforces a particular superposition of triplet ground states; such a measurement crucially leaves the memory qubits themselves in the triplet subspace, now effectively an additional spin-1 site. Using the $\mathbb{Z}_2\times\mathbb{Z}_2$ symmetry of the AKLT state, it is possible to ``insert'' this additional spin-1 site into the chain such that the entire state is the AKLT state with periodic boundary conditions. We defer to Appendix \ref{sec:app:periodicboundaries} for details.

Due to the associated overhead of performing the SWAP test on physical hardware \footnote{Implementation of the SWAP test involves a Toffoli gate, and thus adds significant overhead to the state preparation circuit.}, we incorporate the memory qubits into the chain as a spin-1 site through projection onto the symmetric subspace during post-processing. The projection onto the spin-1 subspace thus serves a dual purpose: (1) it encompasses an error mitigation strategy for all sites, where we post-select for shots where all sites are in valid spin-1 states and (2) it enforces particular boundary conditions, where the memory qubits themselves become a spin site and are ``inserted'' into the AKLT chain by leveraging its $\mathbb{Z}_2\times\mathbb{Z}_2$ symmetry. 

Fig.\ \ref{fig:f5} displays tomography results for AKLT chains of $N=2$ and $N=3$ sites, with panels (a) and (b) displaying Hinton diagrams for the experimentally reconstructed density matrix for each case, with the exact theoretical density matrices shown for comparison in panel (c). We carry out a full set of Pauli measurements for each qubit, requiring 81 distinct measurement circuits  for the $N=2$ (four total qubits) and 729 distinct measurement circuits for the $N=3$ (six total qubits). Each circuit is carried out for $10^5$ shots, and the state of the system is reconstructed according to $\rho = \sum_{i=1}^{4^{2N}}P_i\braket{P_i}/2^{2N}$,
where each $P_i$ is a Pauli operator of weight $2N$, estimated from readout-error corrected \cite{Nation2021} measurement results. The density matrix $\rho$ is then projected onto the spin-1 subspace and McWeeny purification~\cite{google2020hartree,mcweeny1960some}, defined by the iteration
\begin{equation}
    \rho_{n+1} = 3\rho_n^2 - 2\rho_n^3,
    \label{eq:mcweeny}
\end{equation}
is applied in order to extract a maximum-likelihood estimate for a pure state from the reconstructed impure density operator.

For these small systems, the success of projecting the memory qubits into the symmetric and antisymmetric subspace is dependent on $N$: for $N=2$, the symmetric subspace is theoretically guaranteed, while for $N=3$ the triplet is measured with a probability of $1/3$. Upon projection of the memory qubits, we find a rejection rate of $5.1\%$ ($31.0\%$) for $N=2$ ($N=3$), in good agreement with theoretical expectation. Furthermore, we reject an additional $1.1\%$ ($2.7\%$) of shots upon post-selection of valid spin-1 states for all other sites.

The experimentally reconstructed and McWeeny purified density matrix shows excellent agreement for both $N=2$ and $N=3$ sites, with fidelities of $F=0.997$ and $F=0.993$ computed with respect to the exact AKLT ground state with periodic boundary conditions. Without purification, we find $F=0.940$ and $F=0.808$, still demonstrating good agreement with theory, but nonetheless exhibiting the utility of McWeeny purification as a post-processing technique.

\subsection{Measurement-assisted preparation of the AKLT state}\label{ssec:longchain}

\begin{figure*}
\centering
\includegraphics[width=1\linewidth]{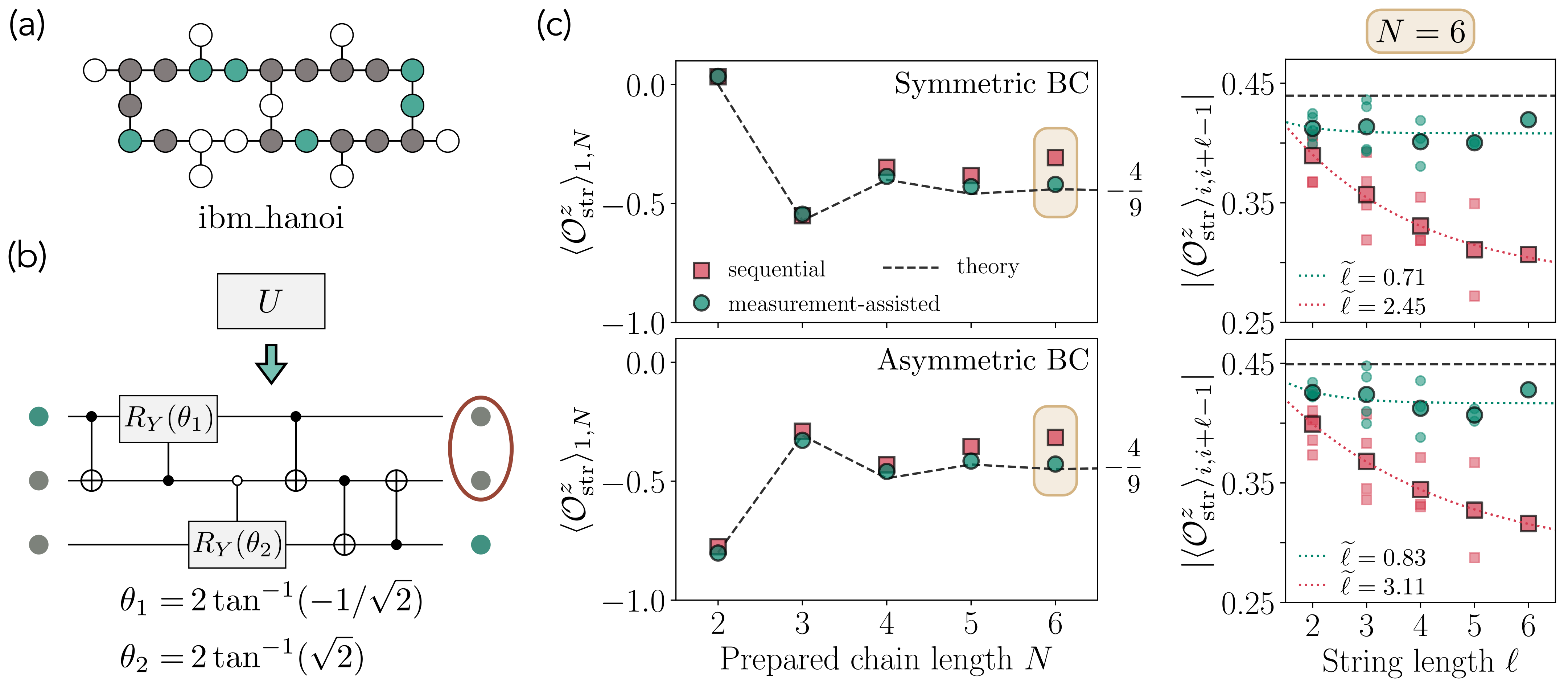}
\caption{Preparation of the AKLT state and measurement of string order on ibm\_hanoi. For all presented data, sampling uncertainties (estimated by bootstrapping measurement results acquired across $10^5$ shots per circuit) were found to lie within the displayed markers and are therefore not shown. (a) Qubit connectivity on ibm\_hanoi. Qubit colors indicate final qubit locations after measurement-assisted preparation of a $N=6$ site AKLT state. Gray, green, and white circles corresponding to site, memory, and unused qubits, respectively. (b) Implementation of $U$, adapted to permute the memory qubit to the far edge, enabling both measurement-assisted and sequential preparation schemes with only linear connectivity. (c) Measured string order $\braket{ \mathcal{O}^z_{\textrm{str}}}_{i,i+\ell-1}$ after sequential (red) and measurement-assisted (green) preparation, compared to theoretical expectation (dashed black line). Left panels display the string order measured from edge-to-edge ($i=1$, $\ell=N$) for a length $N$ chain, with top and bottom distinguishing between post-selection upon measurement of symmetric ($\ket{L}=\ket{R}$) and asymmetric ($\ket{L}=X\ket{R}$) boundary conditions. Right panels show measured values of $|\braket{ \mathcal{O}^z_{\textrm{str}}}_{i,i+\ell-1}|$ for $N=6$ and varying string length $\ell$. Small transparent markers represent individual measurements for varying $i$, while large bold markers indicate their average value for each $\ell$, and dashed lines correspond to their best-fit to an exponential curve $A e^{-\ell/\widetilde{\ell}} +B$ to guide the eye. Best-fit values for $\widetilde{\ell}$ are shown in the legend, with coefficients $A$ and $B$ reported in the Supplemental Material \cite{SM}.}
\label{fig:f6}
\end{figure*}

We now demonstrate the primary result of this work: the deterministic, measurement-assisted preparation of the AKLT state on a noisy quantum processor. Experiments are carried out on the 27-qubit IBM Quantum Falcon processor ibm\_hanoi. To account for the heavy-hex qubit connectivity graph of this device (Fig.\ \ref{fig:f6}a), we adapt the three qubit unitary $U$ such that the memory qubit is additionally swapped with the prepared spin-1 site (Fig.\ \ref{fig:f6}b), and is therefore in a position for either a fusion measurement (measurement-assisted preparation, Fig.\ \ref{fig:f4}b) or a subsequent application of $U$ (sequential preparation, Fig.\ \ref{fig:f2}). This implementation has an associated overhead of two CNOTs per application of $U$, but ultimately allows for the transpilation of both measurement-assisted and sequential preparation schemes with only linear connectivity and without additional swapping on physical hardware.

We prepare AKLT chains of up to $N=6$ sites (12 qubits) along the ``loop'' of qubits along the perimeter of the heavy-hex unit cell, as indicated in Fig.\ \ref{fig:f6}a where gray and green qubits indicate the final positions of spin-1-site and memory qubits. In total, our implementation of measurement-assisted preparation requires $3N + (N\mod 2)$ qubits whereas, by comparison, the sequential preparation requires just $2N + 2$ qubits. We note, however, that it is possible to reduce qubit waste in the former (at the expense of a constant gate-count overhead and non-deterministic prepared chain length) by implementing SWAP test fusion measurements in place of Bell measurements -- for details, see Appendix \ref{sec:app:periodicboundaries}.

\subsubsection{Measurement of string order}
To validate the measurement-assisted preparation of the AKLT state and benchmark it against both sequential preparation and theory, we first measure the string order $\langle \mathcal{O}_{\textrm{str}}^z\rangle_{i,j} = \langle S_i^z \prod_k^{j-1} \textrm{exp}(i \pi S^z_{k}) S_{j}^z \rangle$ between sites $i$ and $j$. This parameter is finite throughout the Haldane phase \cite{Nijs1989} and, at the AKLT point, has the special property that for a chain of fixed length $N$, $\langle \mathcal{O}_{\textrm{str}}^z\rangle_{i,j}$ is constant, independent of $i$ and $j$. Its particular value, however, is dependent upon boundary conditions, and furthermore oscillates with respect to $N$ about the well-known $N\to\infty$ value of $-4/9$ \cite{Nijs1989}. Measurement of string order is straightforward in our encoding, as the eigenstates of $S^z$ are computational basis states. Consequently, estimating the string order amounts to a measurement of all site qubits in the $Z$-basis. 

We batch and execute circuits on ibm\_hanoi using Qiskit Runtime \footnote{For more details, see \url{https://quantum-computing.ibm.com/lab/docs/iql/runtime/}}. Each circuit is run for $10^5$ shots, and measurement-error mitigation \cite{Nation2021} is applied. We post-select on valid spin-1 states, discarding shots where the (encoded) singlet $\ket{11}$ is measured at a spin-1 site. In addition, we reject shots where measured boundary conditions (enforced by the residual edge memory qubits) are inconsistent with the boundaries of the spin-1 chain \cite{SM}. Crucially, we do not post-select on known information about the AKLT state, such as its hidden antiferromagnetic ordering, or spatial inversion symmetry.

Fig.\ \ref{fig:f6}c displays the string order parameter measured for the AKLT state prepared by both measurement-assisted and sequential algorithms. Left panels show expectation values $\langle \mathcal{O}_{\textrm{str}}^z\rangle_{1,N}$ measured edge-to-edge for symmetric (top) and asymmetric (bottom) boundary conditions (BC). The former correspond to triplet ground states with edges spin-1/2s in the same state $|L\rangle = |R\rangle$ (i.e., the $\pm 1$ eigenstates of $S^z_{\textrm{tot}}$), while the latter correspond to edge spin-1/2s in the opposite state $|L\rangle = X|R\rangle$, superpositions of which compose the symmetric (triplet) and antisymmetric (singlet) eigenstates of $S^z_{\textrm{tot}}$ with eigenvalue 0.

While it can be reasonably expected that the constant-depth measurement-assisted scheme will outperform linear-depth sequential preparation for large $N$, we find that even for $N\leq6$ the former produces a state whose estimated string order is markedly improved over the latter. This is notable, as readout error is a primary source of noise in present-day superconducting devices \cite{Chen2019a, Bravyi2021}, and the success of our measurement-assisted preparation algorithm is inherently dependent on reliable fusion and identification of Pauli defects through measurement. This also underlines the importance of measurement-error mitigation for measurement-assisted circuits -- as shown in the Supplemental Material \cite{SM}, we find substantively improved agreement upon inclusion of such strategies.

Top- and bottom-right panels of Fig.\ \ref{fig:f6}c show companion measurements of $|\langle \mathcal{O}_{\textrm{str}}^z\rangle_{i,i+\ell-1}|$ for varying index $i$ and string length $\ell$ for an $N=6$ chain, acquired simultaneously with the data in the left panels. For the sequential preparation, we find that the string order follows a clear exponential decay for increasing string length $\ell$. This can intuitively be understood as follows: for the sequential preparation, the role of the memory qubit is to propagate correlations from site-to-site. Consequently, the length-scale of string order is inherently limited by the decoherence of the memory qubit. Contributing to this decoherence is not only the $T_1$ and $T_2$ of individual qubits (on the order of $\sim 100-200$ $\mu$s for most ibm\_hanoi qubits throughout these experiments), but in addition the cumulative decohering effect of any errors introduced by single- and two-qubit gates while the memory qubit is swapped along the chain and used to prepare spin-1 sites. 

For our best runs of the sequential preparation, exponential fits to the decay of $|\langle \mathcal{O}_{\textrm{str}}^z\rangle_{i,i+\ell-1}|$ for increasing string length $\ell$ yields a characteristic coherence length-scale $\widetilde{\ell}$ on the order of $2.5 -3$ sites. In contrast, our measurement-assisted approach uses each memory qubit to locally prepare a single site, and all correlations are propagated via a single layer of Bell measurements. We find that this approach leads to, at worst, a comparatively slower decay of string order with increasing $\ell$ \cite{SM} or, in the case of Fig.\ \ref{fig:f6}c, no obvious decay up to $\ell=6$.

\subsubsection{Measurement of the entanglement spectrum}
To further validate the prepared state, we measure its entanglement spectrum \cite{Li_2008} and compare against known theoretical results for the AKLT state \cite{Pollmann2010, Fan2004, Fan_2007, Geraedts_2010}. Inspired by arguments presented in Ref.\ \cite{Foss_Feig_2022}, we achieve this through state tomography of the residual boundary memory qubits following preparation of the spin-1 chain. 

To understand the relation between the state of the memory and entanglement of the prepared matrix product state, it is helpful to consider the sequential preparation scheme outlined in Section \ref{ssec:sequentialprep} and illustrated in Fig.\ \ref{fig:f2} (though the argument also holds for our measurement-assisted scheme). For clarity of explanation, we will specialize to the case where we have just one memory qubit whose initial and final state encode right- and left-boundary conditions. 

After preparation of the $i$th spin-1 site, the ``outgoing'' memory qubit encodes complete information to prepare all subsequent sites $j>i$. As a result, the bipartite entanglement between sites $j\leq i$ and the outgoing memory qubit \emph{exactly} coincides with the bipartite entanglement between sites $j\leq i$ and $j>i$ in a fictitiously prepared spin chain \cite{Gopalakrishnan2019, Foss_Feig_2022}. In other words, we can probe the entanglement between sites $j\leq i$ and $j>i$ entirely from the reduced density matrix of the memory. While at first surprising due to the dimension mismatch between the memory and that of all spin-1 sites $j>i$, it is helpful to recall that the dimension of the memory, $D=2$, by construction corresponds to the maximum Schmidt rank at an inter-site entanglement cut, equal to the bond dimension ($D=2$) of the AKLT state. Remarkably, this equivalence allows us to experimentally probe the entanglement content of a semi-infinite (or, in the case of dual parallel memory qubits, infinite) 1D AKLT state via tomography of the memory \cite{Foss_Feig_2022}.

\begin{figure}
\centering
\includegraphics[width=1\linewidth]{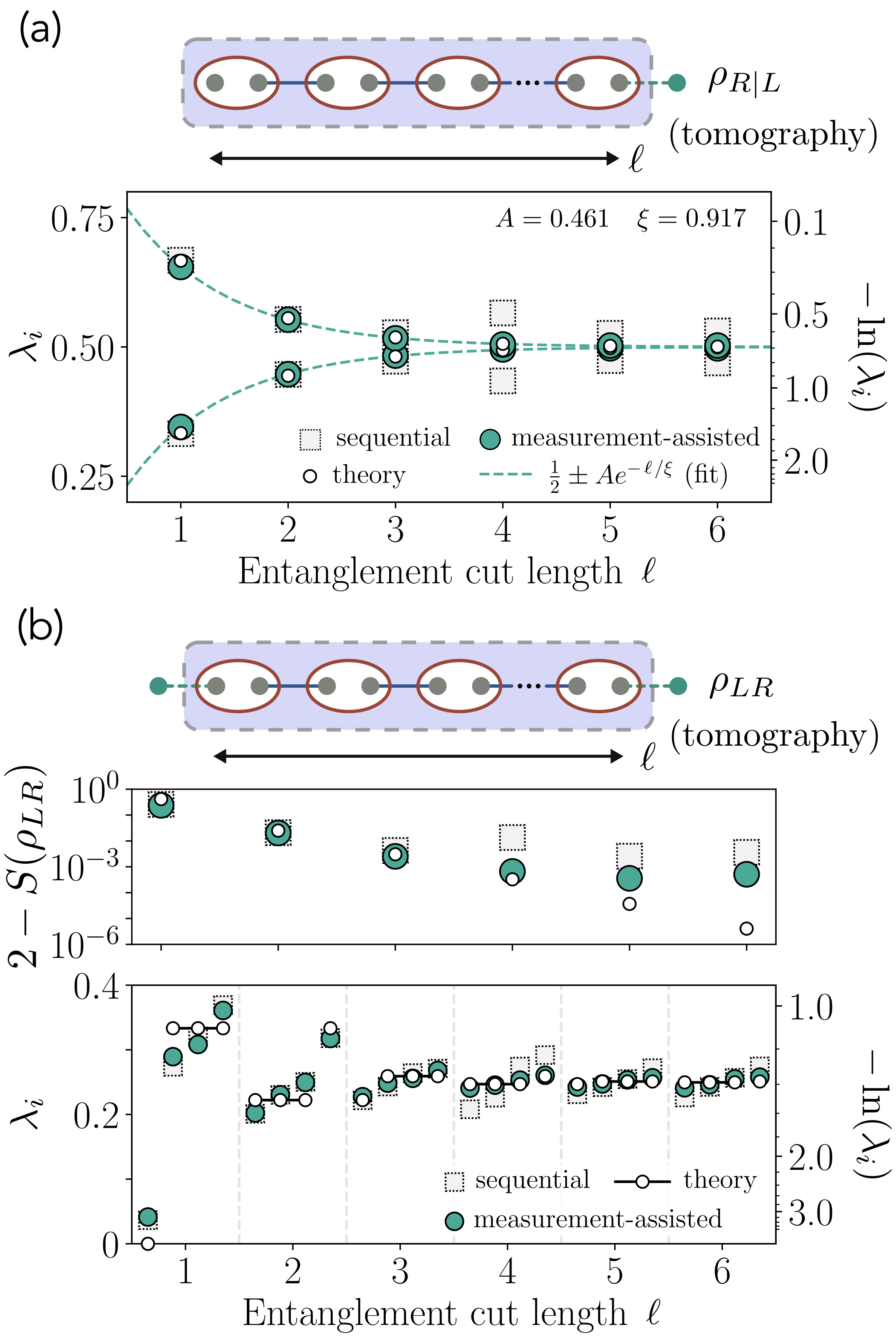}
\caption{Entanglement spectrum for varying entanglement cut length $\ell$, experimentally determined via state tomography of memory qubits. Left axes display eigenvalues $\lambda_i$ of (a) $\rho_{R|L}$ (b) $\rho_{LR}$, while right axes show the corresponding entanglement spectrum $-\ln(\lambda_i)$. As in Fig.\ \ref{fig:f6}, uncertainties were computed via bootstrap methods and found to fall within the displayed markers. (a) Entanglement spectrum estimated from the state of the right qubit given the measured state of the left, $\rho_{R|L}$. Such a measurement emulates the entanglement cut of a semi-infinite spin chain, in analogy to the holographic measurement scheme presented in Ref.~\cite{Foss_Feig_2022}. The dashed green line displays a least-squares best fit of $\frac{1}{2} \pm A e^{-\ell/\xi}$ to experimentally measured eigenvalues $\lambda_i$ for the state prepared via our measurement-assisted approach, with $\xi = 0.9172(4)$ -- in excellent agreement with the exact correlation length of the AKLT state, $\xi_{\textrm{AKLT}}=1/\ln(3) \approx 0.9102$. (b) Experimentally measured entanglement spectrum (bottom) and entanglement entropy (top), extracted from tomographic measurement of both memory qubits ($\rho_{LR}$). This measurement corresponds to a bulk length-$\ell$ entanglement cut of an infinite AKLT chain, expected to (exponentially quickly) converge to a fourfold degenerate entanglement spectrum and entropy $S(\rho_{LR})=2$ for increasing $\ell$ \cite{Fan2004,Fan_2007}.}
\label{fig:f7}
\end{figure}

We experimentally probe two distinct entanglement cuts of our system, as illustrated in Fig.\ \ref{fig:f7}. Following the preparation of a spin-1 chain of length $\ell$, we carry out a complete set of Pauli measurements for the two edge memory qubits, with each measurement circuit repeated for $10^5$ shots. We again post-select upon measurement of valid triplet states at spin-1 sites, and subsequently reconstruct the state of the memory qubits ($\rho_{LR}$) following the procedure outlined in Section \ref{ssec:tomography_small} (though here without McWeeny purification, as we expect a mixed state). This procedure is repeated for lengths up to $\ell=6$. All experiments are carried out on ibm\_hanoi using Qiskit Runtime. In order to mitigate for device parameter variability between calibrations, we batch both measurement-assisted and sequential preparation circuits as a single job.

In Fig.\ \ref{fig:f7}a, we probe the entropy across the bond just after site $N$ given complete knowledge of the left boundary condition $\ket{L}$. This last point is subtle, yet crucial for the validation of the prepared state: without enforcement of a left-boundary condition, such a partition emulates that of an infinite AKLT chain. Consequently, the right memory qubit $\rho_R$ is expected to be in a maximally mixed state independent of $\ell$, yielding the doubly degenerate entanglement spectrum characteristic of the Haldane phase \cite{Pollmann2010}. Experimentally measuring $\rho_R$ to be a maximally mixed state on a noisy device, however, provides little assurance that the memory has not simply decohered and become entangled with the environment. 

To mitigate for this, we introduce and study the impact of boundary effects \cite{Fan_2007} in a semi-infinite AKLT chain by conditioning the state of the right memory qubit on the outcome of the left:
\begin{equation}
    \rho_{R|L} = \langle L|\rho_{LR} | L\rangle,
\end{equation}
where $L = \in \{0,1\}$. Intuitively, the eigenvalues of $\rho_{R|L}$ convey information concerning the correlations between the left and right boundary spin-1/2s, expected to decay exponentially with increasing $\ell$ at a rate corresponding to the correlation length of the AKLT state, $\xi_{\textrm{AKLT}}=1/\ln(3)$ \cite{Geraedts_2010}.

Fig.\ \ref{fig:f7}a displays the measured eignvalues $\lambda_i$ of $\rho_{R|L}$ (and corresponding entanglement spectrum $-\ln(\lambda_i)$) as a function of $\ell$, averaged over left boundary conditions $L$. For the AKLT state prepared via measurement-assisted preparation, we find excellent agreement with theory and observe a clear exponential decay toward a two-fold degeneracy for large $\ell$. To verify, we perform a least-squares fit of $\lambda_i$ to $\frac{1}{2} \pm A e^{-\ell/\xi}$. Remarkably, we find $\xi=0.9172(4)$, differing from the well-known value of $\xi_{\textrm{AKLT}}=1/\ln(3)\approx 0.9102$ by just over 0.6\%. We additionally find a sensible best-fit coefficient $A = 0.461$, whose deviation from $1/2$ indicates imperfect preparation of the initial singlet state between memory qubits. We emphasize that we find this exceptional agreement with theory to be repeatable: two additional runs that yield measured correlation lengths of $\xi=0.9221(16)$ and $\xi = 0.9102(10)$ are included in the Supplemental Material \cite{SM}.

By contrast, eigenvalues $\lambda_i$ measured upon sequential preparation of the AKLT state agree less favorably with theory for $\ell>3$, and overall diverge from the expected trend $\frac{1}{2}(1\pm e^{-\ell/\xi_{\textrm{AKLT}}})$. We emphasize that both the measurement-assisted and sequential preparation circuits were executed within a single job and, in principle, were subject to comparable single- and multi-qubit errors (though each involves a different number of qubits, and the comparison is therefore imperfect). For all $\ell\geq4$, we find the right memory qubit to be biased toward the ground state $\ket{0}$, independent of the measured state of the left, suggesting the likely role of relaxation errors. This finding is in contrast with the measurement-assisted preparation scheme, where $\ket{0}$ and $\ket{1}$ are approximately equally likely for large $\ell$. Finally, we note that while agreement appears to improve for $\ell=5$ and $\ell = 6$, we reemphasize that caution must be exercised in its interpretation -- disagreement for $\ell = 4$ necessarily implies imperfect preparation of $\ell > 4$ for a sequentially prepared state. As a result, the apparent trend trend gives little confidence that the agreement for $\ell > 4$ is not in part due to decoherence of the memory, nor does it suggest that this enhanced agreement will persist.

As shown in Fig.\ \ref{fig:f7}b, we also measure the eigenvalues of the composite density matrix $\rho_{LR}$ of both memory qubits for the same run displayed in (a). Such a measurement emulates a partitioning that encircles $\ell$ sites from the bulk of an infinite MPS, expected to reveal a four-fold degeneracy in the case of the AKLT state for $\ell \gg \xi_{\textrm{AKLT}}$. For finite entanglement cuts, boundary effects lift this fourfold degeneracy, and instead three of the eigenvalues are exactly degenerate, with the fourth approaching degeneracy for large $\ell$ \cite{Fan2004,Fan_2007}. Intuitively, this is understood as a biasing of the edge states toward the triplet (singlet) for odd (even) chains.

We find that the measured entanglement spectrum agrees favorably with theoretical expectation, with the state prepared via the measurement-assisted circuit once again outperforming the sequential preparation past $\ell = 3$. In the upper panel, we compare the entanglement entropy $S(\rho_{LR})$ with the theoretical prediction of Ref.\ \cite{Fan2004}. We observe good agreement with the expected exponential trend for $\ell < 4$, past which the measured value of $2 - S(\rho_{LR})$ appears to plateau with minimum measured values of $3.6\times10^{-4}$ and $2.6\times10^{-3}$ for the measurement-assisted and sequential preparation, respectively. This observed ``noise floor'' can be understood as arising from qubit decoherence, readout errors and gate errors, all of which can all appreciably lift the expected degenaracy of a SPT prepared on a noisy quantum computer \cite{Choo2018}, but which nonetheless appear to have lesser impact on the measurement-based preparation as compared to its sequential counterpart.

\subsection{Quantum teleportation with the AKLT state} \label{ssec:teleportation}

While our focus has thus far narrowed on the task of state preparation, we reemphasize the practical utility of the AKLT state as a resource state for MBQC \cite{Brennen2008, Kaltenbaek2010}. In this section, we demonstrate that our measurement-assisted scheme prepares the AKLT state with sufficient coherence for such applications. In particular, we carry out the simplest possible version of a MBQC protocol: quantum teleportation.

\begin{figure*}
\centering
\includegraphics[width=1\linewidth]{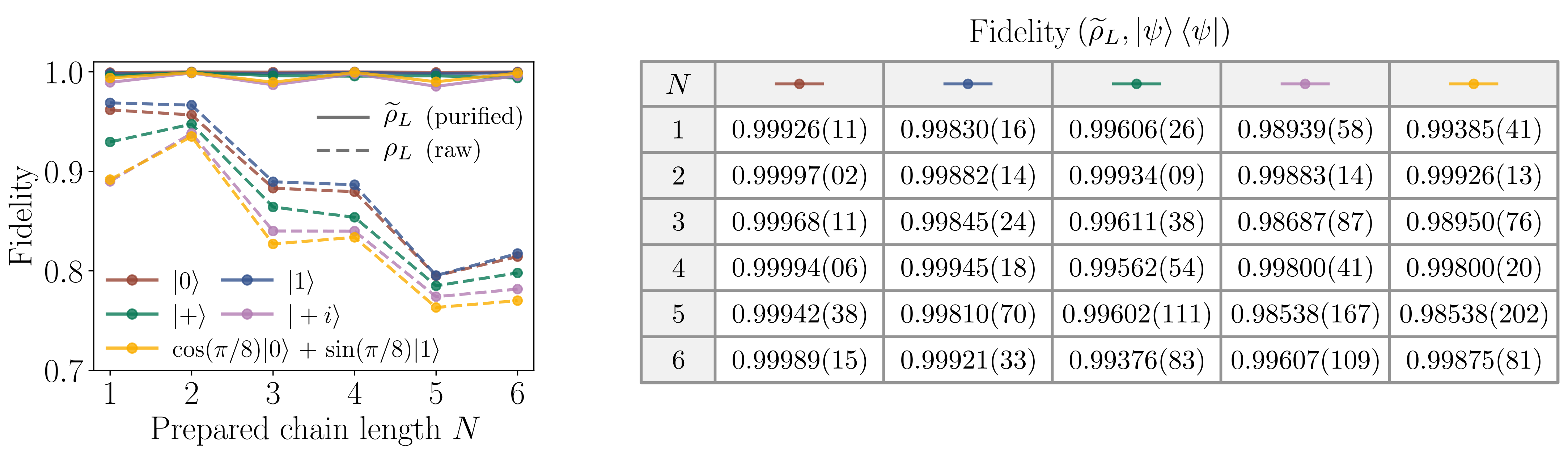}
\caption{Quantum teleportation fidelity as a function of AKLT chain length $N$ for an ensemble of target states. AKLT chains are prepared using our measurement-assisted scheme and subsequently used as a quantum wire to teleport a chosen target pure state $\ket{\psi}$ from the right chain edge to the left. The fidelity between the target and (tomographically reconstructed) received message is plotted both for the raw ($\rho_L$, markers with dashed lines) and McWeeny purified ($\widetilde{\rho}_L$, markers with solid lines) received message, with colors differentiating intended target states. Fidelities between the target and purified received message are additionally tabulated in the right panel, with digits in parentheses corresponding to 95\% confidence intervals computed via 50 bootstrap numerical experiments using $10^5$ shots per circuit.}
\label{fig:f8}
\end{figure*}

We note that the same core mechanism underlies both quantum teleportation and MBQC with the AKLT state -- the former can be simply viewed as a particular instance of the latter, with only identity operations applied. However, MBQC requires feed-forward to adapt measurement bases to past measurement outcomes \cite{Raussendorf2001}, while teleportation can be achieved without feed-forward. The implementation of feed-forward is a challenging and active area of development for current quantum devices \cite{Corcoles_2021, Gaebler2021}. We therefore focus on teleportation here, leaving a general demonstration of MBQC as a future direction.

The full details of our teleportation protocol are described in Appendix \ref{sec:app:teleportation}. In brief, the core idea is to teleport a quantum state $|\psi\rangle$ from Alice to Bob, using the AKLT state as a ``quantum wire'' \cite{Gross2007,Gross_2010}. Beginning with the resource state in Eq.~\ref{eq:AKLTpar} (i.e., the AKLT state with boundaries still entangled with memory qubits), Alice first prepares the right spin-1/2 edge state in the target state $\ket{\psi}$ by measuring the right memory qubit in an appropriate basis and post-selecting on the outcome. 

Next, all spin-1 sites are measured in the basis
$\{\ket{x},\ket{y},\ket{z}\}$, where
\begin{equation}
    \begin{split}
        \ket{x} &= (\ket{+} - \ket{-})/\sqrt{2} \\
        \ket{y} &= (\ket{+} + \ket{-})/\sqrt{2} \\
        \ket{z} &= \ket{\bar{0}},
    \end{split}
\end{equation}
such that the local tensors $A^m$ collapse onto one of three Pauli operators: $A^x=\sqrt{2/3}\,X$, $A^y=i\sqrt{2/3}\,Y$, or $A^z=-\sqrt{2/3}\,Z$. After all spin-1 sites have been measured, it can then be shown that the left memory qubit (which we assume to be in Bob's possession) is in the state $\ket{L} = \Lambda\ket{\psi}$, where $\Lambda\propto\{I,X,Y,Z\}$ is a known Pauli byproduct operator which can be removed (or accounted for upon measurement of $\ket{L}$) by Bob. Thus, the target state $\ket{\psi}$ has been teleported from Alice to Bob up to the known Pauli byproduct operator $\Lambda$.

We carry out this protocol on ibm\_hanoi using AKLT states of lengths ranging from $N=1$ to $N=6$ prepared via our measurement-assisted scheme. We initialize the target (pure) state $\ket{\psi}$ by measuring the right memory qubit in the basis $\{\ket{\psi},\ket{\psi_\perp}\}$ and post-selecting on the desired outcome. To assess the fidelity of the teleportation, we tomographically measure the state of the left memory qubit $\rho_L$ (taking into account the byproduct operator $\Lambda$), and compute the fidelity between the received and target message, $F(\rho_L, \ket{\psi}\bra{\psi})$. Leveraging the knowledge that the intended message is a pure state, we also compute the fidelity $F(\widetilde{\rho}_L, \ket{\psi}\bra{\psi})$, where $\widetilde{\rho}_L$ is the maximum-likelihood estimate for the received message obtained via McWeeny purification \cite{google2020hartree, mcweeny1960some} of the raw tomographically reconstructed state $\rho_L$ (see Eq.~\ref{eq:mcweeny}). 

While here we restrict our investigation to the teleportation of pure states, we note that mixed states can also be teleported using the AKLT state with a slight variation of our protocol. For example, Alice could first entangle her memory qubit with another subsystem in her possession. In this scenario, McWeeny purification cannot be used to boost the teleportation fidelity unless Bob additionally measures the subsystem with which the teleported qubit is entangled.

Each circuit is repeated for $10^5$ shots, not accounting for post-selection upon successful initialization of the target state which roughly halves the total shot count. As in the case of the above discussed experiments, we additionally post-select on measurement of valid spin-1 states, rejecting any shots where the singlet $\ket{s}=\ket{11}$ is measured (see the Supplemental Material \cite{SM} for rejection rates). We note that Pauli defects arising from fusion measurements do not need to be corrected in this scheme, and instead can be incorporated into $\Lambda$. In carrying out our teleportation experiments, we choose to fix the physical qubit location of the right (target) memory qubit for all $N$, while varying the physical qubit assignment of the left (receiving) memory qubit.

Fig.\ \ref{fig:f8} shows the teleportation fidelity for a selection of target states and AKLT chain lengths. Markers with dashed and solid lines indicate the fidelity computed using raw ($\rho_L$) and McWeeny purified ($\widetilde{\rho}_L$) density matrices, respectively. We additionally tabulate the latter fidelities in the right-hand table, with sampling uncertainties for the final digits indicated in parentheses. These uncertainties correspond to 95\% confidence intervals, estimated by bootstrapping experimental measurement counts \footnote{We note that while estimated uncertainties are smaller than would be expected for $10^5$ samples per circuit, this is a consequence of McWeeny purification. In contrast, we find that uncertainties for raw fidelities (computed using impure density matrices) scale roughly as $1/\sqrt{N_{\textrm{shots}}}$, as expected.}. Corresponding likelihoods are reported in the Supplemental Material \cite{SM}.

We find that the raw teleportation fidelities generally decrease with increasing chain length, following a step-like pattern where even length chains teleport the state as reliably as (and sometimes even more reliably than) the shorter, odd length chains of one fewer site. Given the high fidelity with which small chains can be prepared (see Fig. \ref{fig:f5}), we expect fusion measurements to be a primary source of error for our measurement-assisted preparation scheme. The observed step-like trend supports this expectation, as a length $N$ chain involves $\lfloor (N-1)/2 \rfloor$ fusion measurements. Consequently, the raw teleportation fidelity roughly tracks the number of fusion measurements.

In contrast, the fidelity of the purified state $\widetilde{\rho}_L$ is fairly consistent for all $N$ and $\ket{\psi}$. Because McWeeny purification converges to a density matrix with eigenvalues $\{0,1\}$ while leaving the eigenvectors unchanged, this suggests that incoherent errors are largely responsible for the decreasing raw teleportation fidelity for increasing $N$. Despite some variation for different $N$ and $\ket{\psi}$ -- to be expected due to variable physical qubit assignment for the receiving memory qubit and the  $\ket{\psi}$-dependent gates required to initialize the target -- we find purified teleportation infidelities to fall well below $1\%$ in most cases, demonstrating the ability to faithfully reconstruct a (pure) target state by sampling multiple uses of the teleportation channel and, more generally, the utility of purification techniques for NISQ-era experiments \footnote{We note that error reduction of a similar scale ($\sim 1\%$ error) has been realized by combining post-selection and McWeeny purification in other experiments. See, for example, Ref. \cite{google2020hartree}}.

We emphasize that for our largest chain consisting of $N=6$ sites, the target state is teleported across 16 intermediate qubits with linear connectivity. For this case, we observe raw and McWeeny purified fidelities in excess of 76\% and 99\% respectively for all target states, the former in agreement with a rough estimate of the expected error considering both the total CNOT count and post-selection (see the Supplemental Material \cite{SM} for more information). Crucially, the raw fidelity exceeds the classical teleportation fidelity limit of 2/3 \cite{Massar_1995}, indicating that our measurement-assisted scheme not only prepares the AKLT state efficiently, but with sufficient coherence for its subsequent use as a resource state on a NISQ device.

\section{Summary and Discussion}\label{sec:conclusion}
In this work we have presented a constant-depth protocol to deterministically prepare the paradigmatic spin-1 AKLT state. This result is notable as the AKLT state, having a finite correlation length, cannot be prepared exactly from a product state using a constant-depth circuit composed only of local unitary gates, instead requiring a circuit whose depth scales with system size. Furthermore, all known methods to prepare the AKLT state -- including sequential unitary, \cite{Schoen2005,Huang2015}, dissipative \cite{Zhou2021}, adiabatic \cite{Wei2022c}, and non-deterministic measurement-based \cite{Kaltenbaek2010, Murta2022, Chen2022} strategies -- all require preparation times that scale with system size. Here, we have overcome this limitation by augmenting a constant-depth circuit with fusion measurements.

The core idea behind our approach is to parallelize the sequential preparation of an MPS \cite{Schoen2005} by blending tensor network representations with LOCC-assisted quantum circuits. In particular, this is achieved through preparation and subsequent fusion of small matrix product states through Bell measurement of their bond space ancillae, here termed memory qubits. While the outcome of fusion measurement is inherently probabilistic, we show that the $\mathbb{Z}_2\times \mathbb{Z}_2$ symmetry of the AKLT state, along with knowledge of the measurement result, allows for recovery and deterministic state preparation regardless of measurement result. Importantly, all post-measurement operations correspond to permutations of our local encoding and can thus be handled in post-processing, thereby avoiding the need for feed-forward which is an active area of development on current quantum devices \cite{Corcoles_2021, Gaebler2021}.

We have demonstrated the realizable improvement afforded by our measurement-based preparation scheme by preparing the AKLT state of up to $N=6$ sites (composed of 12 qubits) on the 27 qubit IBM Quantum processor ibm\_hanoi, comparing both measurement-assisted and sequential preparation schemes. We measured both the string order and entanglement spectrum of prepared states and, using these as metrics, find that our measurement-assisted approach reliably outperforms its sequential counterpart even for the relatively small system sizes studied here. 

Finally, we have carried out quantum teleportation experiments using a prepared AKLT state as a quantum wire. In particular, we have illustrated that our measurement-assisted preparation scheme leaves ample circuit depth for subsequent use of the state, with raw teleportation fidelities exceeding 76\% for our longest prepared chains, surpassing the classical teleportation threshold of 2/3 \cite{Massar_1995}. Furthermore, we have shown that teleportation fidelities are enhanced to 99\% when combined with post-processing purification techniques, demonstrating the ability to faithfully reconstruct the intended message upon repeated sampling of the teleportation channel.

While our approach yields a dramatic reduction in circuit depth relative to the purely unitary approach, it is important to note that this does not come without trade-offs. In addition to a reliance on high fidelity measurement, our constant-depth scheme requires an additional two qubits and one entangling gate per fusion measurement compared to the sequential approach. Here, we have demonstrated this trade-off to yield favorable results for small system sizes on an IBM-Q processor. However, the degree to which this holds true on other NISQ-era devices and architectures will be dependent on the strengths and limitations of each particular platform.

More broadly, our work sits firmly in the context of recent work exploring the enhanced state preparation capabilities of finite-depth unitary circuits augmented by LOCC \cite{Piroli_2021, Tantivasadakarn2022a}. In particular, it has been shown that a wide array of long-range entangled states -- such as the GHZ state, toric code and certain non-Abelian topological orders -- are all preparable by augmenting unitary evolution with measurement \cite{Tantivasadakarn2021,Verresen2021,Tantivasadakarn2022,Lu2022}. This work thus not only provides the most efficient known strategy to prepare a specific, nontrivial resource in the AKLT state -- useful for applications ranging from quantum teleportation and MBQC \cite{Verstraete2004,Gross2007a, Brennen2008, Kaltenbaek2010} to blind quantum computation \cite{morimae2015ground} and remote state preparation \cite{liu2014controlled} -- but, more broadly, serves as the first experimental demonstration of deterministic, measurement-assisted preparation of a non-stabilizer state on a NISQ device, to the best of our knowledge. That we find improvement over the known unitary preparation of the AKLT state, even for the small systems sizes studied here, illustrates the clear promise of measurement-based circuit depth reduction strategies for NISQ-era applications.

We conclude by commenting on a few future directions opened by this research. One immediate question is whether or not the 2D AKLT state on a honeycomb lattice, a   universal resource for MBQC \cite{Wei2011} for which no sequential state preparation algorithm is known \cite{Wei2022c}, can be prepared using the techniques developed here. We suspect that it is indeed possible to prepare this state up to local Pauli defects (which have no effect on its utility as a resource for MBQC) using the symmetries of the local tensors in its PEPS construction \cite{Wei2022}, though leave the details to future work. 

Another interesting question is to what extent the methods presented here, which in some sense allow for a speed-up of the sequential preparation algorithm in Ref.\ \cite{Schoen2005}, can be generalized to other tensor network states -- both of higher spatial and bond dimension -- and what restrictions are imposed by the requirement that any defect introduced by a fusion measurement is correctable. While here we have focused on an MPS of bond dimension $D=2$ (and have provided additional $D=2$ examples in the cluster and GHZ states in Appendix \ref{sec:app:GHZcluster}), we suspect that our scheme is universal for the Haldane phase, as it is always possible to decompose the virtual subspace into a ``protected'' subsystem (with $D=2$), and a ``junk'' subsystem, on which the projective representation of $\mathbb{Z}_2\times\mathbb{Z}_2$ acts trivially \cite{Else2012}. For more general symmetries, it is likely that deterministic fusion of tensor network states will necessitate more general measurement schemes. We leave these directions for future work.

\vspace{10pt}
\noindent\emph{Code and data availability} -- Experimental data and code pertaining to this work are available upon reasonable request. 

\noindent\emph{Note added} -- During the preparation of this work, several papers have proposed algorithms to prepare AKLT states using adiabatic \cite{Wei2022c} and non-deterministic measurement-based \cite{Murta2022, Chen2022} approaches, all of which involve preparation times dependent on system size (though with extension to 2D AKLT states in Refs.\ \cite{Wei2022c,Murta2022}). Ref.\ \cite{Chen2022} additionally involves experiments on IBM Quantum devices. We emphasize that our scheme stands in contrast to these, as it is constant-time and entirely deterministic. Separately, we note similarities between our work and Ref. \cite{Lu2022}, which theoretically proposes adaptive, measurement-assisted finite-depth circuits to prepare tensor network states.

\section*{Acknowledgements}
We would like to thank Tyler Ellison for many beneficial conversations pertaining to this work, and for a close reading of the manuscript. We additionally thank Sarah Sheldon and Abhinav Kandala for their counsel in running experiments on IBM Quantum hardware. We thank Tzu-Chieh Wei, Michael DeMarco, Layla Hormozi, Alexander Schuckert, and Andrew Fisher for discussions relevant to this work. We thank the Brookhaven National Laboratory operated IBM-Q Hub for additional support. We acknowledge the use of IBM Quantum services for this work. The views expressed are those of the authors, and do not reflect the official policy or position of IBM or the IBM Quantum team.

This project was primarily supported by the U.S. Department of
Energy, Office of Science, National Quantum Information
Science Research Centers, Co-design Center for Quantum
Advantage under contract number DE-SC0012704.
Eleanor Crane was supported by the Yale-UCL exchange scholarship from
RIGE, as well as ARO MURI W911NF-15-1-0397, and the U.S. Department of Energy, Office of Science, National Quantum Information Science Research Centers, Quantum Systems Accelerator.

\appendix
\section{SWAP-based fusion and deterministic enforcement of periodic boundary conditions}\label{sec:app:periodicboundaries}
In this section, we describe an alternative measurement scheme for fusing AKLT chains using a SWAP test. This strategy is more expensive than using Bell measurement, requiring a Fredkin gate (decomposable into a Toffoli and two CNOTs) and an ancilla. Its advantage is that it is less destructive, projecting the memory qubits onto the symmetric or antisymmetric subspace, but revealing no further information. As we shall see, this -- along with the $\mathbb{Z}_2\times\mathbb{Z}_2$ symmetry of the AKLT state -- can be leveraged to probabilistically include the memory qubits in the prepared chain as additional spin-1 sites. We can also perform a SWAP test on the final edge memory qubits after preparation of the chain to enforce boundary conditions and, there, this same feature allows for the deterministic preparation of the AKLT state with periodic boundary conditions. 

Before measurement, the joint state of two independent AKLT chains can be written as
\begin{equation}
    \begin{split}
        |\Psi\rangle = \sum_{i j k \ell \vec{m}}\langle i| A^{m_1}\ldots A^{m_{n-1}}P^{m_n}\ket{j}\bra{k} A^{m_{n+1}}\ldots & \\
        \times A^{m_{N-1}}P^{m_N} |\ell\rangle|i j k \ell\rangle |\vec{m}\rangle &.
    \end{split}
\end{equation}
Measurement via the SWAP test amounts to application of the symmetric or antisymmetric projection operators: $\mathcal{S} = \ket{s}\bra{s}$ or $\mathcal{P} = \sum_m \ket{m}\bra{m}$, where $\ket{s}$ is the singlet state and $m\in\{+,0,-\}$ labels the triplet states. 

Let us first imagine we measure the intermediary memory qubits (indexed $j$ and $k$) to be in the singlet state. Applying $\mathcal{S}$ to $\ket{\Psi}$, and noting that $\langle s | jk\rangle = S_{jk}$, we find
\begin{equation}
    \begin{split}
        \mathcal{S}|\Psi\rangle &= \sum_{i \ell \vec{m}}\langle i| A^{m_1}\ldots P^{m_n} S A^{m_{n+1}}\ldots P^{m_N} |\ell\rangle |i \ell\rangle |\vec{m}\rangle \ket{s} \\
        &= \sum_{i \ell \vec{m}}\langle i| A^{m_1}\ldots A^{m_n}A^{m_{n+1}}\ldots P^{m_N} |\ell\rangle |i \ell\rangle |\vec{m}\rangle \ket{s},
    \end{split}
\end{equation}
Measurement of the singlet state thus teleports a singlet bond between the edge spin-1/2s, forming a larger AKLT state through fusion of two smaller ones. This is of course no different from measuring the Bell state $|\Psi^-\rangle$.

The more interesting situation is when we project onto the triplet subspace. Applying $\mathcal{P}$ to $\ket{\Psi}$, we find
\begin{equation}
    \begin{split}
        \mathcal{P}|\Psi\rangle = \sum_{i \ell \vec{m}}\sum_{m'}\langle i| A^{m_1}\ldots P^{m_n} P^{m'} A^{m_{n+1}}\ldots & \\
        \times P^{m_N} |\ell\rangle |i \ell\rangle |\vec{m}\rangle \ket{m'}&.
    \end{split}
\end{equation}
This not quite the AKLT state -- in place of a singlet bond, we now have a matrix which is dependent on the (symmetric) state of the memory qubits. We now show that this can be ``patched up'' by invoking the transformation law corresponding to $\mathbb{Z}_2\times\mathbb{Z}_2$ symmetry in Fig.\ \ref{fig:f4}a. Recalling that $S\propto Y$, it can be shown that 
\begin{equation}
    \sum_{mm'} (U_Y)_{mm'} A^{m'} = e^{i\theta_Y} S P^m S,
\end{equation}
where $U_Y = e^{i\pi S_Y}$ and $S_Y$ is the spin-1 operator polarized along the $y-$axis. Applying $U_Y$ to the spin-1 site indexed by $m'$, ignoring the global phase, and reindexing sites to the right of the fusion site, we find
\begin{equation}
    \begin{split}
        U_Y \mathcal{P} |\Psi\rangle = \sum_{i \ell \vec{m}}\sum_{m'}\langle i| A^{m_1}\ldots P^{m_n} S P^{m'} S A^{m_{n+1}}\ldots & \\
            \times P^{m_N} |\ell\rangle |i \ell\rangle |\vec{m}\rangle \ket{m'}&. \\
            = \sum_{i \ell \vec{m}}\sum_{m'}\langle i| A^{m_1}\ldots A^{m_{N}} P^{m_{N+1}} |\ell\rangle |i \ell\rangle |\vec{m}\rangle&,
    \end{split}
\end{equation}
yielding the AKLT state with $N + 1$ sites and boundary conditions still entangled with the unmeased edge memory qubits. The SWAP test can therefore be used to deterministically fuse AKLT chains, where recovery upon measurement of the symmetric subspace is achieved by inserting the memory qubits into the chain as an additional spin-1 site.

As a final note, the above strategy can also be used to deterministically prepare the AKLT state with periodic boundary conditions, albeit with one additional site if the edge memory qubits are measured to be in the symmetric state. This latter feature was used in Section \ref{ssec:tomography_small} to prepare two- and three-site chains using a minimal number of qubits.

\section{Sequential preparation of an MPS using two memory qudits in parallel}\label{sec:app:parallelmemory}

In this section, we present a generalization of the sequential preparation strategy of Section \ref{ssec:generalsequential} which halves the circuit depth at the expense of doubling the dimension of the bond-space ancillary system. In particular, we show that for certain matrix product states (the AKLT state included), it is possible to use two memory qudits of dimension $D$ to ``grow'' an MPS at both boundaries, in parallel. This technique was leveraged for both the sequential and measurement-assisted preparation experiments on IBM Quantum processors, as shown in Fig.\ \ref{fig:f2} and Fig.\ \ref{fig:f4}b.

For simplicity, we will specialize to an MPS with translational invariance. Similar to Section \ref{ssec:generalsequential}, we begin with two memory qudits, each of dimension $D$, and $N$ $d$-dimensional subsystems. Our goal is to prepare the matrix product state,
\begin{equation}
    \ket{\Psi} = \sum_{LR}\sum_{\vec{m}}|L\rangle\otimes |R\rangle\otimes\langle L|A^{m_1}A^{m_2}\ldots A^{m_N}|R\rangle|\vec{m}\rangle,
    \label{eq:app:desiredstate}
\end{equation}
with boundary conditions entangled with the ``left'' and ``right''  memory qudits, indexed by $L$ and $R$. In addition, we assume this MPS to be in left-canonical form. We begin by initially preparing the two memory qudits in the state
\begin{equation}
    |\Lambda\rangle = \sum_{ij}\Lambda_{ij} |i\rangle \otimes |j\rangle,
\end{equation}
and all $d$-dimensional subsystems in the state $|\psi_0\rangle$.

Let us now define two distinct unitaries which act on a $d$-dimensional subsystem and one memory qudit:
\begin{equation}
    \begin{split}
        U_L &= \sum_{m}A^{m} \otimes I\otimes \ket{m}\bra{\psi_0} + C_L \\
        U_R &= \sum_{m_n}I\otimes B^{m} \otimes \ket{m}\bra{\psi_0} + C_R,
    \end{split}
\end{equation}
with constraint $\sum_m B^{m\dagger}B^m = I$ to ensure unitary (note that $\sum_m A^{m\dagger}A^m = I$ is automatically satisfied in left-canonical form).

Without loss of generality, we assume $N$ to be even. Beginning with sites indexed $N/2$ and $N/2 + 1$, sequential application of $U_L$ and $U_R$ to all ``left'' ($j\leq N/2$) and ``right'' ($j \geq N/2 +1$) sites yields
\begin{equation}
    \begin{split}
        |\Psi\rangle &= \sum_{ij}\Lambda_{ij}\sum_{\vec{m}}A^{m_1}A^{m_2}\ldots A^{m_{N/2}}\ket{i} \\ &\otimes B^{m_{N}}B^{m_{N-1}}\ldots B^{m_{N/2+1}}\ket{j}\otimes \ket{\vec{m}}
    \end{split}
\end{equation}
Left-multiplying by the resolution of the identity for both memory qudits $I=\sum_{LR}\ket{L}\bra{L}\otimes \ket{R}\bra{R}$ and noting that $\bra{i} M \ket{j} = \bra{j} M^T \ket{i}$ for any matrix $M$, we can rewrite $\ket{\Psi}$ in the form
\begin{equation}
    \begin{split}
        \ket{\Psi} = &\sum_{LR}\sum_{\vec{m}}\ket{L}\otimes\ket{R}\otimes\bra{L}A^{m_1}A^{m_2}\ldots A^{m_{N/2}} \\
        &\times\Lambda(B^{m_{N/2+1}})^T(B^{m_{N/2+2}})^T\ldots(B^{m_{N}})^T\ket{R}\ket{\vec{m}}
    \end{split}
    \label{eq:app:finalstate}
\end{equation}
where we have defined the matrix $\Lambda = \sum_{ij}\Lambda_{ij}\ket{i}\bra{j}$.

In preparing the AKLT state, we have taken $B^m = A^m$ (i.e., the same unitary $U$ was applied to both ``left'' and ``right'' sites). We are now in a position to elucidate the reason for this choice and its relation to the spatial inversion symmetry of the AKLT state. Similar to the manifestation of  $\mathbb{Z}_2\times\mathbb{Z}_2$ symmetry at the level of the local tensors $A^m$ in Eq.~\ref{eq:SPTfundamental}, spatial inversion symmetry of the AKLT state implies that \cite{Pollmann2010}
\begin{equation}
    (A^m)^T = -Y A^m Y.
\end{equation}
Noting that the Pauli $Y$ operator is proportional to the singlet matrix $S$ and recalling that $A^m = P^m S$ and $S^2 = I$, the final state $|\Psi\rangle$ can be rewritten up to a global phase and normalization factor as
\begin{equation}
    \begin{split}
        \ket{\Psi} = &\sum_{LR}\sum_{\vec{m}}\ket{L}\otimes\ket{R}\otimes\bra{L}A^{m_1}A^{m_2}\ldots A^{m_{N/2}} \\
            &\times\Lambda S A^{m_{N/2+1}} A^{m_{N/2+2}}\ldots P^{m_{N}} \ket{R}\ket{\vec{m}},
    \end{split}
\end{equation}
yielding the AKLT state for $\Lambda = S$, motivating our choice of initial state for the memory qubits in the main text.

More generally, spatial inversion symmetry provides a sufficient condition to grow an MPS in both left and right directions in parallel. To prepare the general state in Eq.~\ref{eq:app:desiredstate}, for example, one possible choice is $(B^m)^T = A^m$, which imposes the condition
\begin{equation}
    \sum_m A^{m*}A^{m*\dagger}=I,
    \label{eq:app:spatialinv}
\end{equation}
equivalent to right-canonical form for real tensors $A^m$. For an MPS with spatial inversion symmetry, the corresponding transformation law \cite{Pollmann2010} $(A^{m})^T = e^{i\theta_\mathcal{I}}U_\mathcal{I}^\dagger A^m U_\mathcal{I}$ can be used to readily prove that Eq.~\ref{eq:app:spatialinv} holds for an MPS in left-canonical form.

\section{Quantum teleportation with the AKLT state} \label{sec:app:teleportation}
In this section we outline the quantum teleportation protocol carried out at the end of Section \ref{ssec:longchain}. For simplicity, we will take as a starting point the $N$-site AKLT state with boundary conditions entangled with the memory qubits,
\begin{equation}
    \ket{\Psi} = \sum_{ij}\sum_{\vec{m}} \bra{i} A^{m_1}A^{m_2}\ldots A^{m_{N-1}}P^{m_{N}}\ket{j}\ket{ij}\ket{\vec{m}},
\end{equation}
prepared either through the sequential or measurement-assisted approach. We note, however, that any state equivalent to $\ket{\Psi}$ up to known Pauli defects is suitable, as it is possible to instead account for such defects in the teleportation byproduct operator. Consequently, Pauli defects arising in our measurement-assisted preparation scheme need not be removed prior to the teleportation protocol, and can instead be incorporated into the teleportation byproduct operator. This feature allows us to carry out the teleportation protocol on IBM Quantum processors without feed-forward.

Following 
Ref.\ \cite{Brennen2008}, we first define a convenient spin-1 basis for quantum teleportation/MBQC:
\begin{equation}
    \begin{split}
        \ket{x} &= (\ket{+} - \ket{-})/\sqrt{2} \\
        \ket{y} &= (\ket{+} + \ket{-})/\sqrt{2} \\
        \ket{z} &= \ket{\bar{0}},
    \end{split}
    \label{eq:teleportationbasis}
\end{equation}
where we have renamed the $m=0$ triplet state for notational purposes. Expressing the tensors $A^m$ in this basis, we find $A^x=\sqrt{2/3}\,X$, $A^y=i\sqrt{2/3}\,Y$, and $A^z=-\sqrt{2/3}\,Z$.

Our protocol begins with measurement of the right memory qubit (here indexed by $j$), thereby enforcing a definite right boundary condition or, equivalently, a definite state for the right edge spin-1/2. In particular, we aim to initialize the edge spin-1/2 to the target state $\ket{\psi}$ by measuring the memory qubit in the basis $\{\ket{\psi},\ket{\psi_\perp}\}$, post-selecting on the desired result. 

We note that post-selection is not strictly necessary in the case where the desired state lies along one of the cardinal axes, as an unsuccessful measurement can be accounted for in the final Pauli byproduct operator. For a general state, however, recovery upon unsuccessful measurement requires access to non-Clifford operations. We assume this not to be the case, and instead post-select on successful initialization of the state $\ket{\psi}$.

After initialization of the target state $\ket{\psi}$, the state of the system may be written in the form
\begin{equation}
\ket{\Psi'} = \sum_{i}\sum_{\vec{m}}\ket{i}\bra{i} A^{m_1}A^{m_2}\ldots P^{m_{N-1}}\ket{\psi}\ket{\vec{m}}\otimes\ket{\psi},
\end{equation}
where we have rearranged the ordering of tensor products. In addition, we use the tensor product operator $\otimes$ to make explicit the portions of the system in a product state.

Next, we measure each spin-1 state in the basis defined in Eq.~\ref{eq:teleportationbasis}. For the encoding used in the main text, $\{$$\ket{+}=\ket{10}$,  $\ket{-}=\ket{01}$, $\ket{\bar{0}}=\ket{00}$, $\ket{s}=\ket{11}$$\}$, this is achieved by first transforming the spin-1 sites with the circuit
\begin{equation}
    \begin{quantikz} 
        \lstick{$\ket{q_0}$} & \targ{} & \ctrl{1} & \targ{} & \qw  \\
        \lstick{$\ket{q_1}$} & \ctrl{-1} & \gate{R_Y(-\pi/2)} & \qw & \qw,
    \end{quantikz}
\end{equation}
which carries out the mapping
\begin{equation}
    \begin{split}
        \ket{x} &\to \ket{10} \hspace{20pt} \ket{y} \to \ket{00} \\
        \ket{z} &\to \ket{01} \hspace{20pt}
        \ket{s} \to \ket{11}
    \end{split}
\end{equation}
for the input state $\ket{q_1 q_0}$. All spin-1 composing qubits are then measured in the computational basis.

Noting that the Pauli group is closed under multiplication, we can use the fact that $A^m$ and $P^m$ are proportional to Pauli operators for all $m\in\{x,y,z\}$ to write the state of the system after measurement of all spin-1 sites as
\begin{equation}
\ket{\Psi''} \propto \sum_{i}\ket{i}\bra{i}\Lambda\ket{\psi}\otimes\ket{\Phi},
\end{equation}
where $\ket{\Phi}$ denotes the state of all measured qubits, and $\Lambda\in\{I,X,Y,Z\}$ is the product of all Pauli matrices corresponding to measurement outcomes up to a global phase. Noting the resolution of identity to the left of $\Lambda$, the unmeasured left memory qubit is in the state $\ket{L} = \Lambda\ket{\psi}$. We have therefore deterministically teleported the state $\ket{\psi}$ from the right memory qubit to the left, up to the known byproduct operator $\Lambda$ which can be removed with a single Pauli operation (or alternatively accounted for in post-processing).

\section{Measurement-assisted preparation of the GHZ and cluster states}
\label{sec:app:GHZcluster}

We demonstrate two further examples of matrix product states in this section -- the cluster state and GHZ state -- which can be prepared using the strategy presented here, i.e., parallel preparation and subsequent deterministic fusion of small resource states through measurement. We emphasize that the possibility for constant-depth measurement-assisted preparation of both these resources is known \cite{Browne_2005, Li_2015, Bartolucci2021, Tantivasadakarn2021, Quek2022}. Nonetheless, we find that the formalism presented here -- the preparation and subsequent fusion of small resource states via Bell measurement and the correction of defects leveraging symmetry -- provides a clear and elucidating preparation protocol. In addition, we find that our preparation of the GHZ state is especially resource efficient, with every involved qubit ultimately encoding a site, in contrast to the recently proposed routine that yields the GHZ state on a subset of the total qubits \cite{Tantivasadakarn2021}.

While here we focus on the 1D cluster and GHZ states for clarity, we note that the general procedure readily extends to higher dimensions.

\subsection{Cluster state}

We begin with the 1D cluster state, the prototypical example of an MPS for MBQC \cite{Raussendorf2001, Gross2007}. It is important to note that the cluster state has zero correlation length, and can therefore be prepared by a constant-depth unitary circuit. However, its measurement-assisted preparation is still of interest, particularly for linear optical quantum computing architectures where Bell measurements are used in lieu of unitary entangling gates \cite{Browne_2005,  Li_2015}. In addition, the measurement-based preparation of cluster states is a core component to the recently developed framework of fusion-based quantum computation \cite{Bartolucci2021}.

The cluster state is most simply prepared by initializing an array of qubits in the $|+\rangle$ state and subsequently performing a controlled-$Z$ gate between adjacent pairs of qubits. Because all of the controlled-$Z$ gates commute, the preparation time is independent of the size of the system. 

As described in Ref.\ \cite{Gross2007}, the cluster state can be written as a matrix product state,
\begin{equation}
    |\Psi\rangle = \sum_{\vec{m}}\langle L|A^{m_1} A^{m_2}\ldots A^{m_N} |R\rangle |\vec{m}\rangle,
\end{equation}
where $A^0 = |+\rangle\langle 0|$ and $A^1 = |-\rangle\langle 1|$. We note that it is often convenient to alternatively express physical indices in the $X$-eigenbasis, where $A^+ \propto H$ and $A^- \propto HZ$, corresponding to the operator basis leveraged for cluster state-based MBQC \cite{Raussendorf2001, Gross2007}.

\begin{figure}
\centering
\includegraphics[width=0.95\linewidth]{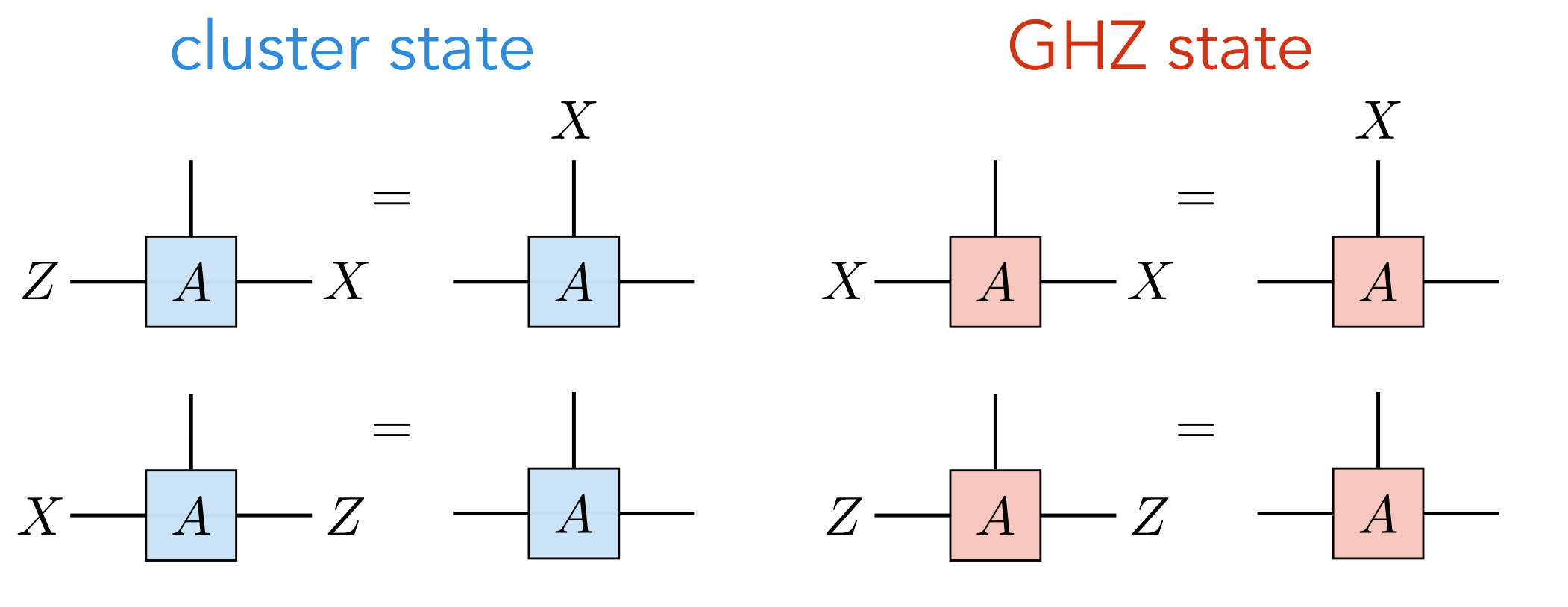}
\caption{Symmetries of the 1D cluster and GHZ states. By leveraging these symmetries, Pauli defects appearing on virtual legs, e.g. arising from the fusion of two states via Bell measurement, can be ``pushed'' to the edge of the prepared state by applying appropriate unitaries to the physical legs.}
\label{fig:f1app}
\end{figure}

Similar to the AKLT state, the measurement-assisted preparation of the cluster state is enabled by Bell measurements that fuse together smaller cluster states prepared in parallel via the unitary preparation scheme outlined above. As with the AKLT state, the possibility to make such a scheme deterministic relies on the $\mathbb{Z}_2\times\mathbb{Z}_2$ symmetry of the state, which we leverage to ``push'' any Pauli defect $B$ to the edge of the chain and remove it. In analogy to Fig.\ \ref{fig:f4}a, we summarize the symmetries of the local tensor $A$ for the cluster state in the left panel of Fig.\ \ref{fig:f1app}. 

As a side note, the $\mathbb{Z}_2\times\mathbb{Z}_2$ symmetry of the cluster state is made apparent upon grouping of pairs of tensors: defining $\widetilde{A}^{n_i} = A^{m_i}A^{m_{i+1}}$ where $n_i$ is a composite index for $m_i$ and $m_{i+1}$, it can be shown that
\begin{equation}
    \sum_{nn'} (U_B)_{nn'}\widetilde{A}^{n'} = B \widetilde{A}^n B,
\end{equation}
for any Pauli $B \in \{I,X,Y,Z\}$, where $U_B$ is some unitary acting on the physical legs, e.g., $U_X = I\otimes X$ and $U_Z = X\otimes I$. This relation is in direct analogy to the symmetry of the tensors composing the AKLT state in Eq.~\ref{eq:SPTfundamental}.

\subsection{GHZ state}
The GHZ state 
\begin{equation}
    \ket{\Psi} = \frac{1}{\sqrt{2}}(\ket{000\ldots 0} + \ket{111\ldots 1})
\end{equation}
provides another example of a resource state for which our measurement-assisted strategy enables a preparation speed-up. Employing only local unitary operations, preparation of the GHZ state necessitates a linear-depth circuit: first prepare a single qubit in the $\ket{+}$ state and all others in the $\ket{0}$ state, and sequentially apply $N$ CNOT gates, in each case using the target qubit of the previous CNOT as the control of the next. Augmenting unitary gates with measurements and feed-forward, however, enables a constant-time preparation protocol for the GHZ state.

Similar to the preparation of the AKLT and cluster states, our procedure begins with the preparation of small GHZ states in parallel using the sequential approach outlined above. Again taking advantage of its MPS representation, we define the local tensors
\begin{equation}
    A^0 = \left[\begin{matrix} 1 & 0 \\ 0 & 0\end{matrix}\right] \hspace{20pt} A^1 = \left[\begin{matrix} 0 & 0 \\ 0 & 1\end{matrix}\right],
\end{equation}
such that the GHZ state may be expressed as an MPS in the form of Eq.~\ref{eq:genericMPS}. We note that the GHZ state has the property that the tensor $A$ is nonzero only when the physical index and both virtual indices are the same, i.e., $\bra{i}A^m\ket{j}=\delta_{im}\delta_{jm}$. Consequently, there is no need for independent ``site'' and ``memory'' qubits as in the case of the AKLT state; the edge sites of the GHZ state may be used as the ``memory'' to prepare subsequent sites.

As with the AKLT and cluster states, fusion of independently prepared GHZ states with Bell measurements yields a larger GHZ state with an intermediary Pauli defect $B$. Given knowledge of the measurement outcome, the defect $B$ can be removed using a single layer of Pauli gates by leveraging the symmetries of the local tensor $A$ outlined in Fig.\ \ref{fig:f1app}. 

To clarify our earlier remark about including the measured qubits in the final GHZ state, we note that the removal of the defect $B~=~\{I,X,Y,Z\}$ is equivalent to carrying out the mapping
\begin{equation}
    B_{ij}\ket{ij}\to I_{ij} \ket{ij}
\end{equation}
on the measured Bell pair. Noting the fact that 
\begin{equation}
    I = \sum_{m_i,m_{i+1}} A^{m_i}A^{m_{i+1}},
    \label{eq:GHZidentity}
\end{equation} this in effect allows us to reincorporate the measured memory qubits (without additional swapping) as an additional pair of GHZ sites, at the expense of two CNOT gates (per fusion measurement) to re-entangle the measured qubits. This can alternatively be viewed (perhaps more simply) as the ability to reinitialize the measured qubits to the ground the state $\ket{00}$ (given the measurement outcome), after which a single layer of the sequential GHZ preparation protocol is carried out, using the recycled qubits to form additional sites without additional swapping. For an example circuit diagram of the measurement-assisted GHZ protocol described here, see the Supplemental Material \cite{SM}.

\ifarXiv
    

\section*{Supplementary material}

# I. EXAMPLE AKLT STATE PREPARATION CIRCUITS

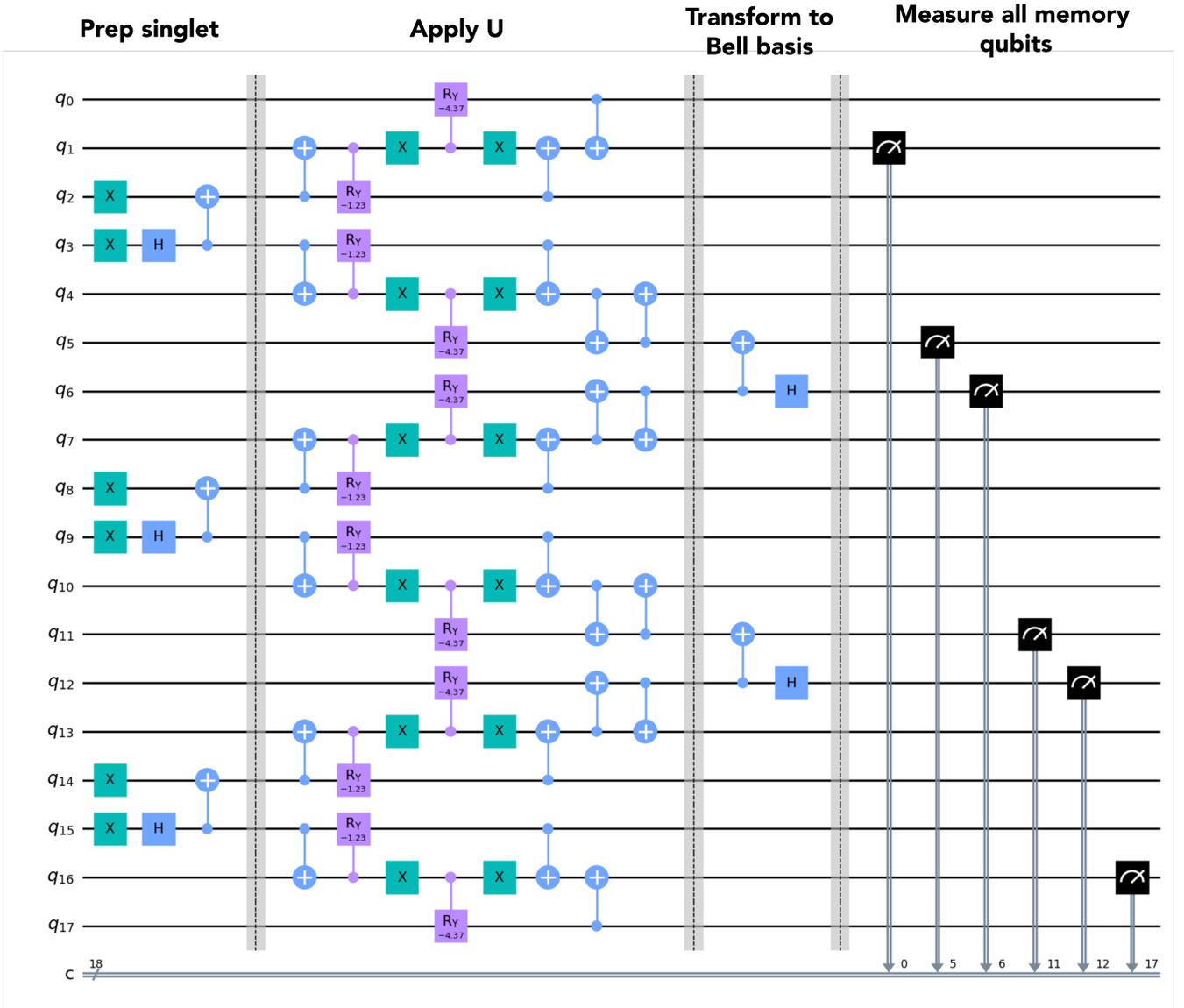

FIG. S1. Measurement-assisted circuit to prepare the $N = 6$ site AKLT state. Barriers have been included for visualization purposes only, and are removed prior to transpilation for physical devices. Steps include (1) preparation of memory qubits in the singlet state, (2) a single layer of application of $U$ to prepare spin–1 sites, and finally, (3) fusion measurements in the Bell basis. As the outermost memory qubits do not take part in fusion measurements, we apply a modified form of $U$ at the edges to minimize swaps; at the edges of the circuit, edge memory qubits are encoded by physical qubits $q_1$ and $q_{16}$ (as opposed to $q_0$ and $q_{17}$). Separate from fusion measurements, edge memory qubits are measured to probabilistically enforce boundary conditions.

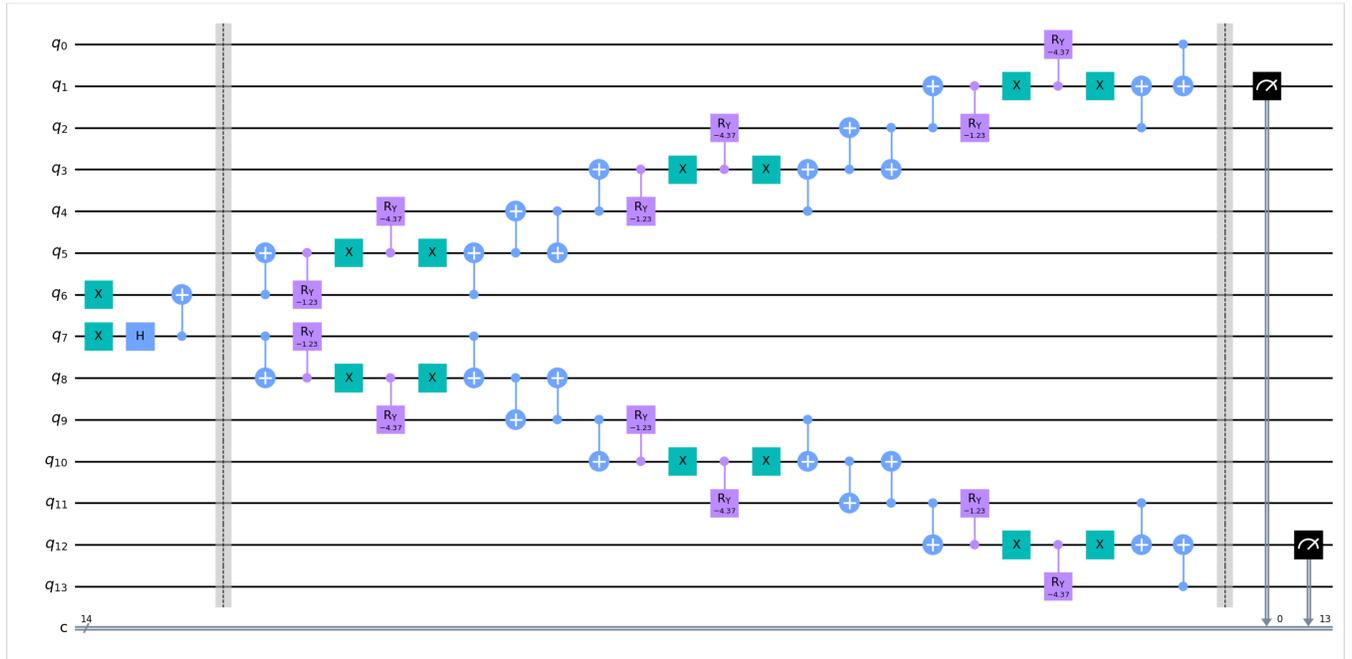

FIG. S2. Circuit to sequentially prepare the $N = 6$ site AKLT state. A single pair of memory qubits are initially prepared in the singlet state, and subsequently used to sequentially prepare all spin–1 sites. As above, we apply a modified form of $U$ at the edges to minimize swaps, and edge memory qubits are measured to enforce boundary conditions.

## A. Scaling comparison between measurement-assisted and sequential circuits

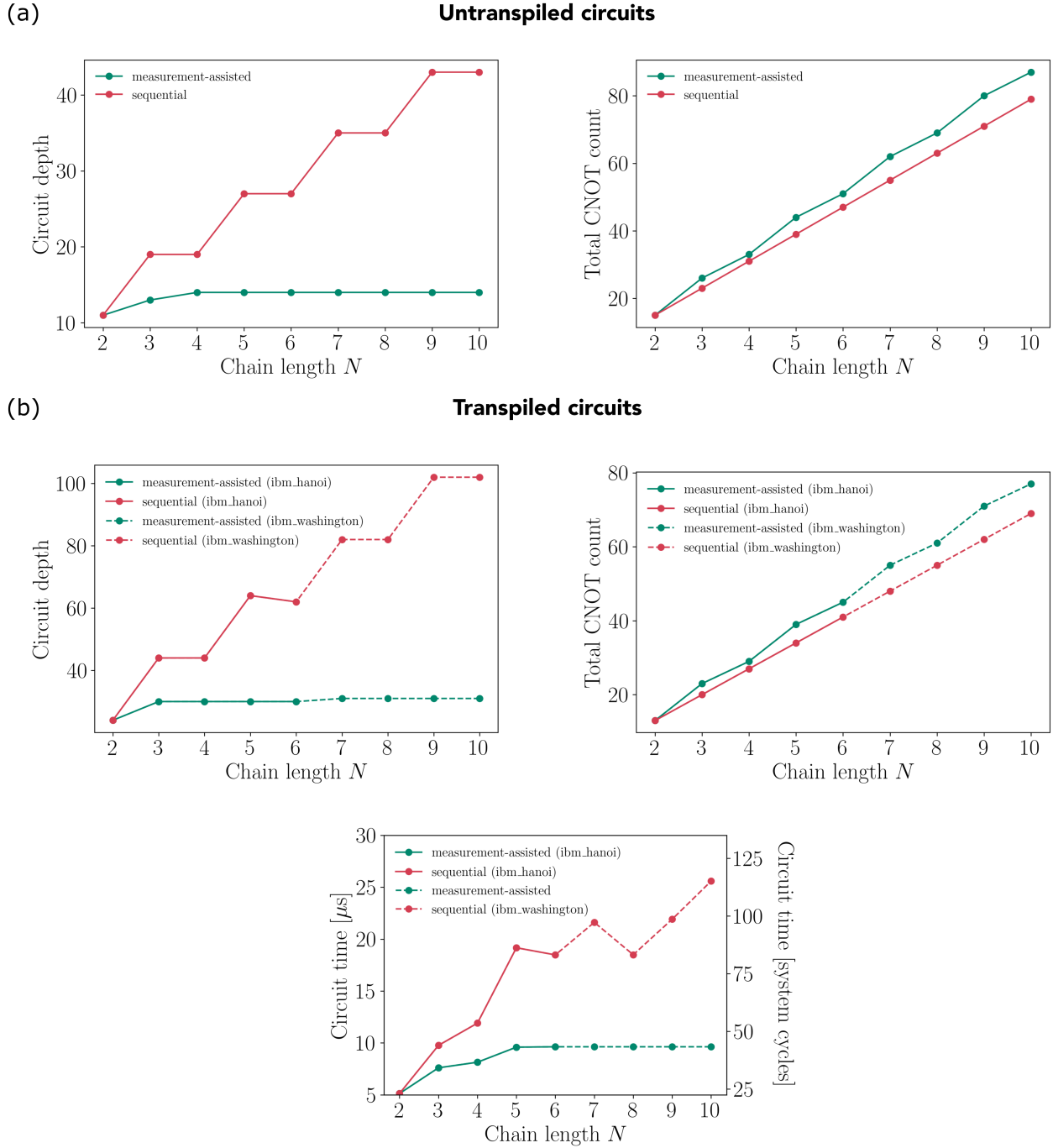

FIG. S3. Scaling of various circuit properties as a function of chain length $N$. (a) Circuit depth and total CNOT count for both measurement-assisted and sequential preparation circuits before transpilation. (b) Circuit depth, total CNOT count, and estimated circuit execution time for transpiled circuits. In particular, we transpile circuits (with maximum optimization and a custom intial physical qubit assignment) both for ibm_hanoi and, to extend our scaling analysis, the 127 qubit device ibm_washington. Circuit execution times include measurement, and are estimated from scheduled circuits using Qiskit's built-in 'ALAP' scheduling method. For both devices, the system cycle time is $dt = 2/9$ $\mu s$.

## II. POST-PROCESSING DETAILS

### A. Rejection rates

#### 1. String order measurements (Fig. 6)

| $N$ | 2 | 3 | 4 | 5 | 6 |
|---|---|---|---|---|---|
| Measurement-assisted | 13.6% | 8.6% | 11.8% | 11.3% | 20.0% |
| Sequential | 10.5% | 12.7% | 16.9% | 17.5% | 21.0% |

TABLE I. Post-selection rejection rates for data used in Fig. 6 of the main text. We post-select both on successful measurement of valid spin–1 states (i.e., rejecting shots where a singlet is measured at any site). We additionally post-select on consistency between measured boundary conditions and observed edge spin–1/2 states (in short, measuring the left memory qubit in the state $|0\rangle$ theoretically ensures that, starting from the left edge and moving right-ward, a "+" spin-1 site will be observed before a "−" spin-1 site. Importantly, we do not enforce any known properties of the AKLT state, such as its hidden topological order. The rejection rates shown here are averaged over all boundary conditions.

#### 2. Entanglement spectrum (Fig. 7)

| $N$ | 1 | 2 | 3 | 4 | 5 | 6 |
|---|---|---|---|---|---|---|
| Measurement-assisted | 1.7% | 7.1% | 5.6% | 7.1% | 14.6% | 20.0% |
| Sequential | 1.8% | 7.2% | 4.5% | 7.0% | 7.5% | 12.0% |

TABLE II. Post-selection rejection rates for data used in Fig. 7 of the main text. We post-select on successful measurement of valid spin–1 states at all sites. Displayed rejection rates are averaged over all measured boundary conditions and all tomography measurement circuits.

#### 3. Teleportation (Fig. 8)

| $N$ | 1 | 2 | 3 | 4 | 5 | 6 |
|---|---|---|---|---|---|---|
| Measurement-assisted | 4.2% | 6.2% | 13.4% | 19.4% | 20.2% | 23.6% |

TABLE III. Post-selection rejection rates for data used in Fig. 8 of the main text. We post-select on successful measurement of a valid spin–1 state. Displayed rejection rates are averaged over all measured all tomography measurement circuits and target states.

## B. Defect removal without feed-forward

As discussed in the main text, we leverage the fact that the symmetry operators $U_B$ merely permute spin–1 basis states to remove defects in post-processing, supplanting the need for feed-forward. Here, we explain how this is achieved with a simple example.

Measurement data from circuits executed on IBM Quantum hardware is returned in the form of a `Counts` object (see https://qiskit.org/documentation/stubs/qiskit.result.Counts.html for more information). This object lists, as a python `dict()`, each measurement outcome and the corresponding number of observations (or, in the case of measurement-error-mitigated results, their corresponding quasiprobabilities).

For simplicity, let us specialize to the case where we measure all qubits in the $Z$ basis (equivalent to a measurement of spin–1 sites in the $S^z$ eigenbasis in our encoding) for a $N = 4$ site chain prepared using our measurement-assisted scheme. As an example, let us take the possible measurement outcome

$$\textcolor{red}{0} \ 10 \ 00 \ \textcolor{red}{01} \ 10 \ 01 \ \textcolor{red}{1}, \tag{1}$$

where we have inserted spaces for readability. Red and black bits correspond to measurement outcomes for memory and spin–1 site composing) qubits, respectively. Red bits at the left and right edges correspond to measurement outcomes for edge memory qubits, while those in the middle denote the fusion measurement result.

Recalling our spin–1 encoding and taking into account the Bell measurement result, this configuration can be "decoded" as

$$\textcolor{red}{0} \ + \ 0 \ \textcolor{red}{X} \ + \ - \ \textcolor{red}{1}, \tag{2}$$

where black symbols represent the spin–1 state at each site. To remove the defect $X$, it is sufficient to apply $U_B = |+\rangle\langle-| + |0\rangle\langle0| + |+\rangle\langle-|$ to all spin–1 sites to the left, and apply $X$ to the left-most edge memory qubit. We remark that we can equivalently correct the state by applying gates to the right. Here, we make the former choice.

In our encoding, these operators will merely amount to bit flips at the level of our classical bits, or equivalently spin-flips at the level of the spin–1 sites, up to a global phase. Equivalently, it amounts to a transformed encoding for all bits to the left of the defect. Removing the defect $X$ therefore yields

$$\textcolor{red}{1} \ - \ 0 \ + \ - \ \textcolor{red}{1}, \tag{3}$$

recovering the expected hidden antiferromagnetic ordering of the AKLT state. For longer chains (and measurements in the $S^z$ basis), it is sufficient to determine the parity of the total number of defects to the right of each spin–1 site that do not commute with $Z$. If the parity is odd, the spin–1 site needs to be flipped, while if even, no action needs to be taken. We note that this scheme can be adapted for spin sites measured in the $S^x$ or $S^y$ basis.

## III. ADDITIONAL EXPERIMENTAL DATA AND ANALYSIS

### A. Tomography

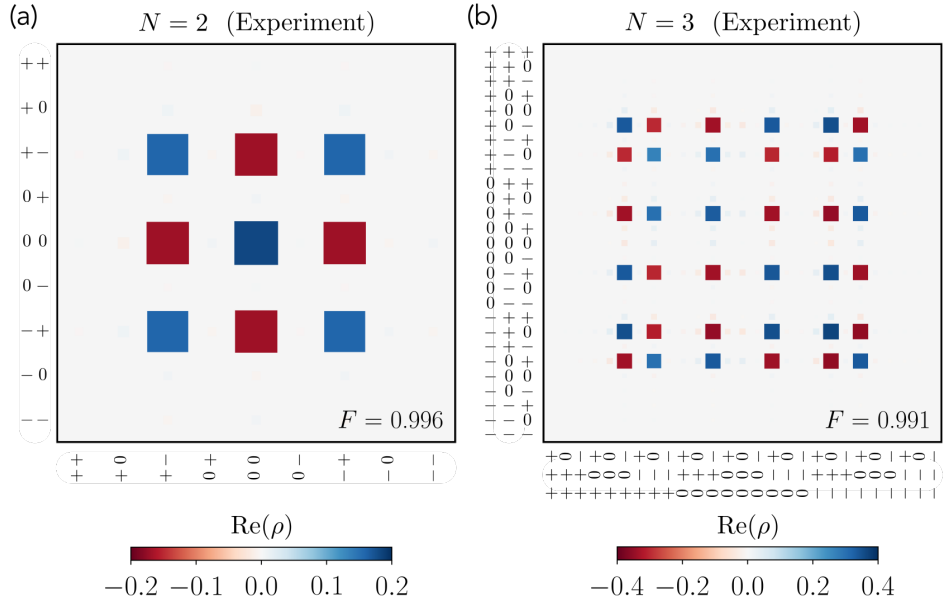

FIG. S4. An additional example of tomography for small chains, carried out on ibm_hanoi. Experimental details can be found in Fig. 5 of the main text. Displayed Hinton diagrams and fidelity correspond to McWeeny purified density matrices. Without purification, we find (a) $F = 0.960$ and (b) $F = 0.780$.

## B. String order

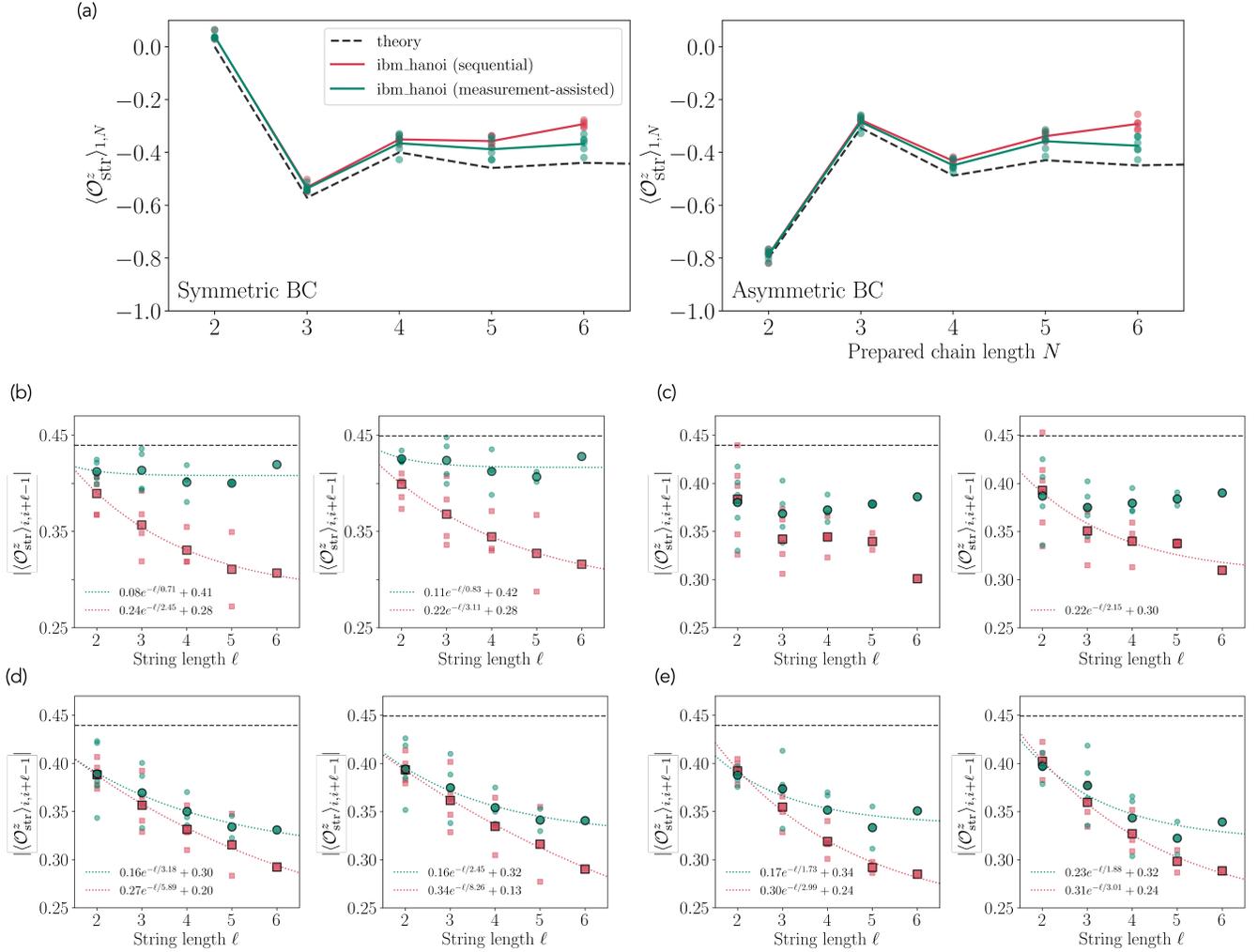

FIG. S5. String order parameter for an ensemble of runs on ibm_hanoi, demonstrating a spread of observed results. For figure details, see Fig. 6 of the main text. (a) String order parameter measured edge-to-edge for prepared chains of length $N$, with the left and right panel displaying data for post-selection upon measurement of symmetric and asymmetric boundary conditions (BC). Individual markers show data points corresponding to individual runs, while solid lines indicate the average for both sequential (red) and measurement-assisted (green) preparation. In total, we display 5 independent runs of sequential preparation, and 6 using our measurement-assisted approach. (b − e) String order of length $\ell$ for prepared chains of length $N$. For all panels, left and right panels display data for symmetric and asymmetric BC, respectively. Panel (b) corresponds to the data shown in the main text for reference. As in Fig. 6, small transparent markers correspond to independent measurements at different locations on the chain, while solid outlined markers correspond to their average. We display parameters corresponding to a best-fit to an exponential curve in the legend. In general, we find the data to fit an exponential decay well, though we show one example in panel (c) in which this is not the case and no converged best-fit solution was found.

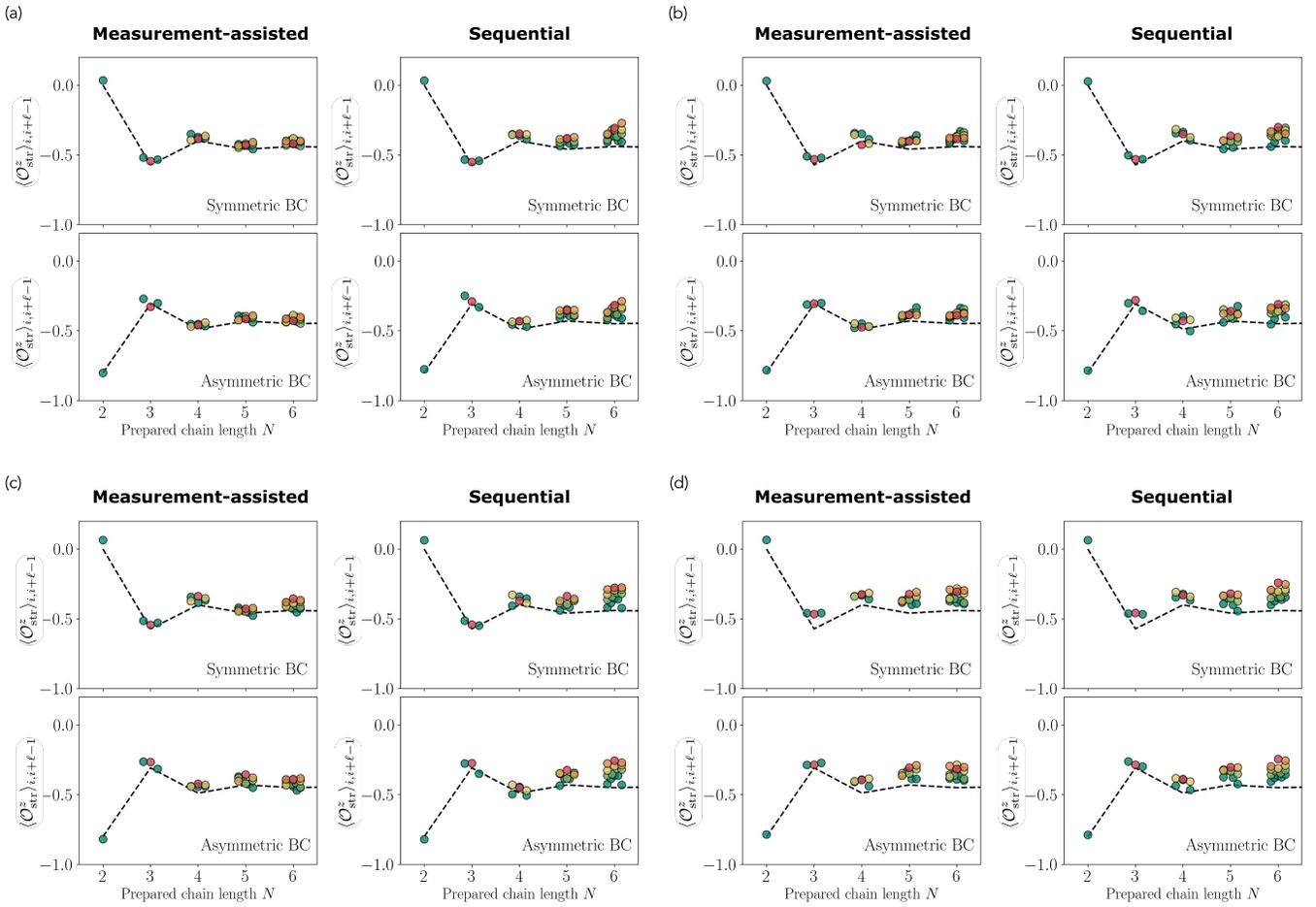

FIG. S6. String order parameter $\langle O_{\text{str}}^z \rangle_{i,i+\ell-1}$ for all $i$ and $\ell$. For all plots, each individual marker correspond to a particular choice of $i$, $\ell$ and $N$. All values of $\langle O_{\text{str}}^z \rangle_{i,i+\ell-1}$ are theoretically expected to lie on the dashed curve for an ideal AKLT state. All measurements for a particular $N$ are binned along the $x$-axis, and artificially shifted left or right depending on $i$. Finally, colors interpolate from green to red, where the former corresponds to $\ell = 2$ and the latter to $\ell = N$. For example, for $N = 4$, there are three possible measurements for a string of length $\ell = 2$: From the first site to the second ($i = 1$), the second to the third ($i = 2$), and the third to the fourth ($i = 3$). These three measurements correspond to the three green markers centered at $N = 4$. Data in panel (a) corresponds to our best runs for both measurement-assisted and sequential preparations. For panels (b)–(d), we compare string order parameters measured for states prepared via measurement-assisted and sequential circuits run together as a single job, with the aim of achieving a fair comparison by minimizing variation in error rates between experiments.

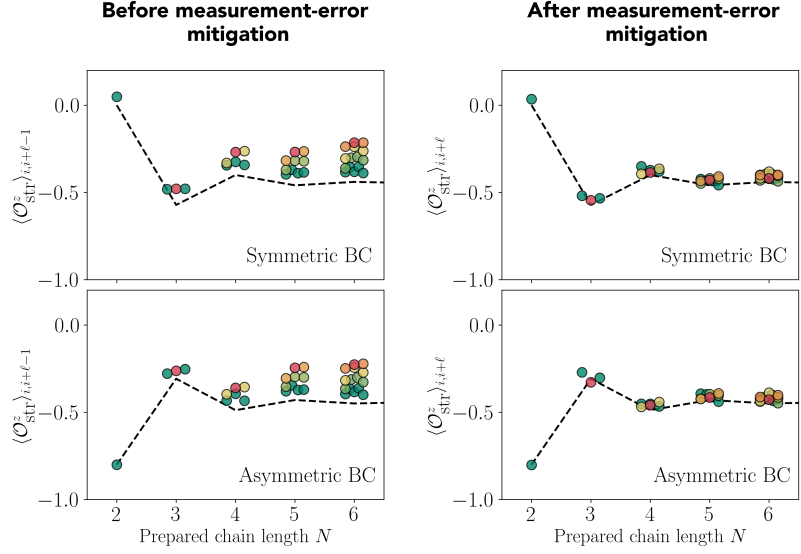

FIG. S7. The impact of measurement-error mitigation on string order. See Fig. S6 for visualization details. As explained in the main text, all experiments utilize measurement error mitigation as a built-in option of Qiskit Runtime. This has a noticeable improvement on the quality of our data, as can be seen clearly from this comparison between string order measurements before (left) and after (right) mitigation is applied. Because the string order is a nonlocal observable involving many qubits across the chain, we find that measurement-error mitigation is particularly important for large string length $\ell$. We note that the displayed (mitigated) string order measurements correspond to the data used to produce Fig. 6 of the main text.

## C. Entanglement spectrum

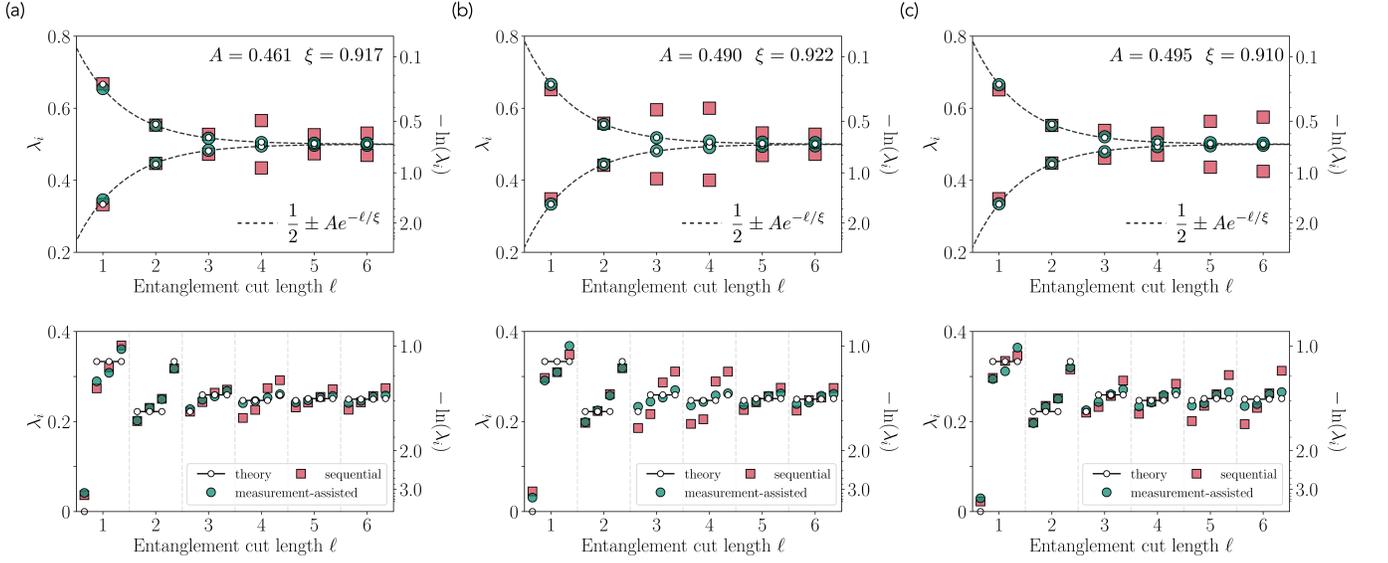

FIG. S8. Additional examples of entanglement entropy as presented in Fig. 7 of the main text. As in the main text, top panels correspond to eigenvalues of $\rho_{R|L}$, while bottom panels correspond to eigenvalues of $\rho_{LR}$. For all panels, data for sequential and measurement-assisted approaches were acquired using circuits batched as a single job. For reference, panel (a) corresponds to the data presented in the main text. In all cases, we find that the measurement-assisted approach outperforms its sequential counterpart. As explained in the main text, we extract estimates of the correlation length from exponential fits to the eigenvalues of $\rho_{L|R}$, finding values of $\xi = \{0.9172(4), 0.9221(16), 0.9102(10)\}$, all in terrific agreement with the exact correlation length of the AKLT state $\xi_{\text{AKLT}} = 1/\ln(3) \approx 0.9102$. In contrast, the sequential preparation yields a trend that deviates from the expected exponential decay, with noise evidently impacting measured values of $\lambda_i$ for $\ell \gtrsim 3$.

## D. Teleportation

### Likelihood $\text{Tr}(\rho, \widetilde{\rho})$

| $N$ | | | | | |
|---|---|---|---|---|---|
| 1 | 0.926 | 0.942 | 0.870 | 0.805 | 0.802 |
| 2 | 0.915 | 0.936 | 0.899 | 0.882 | 0.875 |
| 3 | 0.780 | 0.793 | 0.750 | 0.717 | 0.692 |
| 4 | 0.774 | 0.787 | 0.733 | 0.707 | 0.695 |
| 5 | 0.632 | 0.633 | 0.618 | 0.605 | 0.586 |
| 6 | 0.663 | 0.669 | 0.640 | 0.613 | 0.593 |

FIG. S9. Companion to the fidelity table in the right panel of Fig. 8 of the main text. Likelihood $\text{Tr}(\rho, \widetilde{\rho})$ for the maximum-likelihood pure state $\widetilde{\rho}$ obtained via McWeeny purification of the raw, tomographically reconstructed density matrix $\rho$.

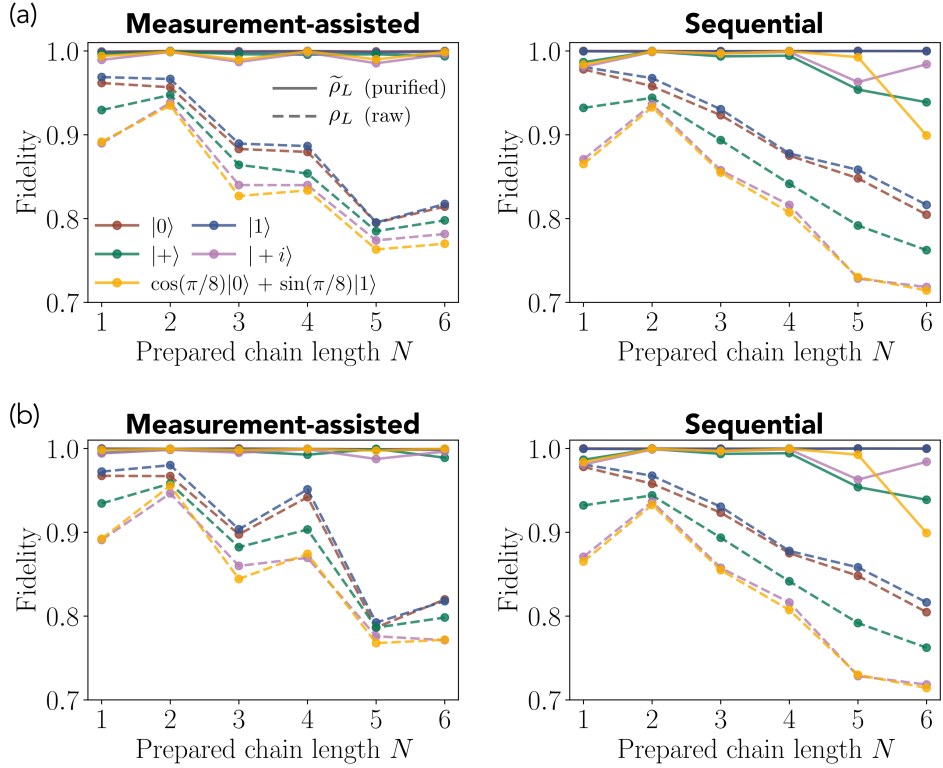

FIG. S10. Further examples of teleportation data – for details, see Fig. 8 of the main text. Left panels display raw and McWeeny purified teleportation fidelities using AKLT chains prepared via our measurement-assisted method, while right panels show fidelities using chains prepared with the sequential unitary approach. As in the case of other experiments, measurement-assisted and sequential circuits are batched within a single job to reduce error rate variation between experiments. The left panel of (a) displays the data shown in the main text for reference.

*Teleportation fidelity estimate*

As discussed in the main text (and evidenced by Fig. S10), we measure raw teleportation fidelities $\gtrsim 0.75$ for all messages using AKLT states prepared via our measurement-assisted scheme. Here, we show that this is roughly in line with a crude error estimate taking CNOT errors to be the dominant source of teleportation infidelity, and additionally considering applied post-selection tactics. As we are only interested in a rough estimate of fidelity, for simplicity we will ignore state-preparation and measurement errors (i.e., neglect the former and assume that measurement-error mitigation perfectly negates the latter), as well as single qubit gate errors and idle-time decoherence.

In total, the (optimized) $N = 6$ teleportation circuit consists of 59 CNOT gates, including pre-measurement basis transformations. Assuming a CNOT error rate of $\epsilon = 0.01$ (consistent with calibration data at the time of experiment), and further assuming the worst-case assumption that all error states correspond to a received message orthogonal to the intended message, the probability that no errors occur $p = (1 - \epsilon)^{59} \approx 0.55$ provides a rough estimate for the raw teleportation fidelity. Taking into account our post-selection strategy, from Table III we discard 23.6% of shots for the $N = 6$ teleportation experiment analyzed in the main text. Assuming these to correspond to shots where CNOT errors occur, post-selection increases the proportion of no-error shots (and the corresponding estimate of teleportation fidelity) to

$$F(N = 6) \approx \frac{(1 - \epsilon)^{59}}{1 - 0.236} \approx 0.72, \tag{4}$$

in rough agreement with observed raw fidelities for $N = 6$. A similar computation agrees well for other chain lengths, which we tabulate below for the experimental data shown in Fig. 8 of the main text. We emphasize that these numbers merely provide a crude estimate, and the additional sources of error neglected (such as idle-time relaxation errors) will invariably impact the final teleportation fidelity.

| $N$ | 1 | 2 | 3 | 4 | 5 | 6 |
|---|---|---|---|---|---|---|
| CNOT count | 10 | 19 | 30 | 39 | 50 | 59 |
| Rejection rate | 0.042 | 0.062 | 0.134 | 0.194% | 0.20 | 0.236 |
| Fidelity estimate | 0.94 | 0.88 | 0.85 | 0.84 | 0.76 | 0.72 |

TABLE IV. Rough estimates of teleportation fidelity (to be compared with raw fidelities in Fig. S10 taking into account the total number of CNOTs and the overall rejection rate upon application of our post-selection strategy.

## IV. EXAMPLE PREPARATION CIRCUIT FOR THE SIX QUBIT GHZ STATE

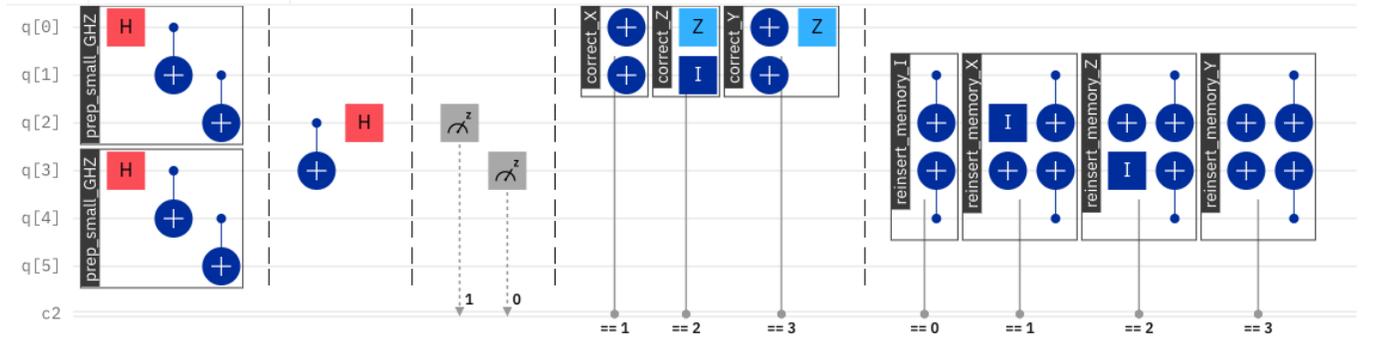

FIG. S11. A constant-depth circuit to prepare the GHZ state using measurements and feed-forward, as described in Appendix D. This figure was generated using the IBM Quantum Composer (https://quantum-computing.ibm.com/composer/docs/iqx/). From left-to-right, barrier-separated stages of the circuit correspond to (1) sequential generation of two small, 3-qubit GHZ states, (2–3) fusion measurement of independent chains, (4) correction of defects conditioned on the outcome of fusion measurement, and (5) recycling and reinserting the measured memory qubits to enlarge the state. We note that without step (5), all unmeasured qubits will be in a four qubit GHZ state. With this step, however, the preparation becomes particularly qubit efficient, yielding a six qubit GHZ state. We note that this circuit can be generalized for any size GHZ state without additional circuit depth – one merely needs to parallelize this approach, similar to the measurement-assisted preparation of the AKLT state in Fig. S1

. .


\fi


\begin{thebibliography}{71}%
\makeatletter
\providecommand \@ifxundefined [1]{%
 \@ifx{#1\undefined}
}%
\providecommand \@ifnum [1]{%
 \ifnum #1\expandafter \@firstoftwo
 \else \expandafter \@secondoftwo
 \fi
}%
\providecommand \@ifx [1]{%
 \ifx #1\expandafter \@firstoftwo
 \else \expandafter \@secondoftwo
 \fi
}%
\providecommand \natexlab [1]{#1}%
\providecommand \enquote  [1]{``#1''}%
\providecommand \bibnamefont  [1]{#1}%
\providecommand \bibfnamefont [1]{#1}%
\providecommand \citenamefont [1]{#1}%
\providecommand \href@noop [0]{\@secondoftwo}%
\providecommand \href [0]{\begingroup \@sanitize@url \@href}%
\providecommand \@href[1]{\@@startlink{#1}\@@href}%
\providecommand \@@href[1]{\endgroup#1\@@endlink}%
\providecommand \@sanitize@url [0]{\catcode `\\12\catcode `\$12\catcode
  `\&12\catcode `\#12\catcode `\^12\catcode `\_12\catcode `\%12\relax}%
\providecommand \@@startlink[1]{}%
\providecommand \@@endlink[0]{}%
\providecommand \url  [0]{\begingroup\@sanitize@url \@url }%
\providecommand \@url [1]{\endgroup\@href {#1}{\urlprefix }}%
\providecommand \urlprefix  [0]{URL }%
\providecommand \Eprint [0]{\href }%
\providecommand \doibase [0]{https://doi.org/}%
\providecommand \selectlanguage [0]{\@gobble}%
\providecommand \bibinfo  [0]{\@secondoftwo}%
\providecommand \bibfield  [0]{\@secondoftwo}%
\providecommand \translation [1]{[#1]}%
\providecommand \BibitemOpen [0]{}%
\providecommand \bibitemStop [0]{}%
\providecommand \bibitemNoStop [0]{.\EOS\space}%
\providecommand \EOS [0]{\spacefactor3000\relax}%
\providecommand \BibitemShut  [1]{\csname bibitem#1\endcsname}%
\let\auto@bib@innerbib\@empty
\bibitem [{\citenamefont {Altman}\ \emph {et~al.}(2021)\citenamefont {Altman},
  \citenamefont {Brown}, \citenamefont {Carleo}, \citenamefont {Carr},
  \citenamefont {Demler}, \citenamefont {Chin}, \citenamefont {DeMarco},
  \citenamefont {Economou}, \citenamefont {Eriksson}, \citenamefont {Fu} \emph
  {et~al.}}]{Altman_2021}%
  \BibitemOpen
  \bibfield  {author} {\bibinfo {author} {\bibfnamefont {E.}~\bibnamefont
  {Altman}}, \bibinfo {author} {\bibfnamefont {K.~R.}\ \bibnamefont {Brown}},
  \bibinfo {author} {\bibfnamefont {G.}~\bibnamefont {Carleo}}, \bibinfo
  {author} {\bibfnamefont {L.~D.}\ \bibnamefont {Carr}}, \bibinfo {author}
  {\bibfnamefont {E.}~\bibnamefont {Demler}}, \bibinfo {author} {\bibfnamefont
  {C.}~\bibnamefont {Chin}}, \bibinfo {author} {\bibfnamefont {B.}~\bibnamefont
  {DeMarco}}, \bibinfo {author} {\bibfnamefont {S.~E.}\ \bibnamefont
  {Economou}}, \bibinfo {author} {\bibfnamefont {M.~A.}\ \bibnamefont
  {Eriksson}}, \bibinfo {author} {\bibfnamefont {K.-M.~C.}\ \bibnamefont {Fu}},
  \emph {et~al.},\ }\href {https://doi.org/10.1103/prxquantum.2.017003}
  {\bibfield  {journal} {\bibinfo  {journal} {{PRX} Quantum}\ }\textbf
  {\bibinfo {volume} {2}},\ \bibinfo {pages} {017003} (\bibinfo {year}
  {2021})}\BibitemShut {NoStop}%
\bibitem [{\citenamefont {Zeng}\ \emph {et~al.}(2019)\citenamefont {Zeng},
  \citenamefont {Chen}, \citenamefont {Zhou},\ and\ \citenamefont
  {Wen}}]{Zeng2019}%
  \BibitemOpen
  \bibfield  {author} {\bibinfo {author} {\bibfnamefont {B.}~\bibnamefont
  {Zeng}}, \bibinfo {author} {\bibfnamefont {X.}~\bibnamefont {Chen}}, \bibinfo
  {author} {\bibfnamefont {D.-L.}\ \bibnamefont {Zhou}},\ and\ \bibinfo
  {author} {\bibfnamefont {X.-G.}\ \bibnamefont {Wen}},\ }\href
  {https://doi.org/10.1007/978-1-4939-9084-9} {\emph {\bibinfo {title} {Quantum
  Information Meets Quantum Matter}}}\ (\bibinfo  {publisher} {Springer New
  York},\ \bibinfo {year} {2019})\BibitemShut {NoStop}%
\bibitem [{\citenamefont {Raussendorf}\ and\ \citenamefont
  {Briegel}(2001)}]{Raussendorf2001}%
  \BibitemOpen
  \bibfield  {author} {\bibinfo {author} {\bibfnamefont {R.}~\bibnamefont
  {Raussendorf}}\ and\ \bibinfo {author} {\bibfnamefont {H.~J.}\ \bibnamefont
  {Briegel}},\ }\href {https://doi.org/10.1103/physrevlett.86.5188} {\bibfield
  {journal} {\bibinfo  {journal} {Physical Review Letters}\ }\textbf {\bibinfo
  {volume} {86}},\ \bibinfo {pages} {5188} (\bibinfo {year}
  {2001})}\BibitemShut {NoStop}%
\bibitem [{\citenamefont {Briegel}\ \emph {et~al.}(2009)\citenamefont
  {Briegel}, \citenamefont {Browne}, \citenamefont {DÃŒr}, \citenamefont
  {Raussendorf},\ and\ \citenamefont {den Nest}}]{Briegel_2009}%
  \BibitemOpen
  \bibfield  {author} {\bibinfo {author} {\bibfnamefont {H.~J.}\ \bibnamefont
  {Briegel}}, \bibinfo {author} {\bibfnamefont {D.~E.}\ \bibnamefont {Browne}},
  \bibinfo {author} {\bibfnamefont {W.}~\bibnamefont {DÃŒr}}, \bibinfo {author}
  {\bibfnamefont {R.}~\bibnamefont {Raussendorf}},\ and\ \bibinfo {author}
  {\bibfnamefont {M.~V.}\ \bibnamefont {den Nest}},\ }\href
  {https://doi.org/10.1038/nphys1157} {\bibfield  {journal} {\bibinfo
  {journal} {Nature Physics}\ }\textbf {\bibinfo {volume} {5}},\ \bibinfo
  {pages} {19} (\bibinfo {year} {2009})}\BibitemShut {NoStop}%
\bibitem [{\citenamefont {Terhal}(2015)}]{Terhal_2015}%
  \BibitemOpen
  \bibfield  {author} {\bibinfo {author} {\bibfnamefont {B.~M.}\ \bibnamefont
  {Terhal}},\ }\href {https://doi.org/10.1103/revmodphys.87.307} {\bibfield
  {journal} {\bibinfo  {journal} {Reviews of Modern Physics}\ }\textbf
  {\bibinfo {volume} {87}},\ \bibinfo {pages} {307} (\bibinfo {year}
  {2015})}\BibitemShut {NoStop}%
\bibitem [{\citenamefont {Degen}\ \emph {et~al.}(2017)\citenamefont {Degen},
  \citenamefont {Reinhard},\ and\ \citenamefont {Cappellaro}}]{Degen_2017}%
  \BibitemOpen
  \bibfield  {author} {\bibinfo {author} {\bibfnamefont {C.~L.}\ \bibnamefont
  {Degen}}, \bibinfo {author} {\bibfnamefont {F.}~\bibnamefont {Reinhard}},\
  and\ \bibinfo {author} {\bibfnamefont {P.}~\bibnamefont {Cappellaro}},\
  }\href {https://doi.org/10.1103/revmodphys.89.035002} {\bibfield  {journal}
  {\bibinfo  {journal} {Reviews of Modern Physics}\ }\textbf {\bibinfo {volume}
  {89}},\ \bibinfo {pages} {035002} (\bibinfo {year} {2017})}\BibitemShut
  {NoStop}%
\bibitem [{\citenamefont {Bravyi}\ \emph {et~al.}(2006)\citenamefont {Bravyi},
  \citenamefont {Hastings},\ and\ \citenamefont {Verstraete}}]{Bravyi2006}%
  \BibitemOpen
  \bibfield  {author} {\bibinfo {author} {\bibfnamefont {S.}~\bibnamefont
  {Bravyi}}, \bibinfo {author} {\bibfnamefont {M.~B.}\ \bibnamefont
  {Hastings}},\ and\ \bibinfo {author} {\bibfnamefont {F.}~\bibnamefont
  {Verstraete}},\ }\href {https://doi.org/10.1103/physrevlett.97.050401}
  {\bibfield  {journal} {\bibinfo  {journal} {Physical Review Letters}\
  }\textbf {\bibinfo {volume} {97}},\ \bibinfo {pages} {050401} (\bibinfo
  {year} {2006})}\BibitemShut {NoStop}%
\bibitem [{\citenamefont {Chen}\ \emph {et~al.}(2010)\citenamefont {Chen},
  \citenamefont {Gu},\ and\ \citenamefont {Wen}}]{Chen_2010}%
  \BibitemOpen
  \bibfield  {author} {\bibinfo {author} {\bibfnamefont {X.}~\bibnamefont
  {Chen}}, \bibinfo {author} {\bibfnamefont {Z.-C.}\ \bibnamefont {Gu}},\ and\
  \bibinfo {author} {\bibfnamefont {X.-G.}\ \bibnamefont {Wen}},\ }\href
  {https://doi.org/10.1103/physrevb.82.155138} {\bibfield  {journal} {\bibinfo
  {journal} {Physical Review B}\ }\textbf {\bibinfo {volume} {82}},\ \bibinfo
  {pages} {155138} (\bibinfo {year} {2010})}\BibitemShut {NoStop}%
\bibitem [{\citenamefont {Pollmann}\ \emph {et~al.}(2012)\citenamefont
  {Pollmann}, \citenamefont {Berg}, \citenamefont {Turner},\ and\ \citenamefont
  {Oshikawa}}]{Pollmann2009}%
  \BibitemOpen
  \bibfield  {author} {\bibinfo {author} {\bibfnamefont {F.}~\bibnamefont
  {Pollmann}}, \bibinfo {author} {\bibfnamefont {E.}~\bibnamefont {Berg}},
  \bibinfo {author} {\bibfnamefont {A.~M.}\ \bibnamefont {Turner}},\ and\
  \bibinfo {author} {\bibfnamefont {M.}~\bibnamefont {Oshikawa}},\ }\href
  {https://doi.org/10.1103/PhysRevB.85.075125} {\bibfield  {journal} {\bibinfo
  {journal} {Physical Review B}\ }\textbf {\bibinfo {volume} {85}},\ \bibinfo
  {pages} {075125} (\bibinfo {year} {2012})},\ \Eprint
  {https://arxiv.org/abs/0909.4059v3} {0909.4059v3} \BibitemShut {NoStop}%
\bibitem [{\citenamefont {Verstraete}\ and\ \citenamefont
  {Cirac}(2004)}]{Verstraete2004}%
  \BibitemOpen
  \bibfield  {author} {\bibinfo {author} {\bibfnamefont {F.}~\bibnamefont
  {Verstraete}}\ and\ \bibinfo {author} {\bibfnamefont {J.~I.}\ \bibnamefont
  {Cirac}},\ }\href {https://doi.org/10.1103/physreva.70.060302} {\bibfield
  {journal} {\bibinfo  {journal} {Physical Review A}\ }\textbf {\bibinfo
  {volume} {70}},\ \bibinfo {pages} {060302} (\bibinfo {year}
  {2004})}\BibitemShut {NoStop}%
\bibitem [{\citenamefont {Gross}\ and\ \citenamefont
  {Eisert}(2007)}]{Gross2007a}%
  \BibitemOpen
  \bibfield  {author} {\bibinfo {author} {\bibfnamefont {D.}~\bibnamefont
  {Gross}}\ and\ \bibinfo {author} {\bibfnamefont {J.}~\bibnamefont {Eisert}},\
  }\href {https://doi.org/10.1103/physrevlett.98.220503} {\bibfield  {journal}
  {\bibinfo  {journal} {Physical Review Letters}\ }\textbf {\bibinfo {volume}
  {98}},\ \bibinfo {pages} {220503} (\bibinfo {year} {2007})}\BibitemShut
  {NoStop}%
\bibitem [{\citenamefont {Brennen}\ and\ \citenamefont
  {Miyake}(2008)}]{Brennen2008}%
  \BibitemOpen
  \bibfield  {author} {\bibinfo {author} {\bibfnamefont {G.~K.}\ \bibnamefont
  {Brennen}}\ and\ \bibinfo {author} {\bibfnamefont {A.}~\bibnamefont
  {Miyake}},\ }\href {https://doi.org/10.1103/physrevlett.101.010502}
  {\bibfield  {journal} {\bibinfo  {journal} {Physical Review Letters}\
  }\textbf {\bibinfo {volume} {101}},\ \bibinfo {pages} {010502} (\bibinfo
  {year} {2008})}\BibitemShut {NoStop}%
\bibitem [{\citenamefont {Kaltenbaek}\ \emph {et~al.}(2010)\citenamefont
  {Kaltenbaek}, \citenamefont {Lavoie}, \citenamefont {Zeng}, \citenamefont
  {Bartlett},\ and\ \citenamefont {Resch}}]{Kaltenbaek2010}%
  \BibitemOpen
  \bibfield  {author} {\bibinfo {author} {\bibfnamefont {R.}~\bibnamefont
  {Kaltenbaek}}, \bibinfo {author} {\bibfnamefont {J.}~\bibnamefont {Lavoie}},
  \bibinfo {author} {\bibfnamefont {B.}~\bibnamefont {Zeng}}, \bibinfo {author}
  {\bibfnamefont {S.~D.}\ \bibnamefont {Bartlett}},\ and\ \bibinfo {author}
  {\bibfnamefont {K.~J.}\ \bibnamefont {Resch}},\ }\href
  {https://doi.org/10.1038/nphys1777} {\bibfield  {journal} {\bibinfo
  {journal} {Nature Physics}\ }\textbf {\bibinfo {volume} {6}},\ \bibinfo
  {pages} {850} (\bibinfo {year} {2010})}\BibitemShut {NoStop}%
\bibitem [{\citenamefont {Wei}\ \emph {et~al.}(2011)\citenamefont {Wei},
  \citenamefont {Affleck},\ and\ \citenamefont {Raussendorf}}]{Wei2011}%
  \BibitemOpen
  \bibfield  {author} {\bibinfo {author} {\bibfnamefont {T.-C.}\ \bibnamefont
  {Wei}}, \bibinfo {author} {\bibfnamefont {I.}~\bibnamefont {Affleck}},\ and\
  \bibinfo {author} {\bibfnamefont {R.}~\bibnamefont {Raussendorf}},\ }\href
  {https://doi.org/10.1103/physrevlett.106.070501} {\bibfield  {journal}
  {\bibinfo  {journal} {Physical Review Letters}\ }\textbf {\bibinfo {volume}
  {106}},\ \bibinfo {pages} {070501} (\bibinfo {year} {2011})}\BibitemShut
  {NoStop}%
\bibitem [{\citenamefont {Sch{\"o}n}\ \emph {et~al.}(2005)\citenamefont
  {Sch{\"o}n}, \citenamefont {Solano}, \citenamefont {Verstraete},
  \citenamefont {Cirac},\ and\ \citenamefont {Wolf}}]{Schoen2005}%
  \BibitemOpen
  \bibfield  {author} {\bibinfo {author} {\bibfnamefont {C.}~\bibnamefont
  {Sch{\"o}n}}, \bibinfo {author} {\bibfnamefont {E.}~\bibnamefont {Solano}},
  \bibinfo {author} {\bibfnamefont {F.}~\bibnamefont {Verstraete}}, \bibinfo
  {author} {\bibfnamefont {J.~I.}\ \bibnamefont {Cirac}},\ and\ \bibinfo
  {author} {\bibfnamefont {M.~M.}\ \bibnamefont {Wolf}},\ }\href
  {https://doi.org/10.1103/physrevlett.95.110503} {\bibfield  {journal}
  {\bibinfo  {journal} {Physical Review Letters}\ }\textbf {\bibinfo {volume}
  {95}},\ \bibinfo {pages} {110503} (\bibinfo {year} {2005})}\BibitemShut
  {NoStop}%
\bibitem [{\citenamefont {Huang}\ and\ \citenamefont {Chen}(2015)}]{Huang2015}%
  \BibitemOpen
  \bibfield  {author} {\bibinfo {author} {\bibfnamefont {Y.}~\bibnamefont
  {Huang}}\ and\ \bibinfo {author} {\bibfnamefont {X.}~\bibnamefont {Chen}},\
  }\href {https://doi.org/10.1103/physrevb.91.195143} {\bibfield  {journal}
  {\bibinfo  {journal} {Physical Review B}\ }\textbf {\bibinfo {volume} {91}},\
  \bibinfo {pages} {195143} (\bibinfo {year} {2015})}\BibitemShut {NoStop}%
\bibitem [{\citenamefont {Zhou}\ \emph {et~al.}(2021)\citenamefont {Zhou},
  \citenamefont {Choi},\ and\ \citenamefont {Lukin}}]{Zhou2021}%
  \BibitemOpen
  \bibfield  {author} {\bibinfo {author} {\bibfnamefont {L.}~\bibnamefont
  {Zhou}}, \bibinfo {author} {\bibfnamefont {S.}~\bibnamefont {Choi}},\ and\
  \bibinfo {author} {\bibfnamefont {M.~D.}\ \bibnamefont {Lukin}},\ }\href
  {https://doi.org/10.1103/physreva.104.032418} {\bibfield  {journal} {\bibinfo
   {journal} {Physical Review A}\ }\textbf {\bibinfo {volume} {104}},\ \bibinfo
  {pages} {032418} (\bibinfo {year} {2021})}\BibitemShut {NoStop}%
\bibitem [{\citenamefont {Wei}\ \emph {et~al.}(2022{\natexlab{a}})\citenamefont
  {Wei}, \citenamefont {Malz},\ and\ \citenamefont {Cirac}}]{Wei2022c}%
  \BibitemOpen
  \bibfield  {author} {\bibinfo {author} {\bibfnamefont {Z.-Y.}\ \bibnamefont
  {Wei}}, \bibinfo {author} {\bibfnamefont {D.}~\bibnamefont {Malz}},\ and\
  \bibinfo {author} {\bibfnamefont {J.~I.}\ \bibnamefont {Cirac}},\ }\Eprint
  {https://arxiv.org/abs/2209.01230} {arXiv:2209.01230 [quant-ph]}  (\bibinfo
  {year} {2022}{\natexlab{a}})\BibitemShut {NoStop}%
\bibitem [{\citenamefont {Murta}\ \emph {et~al.}(2022)\citenamefont {Murta},
  \citenamefont {Cruz},\ and\ \citenamefont {FernÃ¡ndez-Rossier}}]{Murta2022}%
  \BibitemOpen
  \bibfield  {author} {\bibinfo {author} {\bibfnamefont {B.}~\bibnamefont
  {Murta}}, \bibinfo {author} {\bibfnamefont {P.~M.~Q.}\ \bibnamefont {Cruz}},\
  and\ \bibinfo {author} {\bibfnamefont {J.}~\bibnamefont
  {FernÃ¡ndez-Rossier}},\ }\Eprint {https://arxiv.org/abs/2207.07725}
  {arXiv:2207.07725 [quant-ph]}  (\bibinfo {year} {2022})\BibitemShut {NoStop}%
\bibitem [{\citenamefont {Chen}\ \emph {et~al.}(2022)\citenamefont {Chen},
  \citenamefont {Shen}, \citenamefont {Lee},\ and\ \citenamefont
  {Yang}}]{Chen2022}%
  \BibitemOpen
  \bibfield  {author} {\bibinfo {author} {\bibfnamefont {T.}~\bibnamefont
  {Chen}}, \bibinfo {author} {\bibfnamefont {R.}~\bibnamefont {Shen}}, \bibinfo
  {author} {\bibfnamefont {C.~H.}\ \bibnamefont {Lee}},\ and\ \bibinfo {author}
  {\bibfnamefont {B.}~\bibnamefont {Yang}},\ }\Eprint
  {https://arxiv.org/abs/2210.13840} {arXiv:2210.13840 [quant-ph]}  (\bibinfo
  {year} {2022})\BibitemShut {NoStop}%
\bibitem [{\citenamefont {Piroli}\ \emph {et~al.}(2021)\citenamefont {Piroli},
  \citenamefont {Styliaris},\ and\ \citenamefont {Cirac}}]{Piroli_2021}%
  \BibitemOpen
  \bibfield  {author} {\bibinfo {author} {\bibfnamefont {L.}~\bibnamefont
  {Piroli}}, \bibinfo {author} {\bibfnamefont {G.}~\bibnamefont {Styliaris}},\
  and\ \bibinfo {author} {\bibfnamefont {J.~I.}\ \bibnamefont {Cirac}},\ }\href
  {https://doi.org/10.1103/physrevlett.127.220503} {\bibfield  {journal}
  {\bibinfo  {journal} {Physical Review Letters}\ }\textbf {\bibinfo {volume}
  {127}},\ \bibinfo {pages} {220503} (\bibinfo {year} {2021})}\BibitemShut
  {NoStop}%
\bibitem [{\citenamefont {Tantivasadakarn}\ \emph
  {et~al.}(2022{\natexlab{a}})\citenamefont {Tantivasadakarn}, \citenamefont
  {Vishwanath},\ and\ \citenamefont {Verresen}}]{Tantivasadakarn2022a}%
  \BibitemOpen
  \bibfield  {author} {\bibinfo {author} {\bibfnamefont {N.}~\bibnamefont
  {Tantivasadakarn}}, \bibinfo {author} {\bibfnamefont {A.}~\bibnamefont
  {Vishwanath}},\ and\ \bibinfo {author} {\bibfnamefont {R.}~\bibnamefont
  {Verresen}},\ }\Eprint {https://arxiv.org/abs/2209.06202} {arXiv:2209.06202
  [quant-ph]}  (\bibinfo {year} {2022}{\natexlab{a}})\BibitemShut {NoStop}%
\bibitem [{\citenamefont {Tantivasadakarn}\ \emph {et~al.}(2021)\citenamefont
  {Tantivasadakarn}, \citenamefont {Thorngren}, \citenamefont {Vishwanath},\
  and\ \citenamefont {Verresen}}]{Tantivasadakarn2021}%
  \BibitemOpen
  \bibfield  {author} {\bibinfo {author} {\bibfnamefont {N.}~\bibnamefont
  {Tantivasadakarn}}, \bibinfo {author} {\bibfnamefont {R.}~\bibnamefont
  {Thorngren}}, \bibinfo {author} {\bibfnamefont {A.}~\bibnamefont
  {Vishwanath}},\ and\ \bibinfo {author} {\bibfnamefont {R.}~\bibnamefont
  {Verresen}},\ }\Eprint {https://arxiv.org/abs/2112.01519} {arXiv:2112.01519
  [cond-mat.str-el]}  (\bibinfo {year} {2021})\BibitemShut {NoStop}%
\bibitem [{\citenamefont {Verresen}\ \emph {et~al.}(2021)\citenamefont
  {Verresen}, \citenamefont {Tantivasadakarn},\ and\ \citenamefont
  {Vishwanath}}]{Verresen2021}%
  \BibitemOpen
  \bibfield  {author} {\bibinfo {author} {\bibfnamefont {R.}~\bibnamefont
  {Verresen}}, \bibinfo {author} {\bibfnamefont {N.}~\bibnamefont
  {Tantivasadakarn}},\ and\ \bibinfo {author} {\bibfnamefont {A.}~\bibnamefont
  {Vishwanath}},\ }\Eprint {https://arxiv.org/abs/2112.03061} {arXiv:2112.03061
  [quant-ph]}  (\bibinfo {year} {2021})\BibitemShut {NoStop}%
\bibitem [{\citenamefont {Tantivasadakarn}\ \emph
  {et~al.}(2022{\natexlab{b}})\citenamefont {Tantivasadakarn}, \citenamefont
  {Verresen},\ and\ \citenamefont {Vishwanath}}]{Tantivasadakarn2022}%
  \BibitemOpen
  \bibfield  {author} {\bibinfo {author} {\bibfnamefont {N.}~\bibnamefont
  {Tantivasadakarn}}, \bibinfo {author} {\bibfnamefont {R.}~\bibnamefont
  {Verresen}},\ and\ \bibinfo {author} {\bibfnamefont {A.}~\bibnamefont
  {Vishwanath}},\ }\Eprint {https://arxiv.org/abs/2209.03964} {arXiv:2209.03964
  [quant-ph]}  (\bibinfo {year} {2022}{\natexlab{b}})\BibitemShut {NoStop}%
\bibitem [{\citenamefont {Lu}\ \emph {et~al.}(2022)\citenamefont {Lu},
  \citenamefont {Lessa}, \citenamefont {Kim},\ and\ \citenamefont
  {Hsieh}}]{Lu2022}%
  \BibitemOpen
  \bibfield  {author} {\bibinfo {author} {\bibfnamefont {T.-C.}\ \bibnamefont
  {Lu}}, \bibinfo {author} {\bibfnamefont {L.~A.}\ \bibnamefont {Lessa}},
  \bibinfo {author} {\bibfnamefont {I.~H.}\ \bibnamefont {Kim}},\ and\ \bibinfo
  {author} {\bibfnamefont {T.~H.}\ \bibnamefont {Hsieh}},\ }\Eprint
  {https://arxiv.org/abs/2206.13527} {arXiv:2206.13527 [cond-mat.str-el]}
  (\bibinfo {year} {2022})\BibitemShut {NoStop}%
\bibitem [{\citenamefont {Bartolucci}\ \emph {et~al.}(2021)\citenamefont
  {Bartolucci}, \citenamefont {Birchall}, \citenamefont {Bombin}, \citenamefont
  {Cable}, \citenamefont {Dawson}, \citenamefont {Gimeno-Segovia},
  \citenamefont {Johnston}, \citenamefont {Kieling}, \citenamefont {Nickerson},
  \citenamefont {Pant}, \citenamefont {Pastawski}, \citenamefont {Rudolph},\
  and\ \citenamefont {Sparrow}}]{Bartolucci2021}%
  \BibitemOpen
  \bibfield  {author} {\bibinfo {author} {\bibfnamefont {S.}~\bibnamefont
  {Bartolucci}}, \bibinfo {author} {\bibfnamefont {P.}~\bibnamefont
  {Birchall}}, \bibinfo {author} {\bibfnamefont {H.}~\bibnamefont {Bombin}},
  \bibinfo {author} {\bibfnamefont {H.}~\bibnamefont {Cable}}, \bibinfo
  {author} {\bibfnamefont {C.}~\bibnamefont {Dawson}}, \bibinfo {author}
  {\bibfnamefont {M.}~\bibnamefont {Gimeno-Segovia}}, \bibinfo {author}
  {\bibfnamefont {E.}~\bibnamefont {Johnston}}, \bibinfo {author}
  {\bibfnamefont {K.}~\bibnamefont {Kieling}}, \bibinfo {author} {\bibfnamefont
  {N.}~\bibnamefont {Nickerson}}, \bibinfo {author} {\bibfnamefont
  {M.}~\bibnamefont {Pant}}, \bibinfo {author} {\bibfnamefont {F.}~\bibnamefont
  {Pastawski}}, \bibinfo {author} {\bibfnamefont {T.}~\bibnamefont {Rudolph}},\
  and\ \bibinfo {author} {\bibfnamefont {C.}~\bibnamefont {Sparrow}},\ }\Eprint
  {https://arxiv.org/abs/2101.09310} {arXiv:2101.09310 [quant-ph]}  (\bibinfo
  {year} {2021})\BibitemShut {NoStop}%
\bibitem [{\citenamefont {Massar}\ and\ \citenamefont
  {Popescu}(1995)}]{Massar_1995}%
  \BibitemOpen
  \bibfield  {author} {\bibinfo {author} {\bibfnamefont {S.}~\bibnamefont
  {Massar}}\ and\ \bibinfo {author} {\bibfnamefont {S.}~\bibnamefont
  {Popescu}},\ }\href {https://doi.org/10.1103/physrevlett.74.1259} {\bibfield
  {journal} {\bibinfo  {journal} {Physical Review Letters}\ }\textbf {\bibinfo
  {volume} {74}},\ \bibinfo {pages} {1259} (\bibinfo {year}
  {1995})}\BibitemShut {NoStop}%
\bibitem [{\citenamefont {Morimae}\ \emph {et~al.}(2015)\citenamefont
  {Morimae}, \citenamefont {Dunjko},\ and\ \citenamefont
  {Kashefi}}]{morimae2015ground}%
  \BibitemOpen
  \bibfield  {author} {\bibinfo {author} {\bibfnamefont {T.}~\bibnamefont
  {Morimae}}, \bibinfo {author} {\bibfnamefont {V.}~\bibnamefont {Dunjko}},\
  and\ \bibinfo {author} {\bibfnamefont {E.}~\bibnamefont {Kashefi}},\ }\href
  {https://doi.org/10.48550/arXiv.1009.3486} {\bibfield  {journal} {\bibinfo
  {journal} {Quantum information \& computation}\ }\textbf {\bibinfo {volume}
  {15}},\ \bibinfo {pages} {200} (\bibinfo {year} {2015})}\BibitemShut
  {NoStop}%
\bibitem [{\citenamefont {Liu}\ and\ \citenamefont
  {Hwang}(2014)}]{liu2014controlled}%
  \BibitemOpen
  \bibfield  {author} {\bibinfo {author} {\bibfnamefont {L.~L.}\ \bibnamefont
  {Liu}}\ and\ \bibinfo {author} {\bibfnamefont {T.}~\bibnamefont {Hwang}},\
  }\href {https://doi.org/10.1007/s11128-014-0757-3} {\bibfield  {journal}
  {\bibinfo  {journal} {Quantum information processing}\ }\textbf {\bibinfo
  {volume} {13}},\ \bibinfo {pages} {1639} (\bibinfo {year}
  {2014})}\BibitemShut {NoStop}%
\bibitem [{\citenamefont {Cirac}\ \emph {et~al.}(2021)\citenamefont {Cirac},
  \citenamefont {P{\'{e}}rez-Garc{\'{\i}}a}, \citenamefont {Schuch},\ and\
  \citenamefont {Verstraete}}]{Cirac_2021}%
  \BibitemOpen
  \bibfield  {author} {\bibinfo {author} {\bibfnamefont {J.~I.}\ \bibnamefont
  {Cirac}}, \bibinfo {author} {\bibfnamefont {D.}~\bibnamefont
  {P{\'{e}}rez-Garc{\'{\i}}a}}, \bibinfo {author} {\bibfnamefont
  {N.}~\bibnamefont {Schuch}},\ and\ \bibinfo {author} {\bibfnamefont
  {F.}~\bibnamefont {Verstraete}},\ }\href
  {https://doi.org/10.1103/revmodphys.93.045003} {\bibfield  {journal}
  {\bibinfo  {journal} {Reviews of Modern Physics}\ }\textbf {\bibinfo {volume}
  {93}},\ \bibinfo {pages} {045003} (\bibinfo {year} {2021})}\BibitemShut
  {NoStop}%
\bibitem [{\citenamefont {Pollmann}\ \emph {et~al.}(2010)\citenamefont
  {Pollmann}, \citenamefont {Turner}, \citenamefont {Berg},\ and\ \citenamefont
  {Oshikawa}}]{Pollmann2010}%
  \BibitemOpen
  \bibfield  {author} {\bibinfo {author} {\bibfnamefont {F.}~\bibnamefont
  {Pollmann}}, \bibinfo {author} {\bibfnamefont {A.~M.}\ \bibnamefont
  {Turner}}, \bibinfo {author} {\bibfnamefont {E.}~\bibnamefont {Berg}},\ and\
  \bibinfo {author} {\bibfnamefont {M.}~\bibnamefont {Oshikawa}},\ }\href
  {https://doi.org/10.1103/physrevb.81.064439} {\bibfield  {journal} {\bibinfo
  {journal} {Physical Review B}\ }\textbf {\bibinfo {volume} {81}},\ \bibinfo
  {pages} {064439} (\bibinfo {year} {2010})}\BibitemShut {NoStop}%
\bibitem [{\citenamefont {Else}\ \emph {et~al.}(2012)\citenamefont {Else},
  \citenamefont {Schwarz}, \citenamefont {Bartlett},\ and\ \citenamefont
  {Doherty}}]{Else2012}%
  \BibitemOpen
  \bibfield  {author} {\bibinfo {author} {\bibfnamefont {D.~V.}\ \bibnamefont
  {Else}}, \bibinfo {author} {\bibfnamefont {I.}~\bibnamefont {Schwarz}},
  \bibinfo {author} {\bibfnamefont {S.~D.}\ \bibnamefont {Bartlett}},\ and\
  \bibinfo {author} {\bibfnamefont {A.~C.}\ \bibnamefont {Doherty}},\ }\href
  {https://doi.org/10.1103/physrevlett.108.240505} {\bibfield  {journal}
  {\bibinfo  {journal} {Physical Review Letters}\ }\textbf {\bibinfo {volume}
  {108}},\ \bibinfo {pages} {240505} (\bibinfo {year} {2012})}\BibitemShut
  {NoStop}%
\bibitem [{\citenamefont {Stephen}\ \emph {et~al.}(2017)\citenamefont
  {Stephen}, \citenamefont {Wang}, \citenamefont {Prakash}, \citenamefont
  {Wei},\ and\ \citenamefont {Raussendorf}}]{Stephen2017}%
  \BibitemOpen
  \bibfield  {author} {\bibinfo {author} {\bibfnamefont {D.~T.}\ \bibnamefont
  {Stephen}}, \bibinfo {author} {\bibfnamefont {D.-S.}\ \bibnamefont {Wang}},
  \bibinfo {author} {\bibfnamefont {A.}~\bibnamefont {Prakash}}, \bibinfo
  {author} {\bibfnamefont {T.-C.}\ \bibnamefont {Wei}},\ and\ \bibinfo {author}
  {\bibfnamefont {R.}~\bibnamefont {Raussendorf}},\ }\href
  {https://doi.org/10.1103/physrevlett.119.010504} {\bibfield  {journal}
  {\bibinfo  {journal} {Physical Review Letters}\ }\textbf {\bibinfo {volume}
  {119}},\ \bibinfo {pages} {010504} (\bibinfo {year} {2017})}\BibitemShut
  {NoStop}%
\bibitem [{\citenamefont {Nielsen}(2006)}]{nielsen2006cluster}%
  \BibitemOpen
  \bibfield  {author} {\bibinfo {author} {\bibfnamefont {M.~A.}\ \bibnamefont
  {Nielsen}},\ }\href {https://doi.org/10.1016/S0034-4877(06)80014-5}
  {\bibfield  {journal} {\bibinfo  {journal} {Reports on Mathematical Physics}\
  }\textbf {\bibinfo {volume} {57}},\ \bibinfo {pages} {147} (\bibinfo {year}
  {2006})}\BibitemShut {NoStop}%
\bibitem [{\citenamefont {Wei}\ \emph {et~al.}(2012)\citenamefont {Wei},
  \citenamefont {Affleck},\ and\ \citenamefont {Raussendorf}}]{Wei2012a}%
  \BibitemOpen
  \bibfield  {author} {\bibinfo {author} {\bibfnamefont {T.-C.}\ \bibnamefont
  {Wei}}, \bibinfo {author} {\bibfnamefont {I.}~\bibnamefont {Affleck}},\ and\
  \bibinfo {author} {\bibfnamefont {R.}~\bibnamefont {Raussendorf}},\ }\href
  {https://doi.org/10.1103/physreva.86.032328} {\bibfield  {journal} {\bibinfo
  {journal} {Physical Review A}\ }\textbf {\bibinfo {volume} {86}},\ \bibinfo
  {pages} {032328} (\bibinfo {year} {2012})}\BibitemShut {NoStop}%
\bibitem [{\citenamefont {Chen}\ \emph {et~al.}(2013)\citenamefont {Chen},
  \citenamefont {Gu}, \citenamefont {Liu},\ and\ \citenamefont
  {Wen}}]{Chen2013}%
  \BibitemOpen
  \bibfield  {author} {\bibinfo {author} {\bibfnamefont {X.}~\bibnamefont
  {Chen}}, \bibinfo {author} {\bibfnamefont {Z.-C.}\ \bibnamefont {Gu}},
  \bibinfo {author} {\bibfnamefont {Z.-X.}\ \bibnamefont {Liu}},\ and\ \bibinfo
  {author} {\bibfnamefont {X.-G.}\ \bibnamefont {Wen}},\ }\href
  {https://doi.org/10.1103/physrevb.87.155114} {\bibfield  {journal} {\bibinfo
  {journal} {Physical Review B}\ }\textbf {\bibinfo {volume} {87}},\ \bibinfo
  {pages} {155114} (\bibinfo {year} {2013})}\BibitemShut {NoStop}%
\bibitem [{\citenamefont {Hagiwara}\ \emph {et~al.}(1990)\citenamefont
  {Hagiwara}, \citenamefont {Katsumata}, \citenamefont {Affleck}, \citenamefont
  {Halperin},\ and\ \citenamefont {Renard}}]{Hagiwara_1990}%
  \BibitemOpen
  \bibfield  {author} {\bibinfo {author} {\bibfnamefont {M.}~\bibnamefont
  {Hagiwara}}, \bibinfo {author} {\bibfnamefont {K.}~\bibnamefont {Katsumata}},
  \bibinfo {author} {\bibfnamefont {I.}~\bibnamefont {Affleck}}, \bibinfo
  {author} {\bibfnamefont {B.~I.}\ \bibnamefont {Halperin}},\ and\ \bibinfo
  {author} {\bibfnamefont {J.~P.}\ \bibnamefont {Renard}},\ }\href
  {https://doi.org/10.1103/physrevlett.65.3181} {\bibfield  {journal} {\bibinfo
   {journal} {Physical Review Letters}\ }\textbf {\bibinfo {volume} {65}},\
  \bibinfo {pages} {3181} (\bibinfo {year} {1990})}\BibitemShut {NoStop}%
\bibitem [{\citenamefont {Mishra}\ \emph {et~al.}(2021)\citenamefont {Mishra},
  \citenamefont {Catarina}, \citenamefont {Wu}, \citenamefont {Ortiz},
  \citenamefont {Jacob}, \citenamefont {Eimre}, \citenamefont {Ma},
  \citenamefont {Pignedoli}, \citenamefont {Feng}, \citenamefont {Ruffieux},
  \citenamefont {Fern{\'{a}}ndez-Rossier},\ and\ \citenamefont
  {Fasel}}]{Mishra_2021}%
  \BibitemOpen
  \bibfield  {author} {\bibinfo {author} {\bibfnamefont {S.}~\bibnamefont
  {Mishra}}, \bibinfo {author} {\bibfnamefont {G.}~\bibnamefont {Catarina}},
  \bibinfo {author} {\bibfnamefont {F.}~\bibnamefont {Wu}}, \bibinfo {author}
  {\bibfnamefont {R.}~\bibnamefont {Ortiz}}, \bibinfo {author} {\bibfnamefont
  {D.}~\bibnamefont {Jacob}}, \bibinfo {author} {\bibfnamefont
  {K.}~\bibnamefont {Eimre}}, \bibinfo {author} {\bibfnamefont
  {J.}~\bibnamefont {Ma}}, \bibinfo {author} {\bibfnamefont {C.~A.}\
  \bibnamefont {Pignedoli}}, \bibinfo {author} {\bibfnamefont {X.}~\bibnamefont
  {Feng}}, \bibinfo {author} {\bibfnamefont {P.}~\bibnamefont {Ruffieux}},
  \bibinfo {author} {\bibfnamefont {J.}~\bibnamefont
  {Fern{\'{a}}ndez-Rossier}},\ and\ \bibinfo {author} {\bibfnamefont
  {R.}~\bibnamefont {Fasel}},\ }\href
  {https://doi.org/10.1038/s41586-021-03842-3} {\bibfield  {journal} {\bibinfo
  {journal} {Nature}\ }\textbf {\bibinfo {volume} {598}},\ \bibinfo {pages}
  {287} (\bibinfo {year} {2021})}\BibitemShut {NoStop}%
\bibitem [{\citenamefont {{d}en Nijs}\ and\ \citenamefont
  {Rommelse}(1989)}]{Nijs1989}%
  \BibitemOpen
  \bibfield  {author} {\bibinfo {author} {\bibfnamefont {M.}~\bibnamefont
  {{d}en Nijs}}\ and\ \bibinfo {author} {\bibfnamefont {K.}~\bibnamefont
  {Rommelse}},\ }\href {https://doi.org/10.1103/physrevb.40.4709} {\bibfield
  {journal} {\bibinfo  {journal} {Physical Review B}\ }\textbf {\bibinfo
  {volume} {40}},\ \bibinfo {pages} {4709} (\bibinfo {year}
  {1989})}\BibitemShut {NoStop}%
\bibitem [{Note1()}]{Note1}%
  \BibitemOpen
  \bibinfo {note} {Because the bulk operators merely alter the sign of $\langle
  O_{\protect \textrm {str}}^z\rangle $ in the $S^z$ basis, the particular
  value of $-4/9$ can be understood as arising from the roughly equal
  probability for each edge spin-1 site to be in the $+1$, $0$, or $-1$
  eigenstate of $S^z$ when $N\gg 1$: out of the $9$ possible combinations, the
  four with $\pm 1$ at both edges contribute $-1$ to $\langle O_{\protect
  \textrm {str}}^z\rangle $. All others have a vanishing
  eigenvalue.}\BibitemShut {Stop}%
\bibitem [{\citenamefont {Schuch}\ \emph {et~al.}(2008)\citenamefont {Schuch},
  \citenamefont {Wolf}, \citenamefont {Verstraete},\ and\ \citenamefont
  {Cirac}}]{Schuch2007}%
  \BibitemOpen
  \bibfield  {author} {\bibinfo {author} {\bibfnamefont {N.}~\bibnamefont
  {Schuch}}, \bibinfo {author} {\bibfnamefont {M.~M.}\ \bibnamefont {Wolf}},
  \bibinfo {author} {\bibfnamefont {F.}~\bibnamefont {Verstraete}},\ and\
  \bibinfo {author} {\bibfnamefont {J.~I.}\ \bibnamefont {Cirac}},\ }\href
  {https://doi.org/10.1103/PhysRevLett.100.030504} {\bibfield  {journal}
  {\bibinfo  {journal} {Physical Review Letters}\ }\textbf {\bibinfo {volume}
  {100}},\ \bibinfo {pages} {030504} (\bibinfo {year} {2008})},\ \Eprint
  {https://arxiv.org/abs/0705.0292} {arXiv:0705.0292 [quant-ph]} \BibitemShut
  {NoStop}%
\bibitem [{\citenamefont {Or{\'u}s}(2014)}]{Orus2013}%
  \BibitemOpen
  \bibfield  {author} {\bibinfo {author} {\bibfnamefont {R.}~\bibnamefont
  {Or{\'u}s}},\ }\href {https://doi.org/10.1016/j.aop.2014.06.013} {\bibfield
  {journal} {\bibinfo  {journal} {Annals of physics}\ }\textbf {\bibinfo
  {volume} {349}},\ \bibinfo {pages} {117} (\bibinfo {year} {2014})},\ \Eprint
  {https://arxiv.org/abs/1306.2164} {arXiv:1306.2164 [cond-mat.str-el]}
  \BibitemShut {NoStop}%
\bibitem [{\citenamefont {Gopalakrishnan}\ and\ \citenamefont
  {Lamacraft}(2019)}]{Gopalakrishnan2019}%
  \BibitemOpen
  \bibfield  {author} {\bibinfo {author} {\bibfnamefont {S.}~\bibnamefont
  {Gopalakrishnan}}\ and\ \bibinfo {author} {\bibfnamefont {A.}~\bibnamefont
  {Lamacraft}},\ }\href {https://doi.org/10.1103/PhysRevB.100.064309}
  {\bibfield  {journal} {\bibinfo  {journal} {Physical Review B}\ }\textbf
  {\bibinfo {volume} {100}},\ \bibinfo {pages} {064309} (\bibinfo {year}
  {2019})},\ \Eprint {https://arxiv.org/abs/1903.11611v1} {1903.11611v1}
  \BibitemShut {NoStop}%
\bibitem [{\citenamefont {Foss-Feig}\ \emph {et~al.}(2022)\citenamefont
  {Foss-Feig}, \citenamefont {Ragole}, \citenamefont {Potter}, \citenamefont
  {Dreiling}, \citenamefont {Figgatt}, \citenamefont {Gaebler}, \citenamefont
  {Hall}, \citenamefont {Moses}, \citenamefont {Pino}, \citenamefont {Spaun},
  \citenamefont {Neyenhuis},\ and\ \citenamefont {Hayes}}]{Foss_Feig_2022}%
  \BibitemOpen
  \bibfield  {author} {\bibinfo {author} {\bibfnamefont {M.}~\bibnamefont
  {Foss-Feig}}, \bibinfo {author} {\bibfnamefont {S.}~\bibnamefont {Ragole}},
  \bibinfo {author} {\bibfnamefont {A.}~\bibnamefont {Potter}}, \bibinfo
  {author} {\bibfnamefont {J.}~\bibnamefont {Dreiling}}, \bibinfo {author}
  {\bibfnamefont {C.}~\bibnamefont {Figgatt}}, \bibinfo {author} {\bibfnamefont
  {J.}~\bibnamefont {Gaebler}}, \bibinfo {author} {\bibfnamefont
  {A.}~\bibnamefont {Hall}}, \bibinfo {author} {\bibfnamefont {S.}~\bibnamefont
  {Moses}}, \bibinfo {author} {\bibfnamefont {J.}~\bibnamefont {Pino}},
  \bibinfo {author} {\bibfnamefont {B.}~\bibnamefont {Spaun}}, \bibinfo
  {author} {\bibfnamefont {B.}~\bibnamefont {Neyenhuis}},\ and\ \bibinfo
  {author} {\bibfnamefont {D.}~\bibnamefont {Hayes}},\ }\href
  {https://doi.org/10.1103/physrevlett.128.150504} {\bibfield  {journal}
  {\bibinfo  {journal} {Physical Review Letters}\ }\textbf {\bibinfo {volume}
  {128}},\ \bibinfo {pages} {150504} (\bibinfo {year} {2022})}\BibitemShut
  {NoStop}%
\bibitem [{Note2()}]{Note2}%
  \BibitemOpen
  \bibinfo {note} {This is only true for for Bell measurement of memory qubits
  belonging to independent chains. If the edge memory qubits of a single chain
  are measured in the Bell basis, i.e. to enforce boundary conditions, the
  exact probability of each outcome is dependent on the length of the chain,
  though tends toward 1/4 for large $N$}\BibitemShut {NoStop}%
\bibitem [{\citenamefont {Kennedy}\ and\ \citenamefont
  {Tasaki}(1992)}]{Kennedy_1992}%
  \BibitemOpen
  \bibfield  {author} {\bibinfo {author} {\bibfnamefont {T.}~\bibnamefont
  {Kennedy}}\ and\ \bibinfo {author} {\bibfnamefont {H.}~\bibnamefont
  {Tasaki}},\ }\href {https://doi.org/10.1103/physrevb.45.304} {\bibfield
  {journal} {\bibinfo  {journal} {Physical Review B}\ }\textbf {\bibinfo
  {volume} {45}},\ \bibinfo {pages} {304} (\bibinfo {year} {1992})}\BibitemShut
  {NoStop}%
\bibitem [{Note3()}]{Note3}%
  \BibitemOpen
  \bibinfo {note} {While $B^\dagger = B$ in this particular case, we leave the
  Hermitian adjoint intact for generality, as this strategy is potentially
  applicable to non-Pauli defects in preparation of other SPTs.}\BibitemShut
  {Stop}%
\bibitem [{SM()}]{SM}%
  \BibitemOpen
  \href@noop {} {}\bibinfo {note} {See Supplemental Material for explicit
  preparation circuits, post-processing details, and additional experimental
  data.}\BibitemShut {Stop}%
\bibitem [{Note4()}]{Note4}%
  \BibitemOpen
  \bibinfo {note} {Implementation of the SWAP test involves a Toffoli gate, and
  thus adds significant overhead to the state preparation circuit.}\BibitemShut
  {Stop}%
\bibitem [{\citenamefont {Nation}\ \emph {et~al.}(2021)\citenamefont {Nation},
  \citenamefont {Kang}, \citenamefont {Sundaresan},\ and\ \citenamefont
  {Gambetta}}]{Nation2021}%
  \BibitemOpen
  \bibfield  {author} {\bibinfo {author} {\bibfnamefont {P.~D.}\ \bibnamefont
  {Nation}}, \bibinfo {author} {\bibfnamefont {H.}~\bibnamefont {Kang}},
  \bibinfo {author} {\bibfnamefont {N.}~\bibnamefont {Sundaresan}},\ and\
  \bibinfo {author} {\bibfnamefont {J.~M.}\ \bibnamefont {Gambetta}},\ }\href
  {https://doi.org/10.1103/prxquantum.2.040326} {\bibfield  {journal} {\bibinfo
   {journal} {{PRX} Quantum}\ }\textbf {\bibinfo {volume} {2}},\ \bibinfo
  {pages} {040326} (\bibinfo {year} {2021})}\BibitemShut {NoStop}%
\bibitem [{\citenamefont {Arute}\ \emph {et~al.}(2020)\citenamefont {Arute},
  \citenamefont {Arya}, \citenamefont {Babbush}, \citenamefont {Bacon},
  \citenamefont {Bardin}, \citenamefont {Barends}, \citenamefont {Boixo},
  \citenamefont {Broughton}, \citenamefont {Buckley}, \citenamefont {Buelland}
  \emph {et~al.}}]{google2020hartree}%
  \BibitemOpen
  \bibfield  {author} {\bibinfo {author} {\bibfnamefont {F.}~\bibnamefont
  {Arute}}, \bibinfo {author} {\bibfnamefont {K.}~\bibnamefont {Arya}},
  \bibinfo {author} {\bibfnamefont {R.}~\bibnamefont {Babbush}}, \bibinfo
  {author} {\bibfnamefont {D.}~\bibnamefont {Bacon}}, \bibinfo {author}
  {\bibfnamefont {J.~C.}\ \bibnamefont {Bardin}}, \bibinfo {author}
  {\bibfnamefont {R.}~\bibnamefont {Barends}}, \bibinfo {author} {\bibfnamefont
  {S.}~\bibnamefont {Boixo}}, \bibinfo {author} {\bibfnamefont
  {M.}~\bibnamefont {Broughton}}, \bibinfo {author} {\bibfnamefont {B.~B.}\
  \bibnamefont {Buckley}}, \bibinfo {author} {\bibfnamefont {D.~A.}\
  \bibnamefont {Buelland}}, \emph {et~al.},\ }\href
  {https://doi.org/10.1126/science.abb9811} {\bibfield  {journal} {\bibinfo
  {journal} {Science}\ }\textbf {\bibinfo {volume} {369}},\ \bibinfo {pages}
  {1084} (\bibinfo {year} {2020})}\BibitemShut {NoStop}%
\bibitem [{\citenamefont {McWeeny}(1960)}]{mcweeny1960some}%
  \BibitemOpen
  \bibfield  {author} {\bibinfo {author} {\bibfnamefont {R.}~\bibnamefont
  {McWeeny}},\ }\href {https://doi.org/10.1103/RevModPhys.32.335} {\bibfield
  {journal} {\bibinfo  {journal} {Reviews of Modern Physics}\ }\textbf
  {\bibinfo {volume} {32}},\ \bibinfo {pages} {335} (\bibinfo {year}
  {1960})}\BibitemShut {NoStop}%
\bibitem [{Note5()}]{Note5}%
  \BibitemOpen
  \bibinfo {note} {For more details, see \protect \url
  {https://quantum-computing.ibm.com/lab/docs/iql/runtime/}}\BibitemShut
  {NoStop}%
\bibitem [{\citenamefont {Chen}\ \emph {et~al.}(2019)\citenamefont {Chen},
  \citenamefont {Farahzad}, \citenamefont {Yoo},\ and\ \citenamefont
  {Wei}}]{Chen2019a}%
  \BibitemOpen
  \bibfield  {author} {\bibinfo {author} {\bibfnamefont {Y.}~\bibnamefont
  {Chen}}, \bibinfo {author} {\bibfnamefont {M.}~\bibnamefont {Farahzad}},
  \bibinfo {author} {\bibfnamefont {S.}~\bibnamefont {Yoo}},\ and\ \bibinfo
  {author} {\bibfnamefont {T.-C.}\ \bibnamefont {Wei}},\ }\href
  {https://doi.org/10.1103/physreva.100.052315} {\bibfield  {journal} {\bibinfo
   {journal} {Physical Review A}\ }\textbf {\bibinfo {volume} {100}},\ \bibinfo
  {pages} {052315} (\bibinfo {year} {2019})}\BibitemShut {NoStop}%
\bibitem [{\citenamefont {Bravyi}\ \emph {et~al.}(2021)\citenamefont {Bravyi},
  \citenamefont {Sheldon}, \citenamefont {Kandala}, \citenamefont {Mckay},\
  and\ \citenamefont {Gambetta}}]{Bravyi2021}%
  \BibitemOpen
  \bibfield  {author} {\bibinfo {author} {\bibfnamefont {S.}~\bibnamefont
  {Bravyi}}, \bibinfo {author} {\bibfnamefont {S.}~\bibnamefont {Sheldon}},
  \bibinfo {author} {\bibfnamefont {A.}~\bibnamefont {Kandala}}, \bibinfo
  {author} {\bibfnamefont {D.~C.}\ \bibnamefont {Mckay}},\ and\ \bibinfo
  {author} {\bibfnamefont {J.~M.}\ \bibnamefont {Gambetta}},\ }\href
  {https://doi.org/10.1103/physreva.103.042605} {\bibfield  {journal} {\bibinfo
   {journal} {Physical Review A}\ }\textbf {\bibinfo {volume} {103}},\ \bibinfo
  {pages} {042605} (\bibinfo {year} {2021})}\BibitemShut {NoStop}%
\bibitem [{\citenamefont {Li}\ and\ \citenamefont {Haldane}(2008)}]{Li_2008}%
  \BibitemOpen
  \bibfield  {author} {\bibinfo {author} {\bibfnamefont {H.}~\bibnamefont
  {Li}}\ and\ \bibinfo {author} {\bibfnamefont {F.~D.~M.}\ \bibnamefont
  {Haldane}},\ }\href {https://doi.org/10.1103/physrevlett.101.010504}
  {\bibfield  {journal} {\bibinfo  {journal} {Physical Review Letters}\
  }\textbf {\bibinfo {volume} {101}},\ \bibinfo {pages} {010504} (\bibinfo
  {year} {2008})}\BibitemShut {NoStop}%
\bibitem [{\citenamefont {Fan}\ \emph {et~al.}(2004)\citenamefont {Fan},
  \citenamefont {Korepin},\ and\ \citenamefont {Roychowdhury}}]{Fan2004}%
  \BibitemOpen
  \bibfield  {author} {\bibinfo {author} {\bibfnamefont {H.}~\bibnamefont
  {Fan}}, \bibinfo {author} {\bibfnamefont {V.}~\bibnamefont {Korepin}},\ and\
  \bibinfo {author} {\bibfnamefont {V.}~\bibnamefont {Roychowdhury}},\ }\href
  {https://doi.org/10.1103/physrevlett.93.227203} {\bibfield  {journal}
  {\bibinfo  {journal} {Physical Review Letters}\ }\textbf {\bibinfo {volume}
  {93}},\ \bibinfo {pages} {227203} (\bibinfo {year} {2004})}\BibitemShut
  {NoStop}%
\bibitem [{\citenamefont {Fan}\ \emph {et~al.}(2007)\citenamefont {Fan},
  \citenamefont {Korepin}, \citenamefont {Roychowdhury}, \citenamefont
  {Hadley},\ and\ \citenamefont {Bose}}]{Fan_2007}%
  \BibitemOpen
  \bibfield  {author} {\bibinfo {author} {\bibfnamefont {H.}~\bibnamefont
  {Fan}}, \bibinfo {author} {\bibfnamefont {V.}~\bibnamefont {Korepin}},
  \bibinfo {author} {\bibfnamefont {V.}~\bibnamefont {Roychowdhury}}, \bibinfo
  {author} {\bibfnamefont {C.}~\bibnamefont {Hadley}},\ and\ \bibinfo {author}
  {\bibfnamefont {S.}~\bibnamefont {Bose}},\ }\href
  {https://doi.org/10.1103/physrevb.76.014428} {\bibfield  {journal} {\bibinfo
  {journal} {Physical Review B}\ }\textbf {\bibinfo {volume} {76}},\ \bibinfo
  {pages} {014428} (\bibinfo {year} {2007})}\BibitemShut {NoStop}%
\bibitem [{\citenamefont {Geraedts}\ and\ \citenamefont
  {S{\o}rensen}(2010)}]{Geraedts_2010}%
  \BibitemOpen
  \bibfield  {author} {\bibinfo {author} {\bibfnamefont {S.~D.}\ \bibnamefont
  {Geraedts}}\ and\ \bibinfo {author} {\bibfnamefont {E.~S.}\ \bibnamefont
  {S{\o}rensen}},\ }\href {https://doi.org/10.1088/1751-8113/43/18/185304}
  {\bibfield  {journal} {\bibinfo  {journal} {Journal of Physics A:
  Mathematical and Theoretical}\ }\textbf {\bibinfo {volume} {43}},\ \bibinfo
  {pages} {185304} (\bibinfo {year} {2010})}\BibitemShut {NoStop}%
\bibitem [{\citenamefont {Choo}\ \emph {et~al.}(2018)\citenamefont {Choo},
  \citenamefont {von Keyserlingk}, \citenamefont {Regnault},\ and\
  \citenamefont {Neupert}}]{Choo2018}%
  \BibitemOpen
  \bibfield  {author} {\bibinfo {author} {\bibfnamefont {K.}~\bibnamefont
  {Choo}}, \bibinfo {author} {\bibfnamefont {C.~W.}\ \bibnamefont {von
  Keyserlingk}}, \bibinfo {author} {\bibfnamefont {N.}~\bibnamefont
  {Regnault}},\ and\ \bibinfo {author} {\bibfnamefont {T.}~\bibnamefont
  {Neupert}},\ }\href {https://doi.org/10.1103/physrevlett.121.086808}
  {\bibfield  {journal} {\bibinfo  {journal} {Physical Review Letters}\
  }\textbf {\bibinfo {volume} {121}},\ \bibinfo {pages} {086808} (\bibinfo
  {year} {2018})}\BibitemShut {NoStop}%
\bibitem [{\citenamefont {C{\'{o}}rcoles}\ \emph {et~al.}(2021)\citenamefont
  {C{\'{o}}rcoles}, \citenamefont {Takita}, \citenamefont {Inoue},
  \citenamefont {Lekuch}, \citenamefont {Minev}, \citenamefont {Chow},\ and\
  \citenamefont {Gambetta}}]{Corcoles_2021}%
  \BibitemOpen
  \bibfield  {author} {\bibinfo {author} {\bibfnamefont {A.~D.}\ \bibnamefont
  {C{\'{o}}rcoles}}, \bibinfo {author} {\bibfnamefont {M.}~\bibnamefont
  {Takita}}, \bibinfo {author} {\bibfnamefont {K.}~\bibnamefont {Inoue}},
  \bibinfo {author} {\bibfnamefont {S.}~\bibnamefont {Lekuch}}, \bibinfo
  {author} {\bibfnamefont {Z.~K.}\ \bibnamefont {Minev}}, \bibinfo {author}
  {\bibfnamefont {J.~M.}\ \bibnamefont {Chow}},\ and\ \bibinfo {author}
  {\bibfnamefont {J.~M.}\ \bibnamefont {Gambetta}},\ }\href
  {https://doi.org/10.1103/physrevlett.127.100501} {\bibfield  {journal}
  {\bibinfo  {journal} {Physical Review Letters}\ }\textbf {\bibinfo {volume}
  {127}},\ \bibinfo {pages} {100501} (\bibinfo {year} {2021})}\BibitemShut
  {NoStop}%
\bibitem [{\citenamefont {Gaebler}\ \emph {et~al.}(2021)\citenamefont
  {Gaebler}, \citenamefont {Baldwin}, \citenamefont {Moses}, \citenamefont
  {Dreiling}, \citenamefont {Figgatt}, \citenamefont {Foss-Feig}, \citenamefont
  {Hayes},\ and\ \citenamefont {Pino}}]{Gaebler2021}%
  \BibitemOpen
  \bibfield  {author} {\bibinfo {author} {\bibfnamefont {J.~P.}\ \bibnamefont
  {Gaebler}}, \bibinfo {author} {\bibfnamefont {C.~H.}\ \bibnamefont
  {Baldwin}}, \bibinfo {author} {\bibfnamefont {S.~A.}\ \bibnamefont {Moses}},
  \bibinfo {author} {\bibfnamefont {J.~M.}\ \bibnamefont {Dreiling}}, \bibinfo
  {author} {\bibfnamefont {C.}~\bibnamefont {Figgatt}}, \bibinfo {author}
  {\bibfnamefont {M.}~\bibnamefont {Foss-Feig}}, \bibinfo {author}
  {\bibfnamefont {D.}~\bibnamefont {Hayes}},\ and\ \bibinfo {author}
  {\bibfnamefont {J.~M.}\ \bibnamefont {Pino}},\ }\href
  {https://doi.org/10.1103/PhysRevA.104.062440} {\bibfield  {journal} {\bibinfo
   {journal} {Physical Review A}\ }\textbf {\bibinfo {volume} {104}},\ \bibinfo
  {pages} {062440} (\bibinfo {year} {2021})},\ \Eprint
  {https://arxiv.org/abs/2108.10932} {arXiv:2108.10932 [quant-ph]} \BibitemShut
  {NoStop}%
\bibitem [{\citenamefont {Gross}\ \emph {et~al.}(2007)\citenamefont {Gross},
  \citenamefont {Eisert}, \citenamefont {Schuch},\ and\ \citenamefont
  {Perez-Garcia}}]{Gross2007}%
  \BibitemOpen
  \bibfield  {author} {\bibinfo {author} {\bibfnamefont {D.}~\bibnamefont
  {Gross}}, \bibinfo {author} {\bibfnamefont {J.}~\bibnamefont {Eisert}},
  \bibinfo {author} {\bibfnamefont {N.}~\bibnamefont {Schuch}},\ and\ \bibinfo
  {author} {\bibfnamefont {D.}~\bibnamefont {Perez-Garcia}},\ }\href
  {https://doi.org/10.1103/physreva.76.052315} {\bibfield  {journal} {\bibinfo
  {journal} {Physical Review A}\ }\textbf {\bibinfo {volume} {76}},\ \bibinfo
  {pages} {052315} (\bibinfo {year} {2007})}\BibitemShut {NoStop}%
\bibitem [{\citenamefont {Gross}\ and\ \citenamefont
  {Eisert}(2010)}]{Gross_2010}%
  \BibitemOpen
  \bibfield  {author} {\bibinfo {author} {\bibfnamefont {D.}~\bibnamefont
  {Gross}}\ and\ \bibinfo {author} {\bibfnamefont {J.}~\bibnamefont {Eisert}},\
  }\href {https://doi.org/10.1103/physreva.82.040303} {\bibfield  {journal}
  {\bibinfo  {journal} {Physical Review A}\ }\textbf {\bibinfo {volume} {82}},\
  \bibinfo {pages} {040303} (\bibinfo {year} {2010})}\BibitemShut {NoStop}%
\bibitem [{Note6()}]{Note6}%
  \BibitemOpen
  \bibinfo {note} {We note that while estimated uncertainties are smaller than
  would be expected for $10^5$ samples per circuit, this is a consequence of
  McWeeny purification. In contrast, we find that uncertainties for raw
  fidelities (computed using impure density matrices) scale roughly as
  $1/\protect \sqrt {N_{\protect \textrm {shots}}}$, as expected.}\BibitemShut
  {Stop}%
\bibitem [{Note7()}]{Note7}%
  \BibitemOpen
  \bibinfo {note} {We note that error reduction of a similar scale ($\sim 1\%$
  error) has been realized by combining post-selection and McWeeny purification
  in other experiments. See, for example, Ref. \cite
  {google2020hartree}}\BibitemShut {NoStop}%
\bibitem [{\citenamefont {Wei}\ \emph {et~al.}(2022{\natexlab{b}})\citenamefont
  {Wei}, \citenamefont {Raussendorf},\ and\ \citenamefont {Affleck}}]{Wei2022}%
  \BibitemOpen
  \bibfield  {author} {\bibinfo {author} {\bibfnamefont {T.-C.}\ \bibnamefont
  {Wei}}, \bibinfo {author} {\bibfnamefont {R.}~\bibnamefont {Raussendorf}},\
  and\ \bibinfo {author} {\bibfnamefont {I.}~\bibnamefont {Affleck}},\ }\Eprint
  {https://arxiv.org/abs/2201.09307} {arXiv:2201.09307 [cond-mat.str-el]}
  (\bibinfo {year} {2022}{\natexlab{b}})\BibitemShut {NoStop}%
\bibitem [{\citenamefont {Browne}\ and\ \citenamefont
  {Rudolph}(2005)}]{Browne_2005}%
  \BibitemOpen
  \bibfield  {author} {\bibinfo {author} {\bibfnamefont {D.~E.}\ \bibnamefont
  {Browne}}\ and\ \bibinfo {author} {\bibfnamefont {T.}~\bibnamefont
  {Rudolph}},\ }\href {https://doi.org/10.1103/physrevlett.95.010501}
  {\bibfield  {journal} {\bibinfo  {journal} {Physical Review Letters}\
  }\textbf {\bibinfo {volume} {95}},\ \bibinfo {pages} {010501} (\bibinfo
  {year} {2005})}\BibitemShut {NoStop}%
\bibitem [{\citenamefont {Li}\ \emph {et~al.}(2015)\citenamefont {Li},
  \citenamefont {Humphreys}, \citenamefont {Mendoza},\ and\ \citenamefont
  {Benjamin}}]{Li_2015}%
  \BibitemOpen
  \bibfield  {author} {\bibinfo {author} {\bibfnamefont {Y.}~\bibnamefont
  {Li}}, \bibinfo {author} {\bibfnamefont {P.~C.}\ \bibnamefont {Humphreys}},
  \bibinfo {author} {\bibfnamefont {G.~J.}\ \bibnamefont {Mendoza}},\ and\
  \bibinfo {author} {\bibfnamefont {S.~C.}\ \bibnamefont {Benjamin}},\ }\href
  {https://doi.org/10.1103/physrevx.5.041007} {\bibfield  {journal} {\bibinfo
  {journal} {Physical Review X}\ }\textbf {\bibinfo {volume} {5}},\ \bibinfo
  {pages} {041007} (\bibinfo {year} {2015})}\BibitemShut {NoStop}%
\bibitem [{\citenamefont {Quek}\ \emph {et~al.}(2022)\citenamefont {Quek},
  \citenamefont {Kaur},\ and\ \citenamefont {Wilde}}]{Quek2022}%
  \BibitemOpen
  \bibfield  {author} {\bibinfo {author} {\bibfnamefont {Y.}~\bibnamefont
  {Quek}}, \bibinfo {author} {\bibfnamefont {E.}~\bibnamefont {Kaur}},\ and\
  \bibinfo {author} {\bibfnamefont {M.~M.}\ \bibnamefont {Wilde}},\ }\Eprint
  {https://arxiv.org/abs/2206.15405} {arXiv:2206.15405 [quant-ph]}  (\bibinfo
  {year} {2022})\BibitemShut {NoStop}%
\end{thebibliography}
\end{document}